\documentclass[prx,twocolumn,longbibliography,superscriptaddress,preprintnumbers]{revtex4-2}
\usepackage[colorlinks,bookmarks=true,citecolor=blue,linkcolor=blue,urlcolor=blue]{hyperref}
\setcitestyle{numbers,square}
\usepackage{amsmath,amssymb,amsthm,amsfonts}
\usepackage{tabularx}
\usepackage{graphicx}
\usepackage{bmpsize}
\usepackage{bm}
\usepackage{color}
\usepackage[version=4]{mhchem}
\usepackage{mathtools,xcolor}
\usepackage[normalem]{ulem}
\usepackage{braket}
\usepackage{siunitx}

\usepackage[T1]{fontenc}
\newcommand{\nn}{\nonumber}
\def\diff{\mathrm d}
\def\sgn{\mathrm {sgn}}

\newcommand{\ov}{\overline}

\renewcommand{\v}[1]{\bm{#1}}

\DeclarePairedDelimiter\abs{\lvert}{\rvert}%

\def\br{{\bm{r}}}

\def\bk{{\bm{k}}}

\def\T{{\mathcal{T}}}
\def\P{{\mathcal{P}}}

\newcommand{\PRLsec}[1]{\textbf{\emph{#1---}}}

\begin{document}
\title{Four Moir\'e materials at One Magic Angle in Helical Quadrilayer Graphene }
\author{Manato Fujimoto}
\affiliation{Department of Physics, Harvard University, Cambridge, Massachusetts 02138, USA}
\affiliation{Department of Applied Physics, The University of Tokyo, Hongo, Tokyo, 113-8656, Japan}
\author{Naoto Nakatsuji}
\affiliation{Department of Physics, Osaka University, Toyonaka, Osaka 563-0024, Japan}
\affiliation{Department of Physics and Astronomy, Stony Brook University, Stony Brook, NY 11794, USA}
\author{Ashvin Vishwanath}
\affiliation{Department of Physics, Harvard University, Cambridge, Massachusetts 02138, USA}
\author{Patrick Ledwith}
\affiliation{Department of Physics, Harvard University, Cambridge, Massachusetts 02138, USA}
\affiliation{Department of Physics, Massachusetts Institute of Technology, Cambridge, Massachusetts 02139, USA}
\date{\today}

\begin{abstract}

We introduce helical twisted quadrilayer graphene (HTQG)—four graphene sheets rotated by the same small angle—as a versatile and experimentally accessible platform for correlated topological matter. HTQG consists of three moir\'e lattices, formed by interference between adjacent graphene layers, that are twisted relative to each other. Lattice relaxation produces four types of large-scale commensurate domains. The domains are characterized by the stacking of the three moir\'e lattices and come in two types: Type-I ``Bernal" stacking and Type-II ``rhombohedral" stacking.
Domain walls between adjacent stackings often host topologically protected edge states, forming networks at the supermoiré and super-supermoiré scales. Remarkably, all four moiré substructures have narrow bands at the same magic angle $\theta \approx 2.3^\circ$, allowing their correlated phases to be simultaneously targeted in device manufacturing. We argue that the Type-I domains are especially suitable for realizing robust superconductivity which emerges from doping topological insulators, and Chern insulators in $C = \pm 2$ bands.

\end{abstract}

\maketitle

\begin{figure*}
\begin{center}
\includegraphics[width=1\hsize]{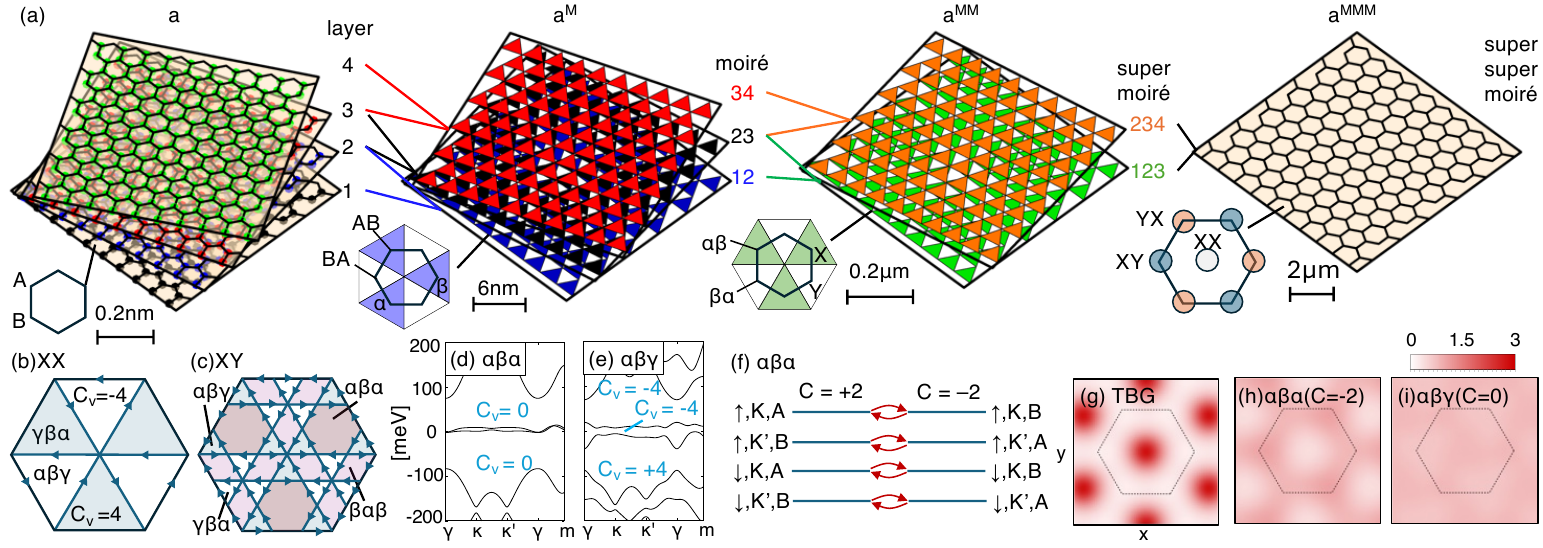}
\caption{(a) Schematic overview of the four graphene layers and their associated moir\'e scales. From left to right: the atomic lattice scale ( $\sim 0.2\,\mathrm{nm} $), the moir\'e scale  $a^M $ (typical TBG period of  $\sim6\,\mathrm{nm} $), the supermoir\'e scale  $a^{MM} $ for HTTG ( $\sim0.1\,\mathrm{\mu m} $), and the super‐supermoir\'e scale  $a^{MMM} $ for HTQG (up to a few micrometers).  
(b,c) Local domains in the XX super-supermoir\'e region are Type-II and carry valley Chern numbers $C_v=+4$ and $-4$ for $\alpha\beta\gamma $ and  $\gamma\beta\alpha $ respectively at charge neutrality. XY includes four domains, $\alpha\beta\gamma $, $\gamma\beta\alpha $, $\alpha\beta\alpha$ and $\beta\alpha\beta$. The arrows indicate the topological chiral boundary modes associated with one of the valleys.
The band structure for and (d)$\alpha\beta\alpha$ and (e)$\alpha\beta\gamma$ are depicted alongside the valley Chern numbers corresponding to the single particle gaps. The valence band of $\alpha \beta \gamma$ has $C=-2$ in the $K$ valley leading to a change in valley Chern number of $C_K-C_{K'}=-8$ when accounting for spin degeneracy.
(f) The Chern-sublattice basis for the narrow bands in the Type-I domains. The single particle dispersion tunnels between symmetry-related Chern sectors, as in TBG.
The real space charge density of the narrow bands is depicted for (g)TBG, (h) $\alpha\beta\alpha$ and (i)$\alpha\beta\gamma$.
}
\label{fig_htqg_summary}
\end{center}
\end{figure*}


Stacking and twisting atomically thin crystals create long-wavelength moiré patterns that provide a wide range of platforms that host
strongly correlated phases. The discovery of correlated insulators and superconductivity in magic angle twisted bilayer graphene (TBG)\cite{cao2018unconventional,cao2018correlated,yankowitz2019tuning,lu2019superconductors,stepanov2020untying,saito2020independent} provided a particularly remarkable starting point. Subsequently, anomalous Hall\cite{sharpe2019emergent,serlin2020intrinsic,nuckolls2020strongly,saito2021hofstadter,choi2021correlation,wu2021chern,pierce2021unconventional} and fractional Chern insulators\cite{xie2021fractional,finneyExtendedFractionalChern2025} have also been observed in TBG aligned to hBN. 
Topological states in TBG are quite fragile, however, due to the concentrated charge density at the AA sites. The concentrated charge density widens the bands through a large Hartree potential\cite{Guineaelectrostatic2018,ardemakercharge2019,ceapinning2019,goodwin2020hartree,Kangcascades2021,pierce2021unconventional,parker2021field} and helps stabilize competing non-topological states, such as a nematic semimetal\cite{PhysRevResearch.3.013033,PhysRevLett.127.027601,PhysRevB.102.205111,bocarslyCoulombInteractionsMigrating2025} and intervalley Kekulé spiral\cite{PhysRevX.11.041063,PhysRevB.108.235128,nuckolls2023quantum,kimImagingIntervalleyCoherent2023}, in the presence of weak heterostrain.

The fragility of topological states in TBG motivated the introduction and realization of helical twisted trilayer graphene (HTTG)\cite{httg_devakul,PhysRevLett.123.026402,PhysRevB.101.224107,PhysRevLett.125.116404,PhysRevLett.127.166802,PhysRevB.107.125423,PhysRevB.107.235425,uri2023superconductivity,becker2023chiral,PhysRevB.108.L081124,PhysRevX.13.041007,PhysRevResearch.5.043079,hao2024robust,PhysRevResearch.6.013165,PhysRevB.110.195112,park2024tunable,PhysRevB.109.125141,craig2024local,PhysRevResearch.6.L022025,PhysRevB.109.205411,PhysRevB.110.075417,min_Evolution,PhysRevB.110.115404,PhysRevB.110.115434,kwan2024fractional,niu2025quantumanomaloushalleffects,hoke2024imaging,xia2025topological} in which three graphene layers are stacked such that each layer is rotated by the same angle relative to the layer below. The three layer helical twist leads to a supermoiré pattern, formed by the relative twisting between the moiré lattice of first and second layers and that of the second and third layers. Lattice relaxation organizes the supermoir\'e into large, $\sim 100$nm, triangular domains related by $C_{2z}$\cite{httg_devakul,PhysRevX.13.041007}. These domains, labeled $\alpha \beta$ and $\beta \alpha$, feature Bernal-type relative displacement between the AA sites of the bilayer moir\'e lattices. The domains host narrow topological bands with distinct valley Chern numbers. The band insulating states of the domain walls therefore produce a network of topologically protected edge states along the domain boundaries, consistent with measurements of local compressibility\cite{hoke2024imaging}. Furthermore AA site displacement smooths out the charge density, thereby leading to narrow bands across the entire doping range\cite{PhysRevB.109.125141} in contrast to TBG. This in turn leads to robust interaction induced topological insulators \cite{httg_devakul,PhysRevB.109.125141}. These states imprint on global transport, leading to anomalous Hall conductance \cite{xia2025topological}, but they can also be probed locally\cite{hoke2024imaging}. However, the symmetry of the HTTG domains is similar to that of hBN-aligned TBG, twisted monolayer-bilayer\cite{polshyn2020electrical,chen2021electrically,he2021competing,zhang2023local,peng2024abundant}, and twisted double bilayer\cite{kuiri2022spontaneous, he2023symmetry}. These systems break $C_{2z}, C_{2x}$ and feature weak or nonexistent superconductivity relative to unaligned TBG\cite{cao2018unconventional,cao2018correlated,yankowitz2019tuning,lu2019superconductors,stepanov2020untying,saito2020independent} and alternating twist multilayers~\cite{khalafMagicAngleHierarchy2019,park2021tunable,hao2021electric,cao2021pauli,park2022robust,zhang2022promotion}. Is it possible to combine advantages of HTTG, uniform charge density and more robust topological states, with the symmetry and topology of TBG?

In this work, we introduce helical twisted quadrilayer graphene (HTQG).  HTQG consists of four graphene sheets with identical successive twists. Like HTTG, lattice relaxation produces large-scale domains in which the the moir\'e lattices formed by adjacent layer pairs are commensurately stacked. We find four types of domains that come in two types. Type-I ($\alpha \beta \alpha, \beta \alpha \beta)$ consists of Bernal stacked moir\'e lattices whereas Type-II ($\alpha\beta\gamma, \gamma \beta \alpha$) consists of rhombohedrally stacked moir\'e lattices. Remarkably, all of these subsystems host narrow bands at the same magic angle $\theta \approx 2.25^\circ$ such that they can be simultaneously targeted when manufacturing a device. 
The Type-I domains realize two nearly flat bands per spin per valley with the same essential symmetries as TBG but doubled Chern number, $C=\pm2$. 
We argue that these symmetries favor a particular class of topological insulators which are especially promising parent states for superconductivity. The Type-II systems feature narrow $C=0$ bands with quantum geometry well beyond a Hubbard-like description. In all cases the real-space charge density is very homogeneous such that the bands should remain narrow and the correlated insulators should remain stable.

\PRLsec{Lattice structure} 
\label{sec_lattice}
We begin by briefly reviewing aspects of TBG, HTTG, and then extending to HTQG. In TBG, a small twist angle between two graphene layers creates a moir\'e period  $a^M \gg a $, where  $a $ is the monolayer lattice constant. By contrast, HTTG consists of three sequentially rotated layers (a “helical twist”) that induce two incommensurate moir\'e superlattices, which combine into a still larger supermoir\'e period  $a^{MM}\gg a^M $ on the order of hundreds of nanometers. One can view this supermoir\'e as arising from the fact that the moir\'e lattice formed by layers12 is itself twisted relative to that of layers23 by angle  $\theta $. Lattice relaxation in HTTG yields two primary “commensurate” domain structures, which have also been experimentally imaged\cite{hoke2024imaging}.

\begin{table}
\begin{tabular}{ c | c | c  }
 2 layers & 3 layers & 4 layers\\ 
\hline 
 AB$=\alpha$ & $\alpha \beta=$X & Type-II $\in$ XX, Types I,II $\in$ XY, YX\\
 $BA=\beta$ & $\beta \alpha = Y$ & Type-I$=(\alpha\beta\alpha,\beta\alpha\beta)$, Type-II$=(\alpha\beta\gamma,\gamma\beta\alpha)$.
\end{tabular}
\caption{Summary of lattice stacking domains formed by lattice relaxation in equal-angle helically twisted graphene.}
\label{tab:stackingnotation}
\end{table}

To explain these ideas, it is helpful to define domain conventions, summarized in Table \ref{tab:stackingnotation}, by starting from the two‐layer scale and building up to HTQG. In TBG, the high‐symmetry moir\'e points are  $\mathcal{C}_6 $‐symmetric AA (perfect alignment) and  $\mathcal{C}_3 $‐symmetric AB/BA (displaced by bond distance). Under lattice relaxation, the large AB and BA domains expand, whereas the AA regions shrink (see Fig.~\ref{fig_htqg_summary}(b)). We label the resulting moir\'e‐scale “AB” domain as “ $\alpha $” and the large “BA” domain as “ $\beta $.”  
In HTTG, the relaxed lattice becomes a patchwork of two supermoir\'e‐scale domains,  $\alpha\beta $ and  $\beta\alpha $, which we call “X” and “Y,” respectively;  $\alpha\alpha $ is diminished by relaxation.

Moving to HTQG, one has two distinct supermoir\'e patterns (from layers123 and layers234) whose interference yields a super‐supermoir\'e period, $a^{MMM}>>a^{MM}$. As shown in Fig.~\ref{fig_htqg_summary}(c,d), three high‐symmetry stackings are possible: “XX,” “XY,” and “YX.” The label XX corresponds to  $\mathcal{C}_6 $ centers (analogous to AA or  $\alpha\alpha $), while XY and YX correspond to  $\mathcal{C}_3 $ centers (akin to AB/BA or  $\alpha\beta/\beta\alpha $). Locally within XX, one finds “ $\alpha\beta\gamma $” and “ $\gamma\beta\alpha $” domains, whereas XY or YX host “ $\alpha\beta\alpha $,” “ $\beta\alpha\beta $,” “ $\alpha\beta\gamma $,” and “ $\gamma\beta\alpha $.” Here,  $\alpha\beta\gamma $ corresponds to ABC stacking in trilayer graphene.

As observed in TBG, lattice relaxation plays a critical role in the small twist angle regime. In-plane lattice relaxation results in the formation of AB/BA domain structures, enhancing the gap in the electronic energy spectrum. For the scale of supermoir\'e period, lattice relaxation becomes even more significant because the displacement of carbon atoms $\delta$ alters the lattice structure with an impact scaling as $O(\delta/\theta^2)$.For instance, in HTTG, lattice reconstruction reduces the area of $\alpha\alpha$ while enlarging the area of $\alpha\beta$ and $\beta\alpha$, resulting in domain structure.\cite{httg_devakul, PhysRevX.13.041007} Thus, lattice relaxation must be considered to accurately predict the twisted two-dimensional multilayer system.

\begin{figure*}
\begin{center}
\includegraphics[width=1\hsize]{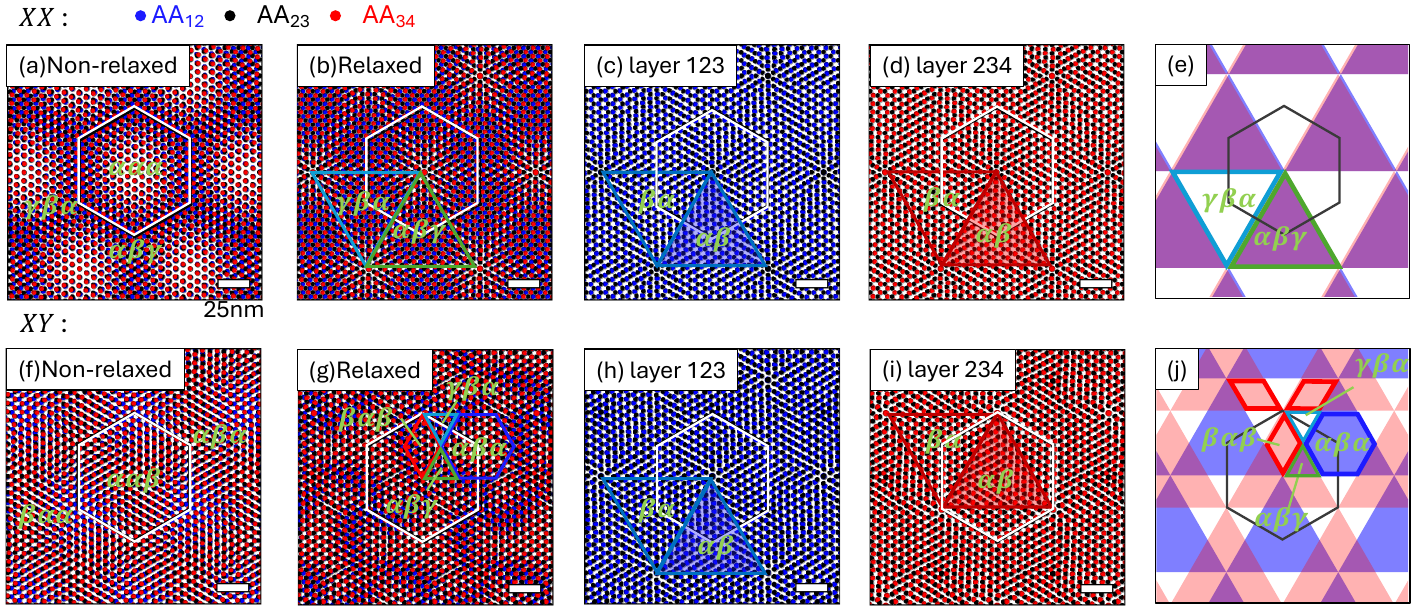}
\caption{The moir\'e superlattice structure for XX is shown for (a) the non-relaxed and (b) the relaxed. The moir\'e structure of three layer out of (b) are depicted in (c) for layer 1, 2 and 3, and (d) for layer 2, 3 and 4.
The blue/black/red points denotes $A_1A_2$/$A_2A_3$/$A_3A_4$.
The shaded region represents $\alpha\beta$ while non-shaded region represents $\beta\alpha$. The blue/red is for layer 1, 2 and 3, and layer 2, 3 and 4, respectively.
(e) is a schematic picture of arrangement of $\alpha\beta$ and $\beta\alpha$.
We slide the blue region relatively for the red region for emphasizing the overlap of the red and the blue region. 
(f)-(j) is the corresponding plots for (a)-(e).
In all figures, white scale bar indicates $25~\rm{nm}$. 
}
\label{fig_lattice_relaxed}
\end{center}
\end{figure*}

In order to find the optimized lattice structure of HTQG, we apply a continuum method\cite{PhysRevB.96.075311,PhysRevB.100.075416,PhysRevB.107.115301,PhysRevX.13.041007} for XX and XY configuration, respectively.
Experimentally realizable sample sizes $\sim \mathrm{\mu m}$ are comparable to the supermoir\'e scale of our target twist angle ($\theta \approx 2^\circ$). 
Here, we select twist angles $(\theta_{1,2},\theta_{2,3},\theta_{3,4})=(2.13^\circ,2.27^\circ,2.13^\circ)$ to obtain commensurate supermoir\'e lattices (see appendix\ref{sec_lattice_app}) as shown in Fig.~\ref{fig_lattice_relaxed}(a) and (f) for XX and XY, respectively.
We minimize the lattice energy, which is the sum of the elastic energy for each layer and the interlayer binding energy of adjacent layers. In this model, we ignore the binding energy between the next-nearest neighbor layers, which might be important only for the small twist angle below $\theta\lesssim 0.3^\circ$\cite{park2024tunable}. The minimization is performed while preserving the three-fold rotational symmetry.  Details of the computational method are provided in the Appendix\ref{sec_lattice_app}.

The upper row of Fig.~\ref{fig_lattice_relaxed} shows the result for the XX configuration.
In Fig.~\ref{fig_lattice_relaxed}(b), the relaxed structure, one notices that the area of $\alpha\alpha\alpha$ is drastically reduced, while the $\alpha\beta\gamma$ and $\gamma\beta\alpha$ form large triangular domains within which the moir\'e is commensurate. This can be understood from the two HTTG-like supermoir\'e's formed by layers 1-3 and layers 2-4, depicted in Figs.\ref{fig_lattice_relaxed}(c),(d), relaxing into $\alpha\beta/\beta\alpha$ domains\cite{httg_devakul,PhysRevX.13.041007}. 
The unshifted stacking of these two relaxed HTTG patterns forms Type-II domains, $\alpha\beta\gamma$ and $\gamma\beta\alpha$ (see Fig.~\ref{fig_lattice_relaxed}(e)).
The lower row of Fig.~\ref{fig_lattice_relaxed} is the corresponding picture for the XY region. The relaxation extends $\alpha\beta\alpha$ to form hexagonal domains, while reducing $\beta\alpha\alpha$ and $\alpha\alpha\beta$. 
Note that $\beta\alpha\beta$, $\alpha\beta\gamma$ and $\gamma\beta\alpha$ regions are not present in the unrelaxed structure but emerge after relaxation. The XY region (and by $C_{2z}$ the YX region) thus features all four domain types, with the Type-I Bernal-like domains ($\alpha\beta\alpha, \beta \alpha \beta$) having the largest area.
Similar to the XX region, the relaxation can be understood through each trilayer supermoir\'e pattern forming $\alpha\beta/\beta\alpha$ domains (Fig.~\ref{fig_lattice_relaxed}(h,i)). The stacking of these supermoir\'e patterns, depicted inn Fig.~\ref{fig_lattice_relaxed}(j) with the appropriate shift for the XY region, reproduces the domains we obtain numerically.

\PRLsec{Model Hamiltonian}
We now study the moir\'e-scale electronic structure of the Type-I ($\alpha \beta \alpha$, $\beta \alpha \beta$) and Type-II ($\alpha \beta \gamma$, $\gamma \beta \alpha$) periodic domains. We will focus on $\alpha \beta \alpha$ and $\alpha \beta \gamma$ as representatives of Types I,II respectively, as the other structures are related to these by $C_{2z}$.
The Hamiltonian for the local domains are given by the Bistritzer-MacDonald continuum model generalized to four layers:
\begin{widetext}
\begin{equation}
\label{eq_Hamiltonian}
\mathcal{H}[\boldsymbol{d}_{1,2},\boldsymbol{d}_{3,4}]=\left(\begin{array}{cccc}
-i v \boldsymbol{\sigma}_{\theta} \cdot \boldsymbol{\nabla} &                        T\left(\bm{r}-\boldsymbol{d}_{1,2}\right)                     &                                0                            &                                0                            \\
                      \text { h.c. }                        & -i v \boldsymbol{\sigma}_{\theta} \cdot \boldsymbol{\nabla} &                        T\left(\bm{r}\right)                     &                                0                            \\
                            0                               &                        \text { h.c. }                       & -i v \boldsymbol{\sigma}_{\theta} \cdot \boldsymbol{\nabla} & T\left(\bm{r}+\boldsymbol{d}_{3,4}\right) \\
                            0                               &                             0                               &                         \text { h.c. }                      & -i v \boldsymbol{\sigma}_{\theta} \cdot \boldsymbol{\nabla}
 \end{array}\right)
\end{equation}
\end{widetext}
Here, 
\begin{equation}
T\left(\mathbf{r}\right)=w\left[\begin{array}{cc}\kappa U_{0}(\mathbf{r}) & U_{-1}(\mathbf{r}) \\ U_{1}(\mathbf{r}) & \kappa U_{0}(\mathbf{r})\end{array}\right] 
\end{equation}
is the interlayer coupling between adjacent layers where $U_{\ell}(\mathbf{r})=\sum_{n=0}^{2} e^{\frac{2 \pi n}{3} \ell n } e^{-i \mathbf{q}_{n} \cdot \mathbf{r}} \left(1+\lambda_{\rm{MDT}} \hat{\boldsymbol{q}}_{n, \perp} \cdot \vec{\nabla}\right)$, with $\bm{q}_{n}=R[(n-1)\pi/3] \frac{4\pi\theta}{3}(0,-1)$ and $\hat{\boldsymbol{q}}_{n, \perp}=\left[\cos \left(\frac{2 \pi}{3} n\right), \sin \left(\frac{2 \pi}{3} n\right)\right]$. The shift of $T\left(\mathbf{r}\right)$ by $\boldsymbol{d}_{1,2/3,4}$ corresponds to the shift of the 1,2 and 3,4 moir\'e lattices respectively, relative to the 2,3 moir\'e. In particular, for $\alpha\beta\gamma$, $(\boldsymbol{d}_{1,2},\boldsymbol{d}_{3,4})=(\boldsymbol{d}_{BA},\boldsymbol{d}_{BA})$ and for $\alpha\beta\alpha$ $(\boldsymbol{d}_{1,2},\boldsymbol{d}_{3,4})=(\boldsymbol{d}_{BA},-\boldsymbol{d}_{BA})$ with $\boldsymbol{d}_{BA}=(\bm{a}^{2,3}_1+\bm{a}^{2,3}_2)/3$. $\bm{a}^{2,3}_i$ is moir\'e superlattice vector formed by layer2 and layer3 (detail definition is in Appendix~\ref{sec_lattice_app}). 
The second term in $U_\ell(\br)$ is a momentum-dependent tunneling which breaks the particle-hole symmetry \cite{PhysRevB.99.205134,fang2019angle,PhysRevResearch.1.013001,PhysRevB.101.195425, kwan2024fractional,xia2025topological} and plays an important role in HTTG \cite{kwan2024fractional}. We choose $\lambda_{\rm MDT} \approx -0.23$~\AA, derived by Ref.\cite{kwan2024fractional}.
The rotation $\boldsymbol{\sigma}_{\theta}=e^{-i \theta \sigma_{z}}\left(\sigma_{x}, \sigma_{y}\right)$ also breaks particle-hole by a small $O(\theta)$ amount; here $\v{\sigma}=(\sigma_x,\sigma_y)$ are Pauli matrices in the graphene sublattice space $(A,B)$ and $v$ is the Dirac velocity. 
This model has a moir\'e translation symmetry with reciprocal lattice vectors $\bm{G}^M_{1,2}$ and lattice vectors $\bm{a}^M_{1,2}$. The $\ell-$th layer's Bloch wavefunction satisfies $\psi_{\bk, \ell}(\br+\bm{a})=e^{i\left(\bk-\bm{K}_{\ell}\right) \cdot \bm{a}} \psi_{\bk, \ell}(\br)$, where $\bm{K}^{\ell} = \bm{K}^{1} + (\ell-1)\bm{q}_1$ are high symmetric points in mBZ. $\bm{K}^{1,4}$ is $\gamma$ point and $\bm{K}^{2,3}$ are $\kappa$ point and $\kappa^{\prime}$ point, respectively. Note that here we write the Hamiltonian for the graphene $K$ valley; the $K'$ valley Hamiltonian is related by time-reversal. We explain the detail of the symmetry of this Hamiltonian in Appendix~\ref{sec_symmetry_app} and argue the fate of the Dirac cones by breaking of symmetries in Appendix~\ref{sec_breaking_symmetry}.

\begin{figure}
\begin{center}
\includegraphics[width=1\hsize]{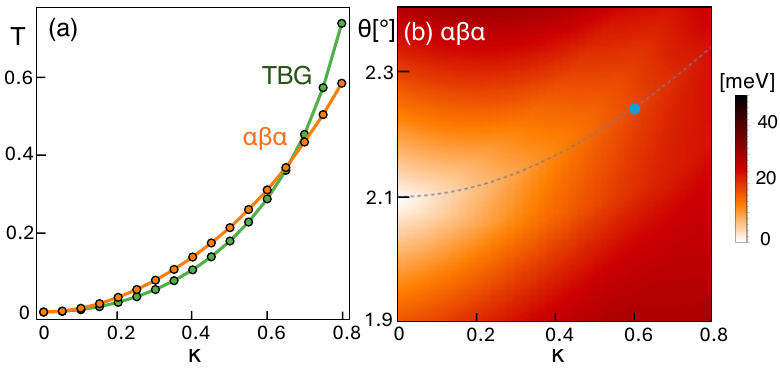}
\caption{(a)The green/orange line represents $T$, deviation from ideal quantum geometry, as a function of $\kappa$ for TBG/$\alpha\beta\alpha$. (b)The bandwidth as a function of $\kappa$ and $\theta$ for $\alpha\beta\alpha$. The cyan dot indicates $(\theta,\kappa)=(2.25^\circ, 0.6)$ which we use in this paper. 
}
\label{fig_BW_TrCond}
\end{center}
\end{figure}

\PRLsec{Type-I($\alpha\beta\alpha$ and $\beta\alpha\beta$):TBG symmetric flat bands}In the following, we compute the band structure with a realistic model of HTQG. For numerical values, we take $v = 8.4 \times 10^5\, \mathrm{m/s}$, $w = 110\, \mathrm{meV}$ and $\kappa = 0.6$, consistent with typical experimental and theoretical estimates for TBG.\cite{PhysRevB.96.075311,PhysRevB.98.224102,PhysRevResearch.1.013001,PhysRevB.99.205134,ledwith2021tb,PhysRevB.101.195425,PhysRevX.8.031087,das2021symmetry,PhysRevLett.125.257602} 
Figure.~\ref{fig_BW_TrCond}(b) shows the average of the bandwidth for the two flat bands, $W$ for $\alpha\beta\alpha$. 
For $\kappa>0$, the magic angle (where bandwidth is minimized) increases. The $\alpha \beta \alpha$ bandwidth is minimized at $\theta = 2.25^\circ$ for $\kappa = 0.6$. Actually, the minimal angle for $\alpha \beta \gamma$ remains very close to that of $\alpha \beta \alpha$. Thus, for the rest of this paper, we fix $\theta = 2.25^\circ$ for $\alpha\beta\alpha$ and $\alpha\beta\gamma$.

Figure~\ref{fig_htqg_summary}(d) shows the band structure obtained for the $\alpha\beta\alpha$ domain. The Dirac cone at the $\gamma$ point remains protected by the $\mathcal{C}_{2x}$ symmetry. The two flat bands are well separated from the remote bands by a gap of about $70\,\mathrm{meV}$, which exceeds the typical Coulomb interaction scale, $\approx40\,\mathrm{meV}$, at this twist angle.  The narrow bandwidth, $W \approx 15\,\mathrm{meV}$, suggests that correlation effects should be robust in these bands.

Because the two nearly flat bands connect adiabatically to the exact zero modes of the chiral limit(see Appendix~\ref{sec_chiral_aba}), $\kappa=0$, one can construct an $\sigma_z$-diagonal (i.e.\ sublattice-polarized) “Chern basis” for these Bloch states \cite{PhysRevB.99.155415,PhysRevX.10.031034,PhysRevB.103.205413}.  We find that a band in this basis carries Chern number $C = +2$, while the other band has $C = -2$.  Moreover, the electron density in these bands is comparatively homogeneous in real space (see Fig.~\ref{fig_htqg_summary}(h), in stark contrast to the strongly AA-centered density found in the magic angle TBG\cite{Trambly2010,PhysRevB.81.165105,PhysRevB.82.121407,PhysRevB.83.045425,PhysRevLett.108.216802,PhysRevB.86.155449,xie2019spectroscopic,kerelsky2019maximized,tilak2021flat}. This homogeneity will suppress the Hartree dispersion~\cite{Guineaelectrostatic2018,ardemakercharge2019,ceapinning2019,goodwin2020hartree,Kangcascades2021,pierce2021unconventional,parker2021field}
and favor a strong-coupling picture of the many-body state in which exchange energy dominates over the single-particle bandwidth. 
The same band structure applies to $\beta\alpha\beta$, related to $\alpha\beta\alpha$ by the symmetry $\mathcal{C}_{2z} \mathcal{T}$.

We note that the system is particularly promising for realizing correlated states at fractional filling, particularly topological charge density waves\cite{polshyn2022topological,xie2021fractional,su2025moire,PhysRevX.15.011045} and fractional Chern insulators (FCIs)\cite{spanton2018observation,xie2021fractional,cai2023signatures,zeng2023thermodynamic,park2023observation,xu2023observation,lu2023fractional}. Realizing FCIs in $\abs{C}>1$ bands\cite{parameswaran2013fractional,bergholtzTOPOLOGICALFLATBAND2013,liuRecentDevelopmentsFractional2022} remains an outstanding goal. In the chiral limit, the $C=\pm 2$ bands have a special analytic property\cite{ledwith2020fractional,PhysRevLett.127.246403,PhysRevLett.128.176404,wang2022hierarchy,ledwith2023vortexability} known as ``ideal quantum geometry" or ``vortexability"\cite{ledwith2023vortexability,Okuma,PhysRevResearch.5.L032048}; multiplication by $z=x +  i\sgn(C) y$, which introduces vortices, does not cause the wavefunction to leave the band of interest. This enables the construction of exact many-body ground states for flat vortexable bands in the limit of short range interactions, in direct analogy with the standard justifications of the Laughlin state in the LLL \cite{TrugmanKivelson1985, VLPokrovsky_1985, ledwith2020fractional, ledwith2023vortexability}. 
For a Chern $|C|$ band the character of these states can be obtained by decomposing\cite{barkeshliTopologicalNematicStates2012,wuBlochModelWave2013,Kumar2014Generalizing} the band into $|C|$ copies of individually vortexable\cite{PhysRevResearch.5.023166,PhysRevResearch.5.023167} $C=\pm1$ bands. One obtains topological charge density waves at filling $\nu = 1/|C|$, translation symmetric FCIs at fillings $\nu = 1/(2|C|s+1)$, and Laughlin-CDW states at fillings $\nu=1/C(2s+1)$. We pause to comment that for $C=2$ a $\nu = \frac{1}{3}$ FCI is found to be similarly robust\cite{PhysRevResearch.5.023166, PhysRevLett.126.026801,niu2025quantumanomaloushalleffects}; its wavefunction generalizes the $(112)$ Halperin state which is the ground state of a spinful LLL in the absence of Zeeman coupling\cite{PhysRevB.41.7910,PhysRevB.45.3418,PhysRevB.49.7515,verdene2007microscopic,PhysRevLett.71.153,PhysRevLett.82.3665}.

Since the above statements are only guaranteed for exactly vortexable bands, we quantify the deviation via
$T=\frac{1}{2\pi}\int_{\rm BZ} d^2\bk \left[\mathrm{tr}g(\bk) - \Omega(\bk)\right] > 0$,
where $\Omega$ is Berry curvature and $g$ is Fubini-Study metric\cite{PhysRevLett.122.106405,PhysRevResearch.2.023237,PhysRevLett.127.246403,varjas2022topological}. The band is vortexable if and only if $T=0$, and $T=2n$ for the $nth$ LL\cite{Mera2021b,PhysRevLett.134.106502,1zg9-qbd6}.
As shown in Fig.~\ref{fig_BW_TrCond}(c), $T$ for $\alpha\beta\alpha$ and TBG have similar dependence for $\kappa$. For both systems $T$ is between that of the LLL and the first LL, but in TBG the concentrated charge density leads to a large interaction induced dispersion that destabilizes zero-field FCIs\cite{parker2021field}. The more homogeneous charge density in $\alpha \beta \alpha$ suggests that FCIs may be more stable; we leave this to future studies involving many-body numerics.

\PRLsec{Type-II($\alpha\beta\gamma$ and $\gamma\beta\alpha$):nonzero gap Chern number}In Fig.~\ref{fig_htqg_summary}(e), we plot the corresponding band structure for the $\alpha\beta\gamma$. We find two bands near the Fermi energy: a narrow conduction band with $C=0$ and a valence band with $C=-2$. We find that the valley Chern number $C_V = C_K - C_{-K}= 2C_K$ at full filling, charge neutrality, and empty filling are $-4,-4,+4$ respectively, where the change of eight comes from occupying the spin degenerate valence bands in both valleys. The valley Chern numbers are opposite for the $\gamma \beta \alpha$ domains as they are related by $C_{2z} T$, such that there should be $\Delta C_v = 8$ edge states on the XX region domain walls. This is twice as many as in HTTG \cite{httg_devakul, PhysRevResearch.6.L022025, PhysRevX.13.041007}. There will also be four edge states in the XY region domain walls, as the Type-I regions adjacent to the Type-II domains have $C_v = 0$ by $C_{2x}$.



Notably, the conduction band with $C = 0$ retains a relatively narrow bandwidth $W \approx 12\,\mathrm{meV}$, implying that strong-correlation effects is important. Moreover, the state has a uniform charge distribution as shown in Fig.~\ref{fig_htqg_summary}(i), leading to the possibility of hosting a different ground state from the Hubbard model, which is associated with a concentrated charge density. 

\PRLsec{Discussion and Outlook}
In this work, we presented a theoretical study of magic-angle HTQG, in which four graphene layers are twisted in the same direction. Our analysis highlights HTQG as a promising platform for studying robust correlated and topological states. Since all domains have the same magic angle, multiple moir\'e systems can effectively be targeted in the same device. Furthermore, the larger magic angle $\theta \approx 2.3^\circ$, relative to $\theta \approx 1.1^\circ$ in TBG, makes the moir\'e subsystems less sensitive to strain and twist angle disorder.

An important next step is the study of correlated phases in each domain. All subregions feature narrow bands that are spin-valley degenerate and far-removed from any Hubbard-like description. A strong coupling analysis\cite{Repellin19,PhysRevX.10.031034,TBGIVGroundState,ledwith2021168646} should then lead to generalized ferromagnetic states at integer filling factors. These states should be more robust than those in TBG\cite{PhysRevX.10.031034,TBGIVGroundState,ledwith2021168646} due to the homogeneous charge density.
In the Type-II regions, for example, we expect the $C=0$ narrow conduction band to fill $\nu$ spin-valley flavors at integer filling $\nu \in [0,4]$. We note that, despite being topologically trivial, this band has uniform density and non-trivial quantum geometry. We leave the largely unexplored~\cite{lin_fractional_2025} correlated physics of such bands to future work.

The strong coupling states of the Type-I domains are especially of interest. These regions have similar symmetry and topology to TBG --- both systems host, within each valley, symmetry-related Chern bands with equal and opposite Chern number. This state of affairs is usually associated to $C_{2z}\T$ symmetry, but particle-hole and $C_{2x}$ also suffice and are preserved in the HTQG Type-I domains. The extra symmetry leads to a larger manifold of generalized ferromagnets beyond spin-valley polarization\cite{PhysRevX.10.031034,TBGIVGroundState,ledwith2021168646} including, e.g., intervalley coherent and quantum spin Hall states\cite{kwanElectronphononCouplingCompeting2024}. 
The single particle dispersion that tunnels between Chern sectors then, through a superexchange-like mechanism, selects states that have opposite occupation between Chern sectors\cite{PhysRevX.10.031034,Khalaf_sciadv2021,ledwith2021168646}. While the ``Chern-antiferromagnetic" (CAF) states and the spin-valley ferromagnets are both strong-coupling topological insulators, their doped states should be contrasted. Indeed, quantum fluctuations of spin-valley ferromagnets are forbidden by symmetry such that the doped electrons essentially form a separate system with fewer internal degrees of freedom. Instead, the CAF orders can fluctuate and mediate interactions between doped electrons. 

In this sense, the CAF states are similar in spirit\cite{ledwith2021168646} to antiferromagnetic order in high temperature superconductors, where spin fluctuations are believed to be foundational for pairing. In addition the CAFs are associated with filling topological bands. One potential mechanism based on CAF states is skyrmion pairing\cite{PhysRevB.101.165141,Khalaf_sciadv2021,PhysRevB.106.035421, Christos_pnas2020,PhysRevX.12.031020,PhysRevB.105.224508}. Here, band topology binds electric charge to skyrmion defects of the symmetry breaking order\cite{sondhiSkyrmionsCrossoverInteger1993,moonSpontaneousInterlayerCoherence1995a} and Chern-antiferromagnetism leads to the binding and pair-condensation of these skyrmions. Superconductors with unconventional Fermi surface pairing enabled by inter-Chern tunneling have been carefully demonstrated as well \cite{sahaySuperconductivityTopologicalLattice2024}. While we leave a careful strong coupling analysis and analysis of pairing possibilities to future work, the above general considerations, specific mechanisms, and phenomenological relationship to TBG make HTQG particularly promising for realizing robust superconductivity.

The supermoir\'e and super-supermoir\'e structure and domain wall networks in HQTG also presents exciting possibilities due to their unique tunability and visibility in local probes \cite{hoke2024imaging}. Thermal cycling in HTG was found to change the shape and size of the supermoir\'e-scale domains without changing the twist angle of the moir\'e domains\cite{hoke2024imaging}. Remarkably, the domains expanded beyond the supermoir\'e lattice constant $a^{MM}=a^M/\theta$, consistent with the generation of a tiny, $\sim 0.01\%$, heterostrain\cite{hoke2024imaging,httg_devakul}. Heterostrain at this scale is negligible for moir\'e scale physics but transformative at the supermoir\'e scale.
It seems likely that the super-supermoir\'e will be even more tunable, enabling the targeting and expansion of favored regions.
The network of topological domain walls can also be directly imaged with local probes, as in HTTG \cite{hoke2024imaging}. The character of the network will depend on the topological states inside the domains, such that topological phase transitions in the domains can cause gapless edge modes to appear or disappear. 
The realization of intervalley coherent states, expected in the Type-I regions of HQTG, provides another route to gapping edge states protected by valley Chern number. Conversely, the absence or presence of gapless domain wall states can be used to constrain the character of the correlated insulators inside the domains.


\begin{acknowledgements}
We thank Daniel Parker, Tomohiro Soejima, Junkai Dong, Eslam Khalaf, Daniele Guerici,  Mikito Koshino,and Trithep Devakul for helpful discussions. M. F. acknowledges support from by JSPS KAKENHI grant no. JP23KJ0339 and JP24K16987, and the Center for the Advancement of Topological Semimetals (CATS), an Energy Frontier Research Center at the Ames National Laboratory. Work at the Ames National Laboratory is supported by the U.S. Department of Energy (DOE), Basic Energy Sciences (BES) and is operated for the U.S. DOE by Iowa State University under Contract No. DE-AC02-07CH11358. N.N is funded by JST CREST (Grant JPMJCR20T3), Japan and thanks JSPS for support from Overseas Research Fellowship. A.V. is supported by a Simons Investigator grant. P.J.L. was supported by the MIT Pappalardo Fellowship.
\end{acknowledgements}

\appendix

\section{Lattice geometry-Rigid lattice}
\label{sec_lattice_app}
\subsection{small twist angle limit}
We here introduce the lattice geometry of HTQG without lattice relaxation. HTQG has the four length-scale patterns, atomic, moir\'e, supermoir\'e and moir\'e of supermoir\'e. Firstly, we define them in small twist angle limit for simplicity.

We consider the four layers of graphene with relative twist angles $\theta_{\ell,\ell+1}$ between layer $\ell$ and $\ell+1$. In this paper, we take the twist angles of each layer $\theta^{\ell}$ as $\theta^{1}=-\theta^{23}/2-\theta^{12}$, $\theta^{2}=-\theta^{23}/2$, $\theta^{3}=+\theta^{23}/2$ and $\theta^{4}=+\theta^{23}/2+\theta^{34}$. The lattice vectors of each graphene layer are given by $\bm{a}^{\ell}_{j}=R(\theta^{\ell})\bm{a}_{j}$, where $\bm{a}_{1}=a(1,0)$ and $\bm{a}_{2}=a(1/2,\sqrt{3}/2)$ are the lattice vector of graphene without rotation, and $R(\theta)$ is the 2-dimensional rotation matrix. The reciprocal lattice vector of each layer are also given by $\bm{G}^{\ell}_{j}=R(\theta^{\ell})\bm{G}_{j}$, where $\bm{G}_{1}=(4\pi/\sqrt{3})(\sqrt{3}/2,-1/2)$ and $\bm{G}_{2}=(4\pi/\sqrt{3})(0,1)$ are the reciprocal lattice vector of non-twisted graphene.The Dirac cones in each layer reside at the corners of the corresponding rotated BZ, $\bm{K}^\ell=-(2\bm{G}_1^\ell+\bm{G}_2^\ell)/3$.

The adjacent layers form moir\'e superlattices described by
\begin{equation}
    \bm{a}^{\ell,\ell+1}_i=R\left(-\frac{\pi}{2}\right)\frac{\bm{a}_i^{\ell} + \bm{a}_i^{\ell+1}}{4\sin(\theta_{\ell,\ell+1}/2)}.
\end{equation}
The corresponding moir\'e reciprocal lattice vectors $\bm{G}^{\ell,\ell+1}_i$ satisfy $\bm{G}^{\ell,\ell+1}_i \cdot \bm{a}^{\ell,\ell+1}_j=2\pi \delta_{ij}$.

The moir\'e superlattice between layer $\ell$ and $\ell+1$ has the three specific local stacking structure with three-fold rotational symmetry called $AA_{\ell,\ell+1}$, $AB_{\ell,\ell+1}$ and $BA_{\ell,\ell+1}$. These local stacking at $\br$ is characterized by a non-rotated bilayer graphene with the displacement vector of adjacent layers $\bm{\delta}_{\ell,\ell+1}(\bm{r})$. For example, $AA_{\ell,\ell+1}$, $AB_{\ell,\ell+1}$ and $BA_{\ell,\ell+1}$ are defined as $\bm{\delta}_{\ell,\ell+1}=\bm{0}, \mp(\bm{a}^{\ell}_{1}+\bm{a}^{\ell}_{2})/3$, respectively.

In the second panel of Fig.~\ref{fig_htqg_summary}(b), the blue shaded triangles correspond to AB stacking regions, while the empty triangles indicate BA stacking regions within each moir\'e pattern. By labeling the triangle centers as $\alpha$ for AB and $\beta$ for BA regions.

In the following, we assume that the twist angles between adjacent layers are equal, $\theta_{1,2}=\theta_{2,3}=\theta_{3,4}=\theta$. In this case, the adjacent moir\'e periods form a large pattern, known as supermoir\'e period,
\begin{equation}
\label{eq_supermoire}
    \bm{a}^{\ell,\ell+1,\ell+2}_i=R\left(-\frac{\pi}{2}\right)\frac{\bm{a}_i^{\ell,\ell+1} + \bm{a}_i^{\ell+1,\ell+2}}{4\sin(\theta/2)},
\end{equation}
for $\ell=1,2$. This supermoir\'e pattern between $\ell$,$\ell+1$ and $\ell+2$ has the three-fold rotational symmetric local stacking structures, $\alpha\alpha_{\ell,\ell+1,\ell+2}$, $\alpha\beta_{\ell,\ell+1,\ell+2}$ and $\beta\alpha_{\ell,\ell+1,\ell+2}$\cite{httg_devakul, PhysRevX.13.041007}. The moir\'e scaled local stacking at $\br$ is identified by a non-rotated pair of moi\'e patterns, analogous to non-rotated bilayer graphene, with the displacement vector of the adjacent moir\'e pattern $\bm{\delta}_{\ell,\ell+1,\ell+2}$. This measure the displacent between $\alpha_{\ell,\ell+1}$ and $\alpha_{\ell+1,\ell+2}$. For example, $\alpha\alpha_{\ell,\ell+1,\ell+2}$, $\alpha\beta_{\ell,\ell+1,\ell+2}$ and $\beta\alpha_{\ell,\ell+1,\ell+2}$ are defined as $\bm{\delta}_{\ell,\ell+1,\ell+2}=\bm{0}, \mp(\bm{a}^{\ell,\ell+1}_{1}+\bm{a}^{\ell,\ell+1}_{2})/3$. In the third panel of Fig.~\ref{fig_htqg_summary}(b), the orange shaded triangles correspond to $\alpha\beta$ stacking regions, while the empty triangles indicate $\beta\alpha$ stacking regions within each supermoir\'e pattern. By labeling the triangle centers as X for $\alpha\beta$ and Y for $\beta\alpha$ regions.

Finally, two supermoir\'e periods lead to super-supermoire period
\begin{equation}
    \bm{a}^{1,2,3,4}_i=R\left(-\frac{\pi}{2}\right)\frac{\bm{a}_i^{1,2,3} + \bm{a}_i^{2,3,4}}{4\sin(\theta/2)}.
\end{equation}
In the fourth panel of Fig.~\ref{fig_htqg_summary}(b), we show the the gray, the blue, the orange shaded region correspond to XX, XY, and YX region. They have three-fold rotational symmetric points. The displacement vector $\boldsymbol{\delta}_{1,2,3,4}$ is used to denote the relative position of $\rm{X}_{1,2,3}$ for $\rm{X}_{2,3,4}$. For XX, the two supermoir\'e pattern are completely overlapped, giving $\boldsymbol{\delta}_{1,2,3,4}=0$. On the other hand, for XY, the $\alpha\beta_{123}$ and $\alpha\beta_{234}$ are overlapped so the displacement vector is given by $\boldsymbol{\delta}_{1,2,3,4}=(\bm{a}^{1,2,3}_{1}+\bm{a}^{1,2,3}_{2})/3$. YX is obtained by the inversion-symmetric counterpart. In Fig.~\ref{fig_lattice_relaxed}(a) and (f), we show the detailed structure of XX and XY. XY has $\alpha\alpha\beta$, $\alpha\beta\beta$ and $\beta\alpha\beta$, and YX has the inversion pair of them. XX has $\alpha\alpha\alpha$, $\alpha\beta\gamma$ and $\gamma\beta\alpha$

\subsection{commensurate moir\'e vectors}
In the previous section we define the supermoir\'e period as \eqref{eq_supermoire} to be applicable for any small twist angle.
For the calculation of lattice relaxation, we consider the commensurate supermoir\'e pattern, where its lattice vector can be defined by
\begin{equation}
\label{eq_com_condition}
    \begin{aligned} 
        \bm{a}^{\ell,\ell+1,\ell+2}_1 &= n_{\ell',\ell'+1} \bm{a}^{\ell',\ell'+1}_{1} + m_{\ell',\ell'+1} \bm{a}^{\ell',\ell'+1}_{2} \\
        \bm{a}^{\ell,\ell+1,\ell+2}_2 &= R(60^\circ)\bm{a}^{\ell,\ell+1,\ell+2}_1,
    \end{aligned}
    \end{equation}
with the integers $n_{\ell',\ell'+1}$ and $m_{\ell',\ell'+1}$. 
In particular, we impose the condition:
\begin{equation}
\bm{a}^{1,2,3}_i=\bm{a}^{2,3,4}_i.
\end{equation}
for aligning two supermoir\'e unitcell.
Solving the above equation for the $\theta^{\ell,\ell+1}$, we obtain two angle pair satisfying Eq.~\eqref{eq_com_condition} as follows,
\begin{equation}
\label{eq_angle_formulas}
     \begin{aligned} 
       \theta_{1,2} &= \theta_{3,4} =\theta(n_{1,2},m_{1,2},n_{2,3},m_{2,3}),\\ 
       \theta_{2,3} &= - \theta(n_{2,3},m_{2,3},n_{1,2},m_{1,2}),
    \end{aligned}
    \end{equation}
    where 
    \begin{widetext}
    \begin{align}\label{eq_angle_formulas_2}
     &\theta(n_{1,2},m_{1,2},n_{2,3},m_{2,3}) =\notag\\
     & \quad 2 \tan^{-1}\frac{\sqrt{3}\left\{m_{1,2} \left(2n_{2,3}+m_{2,3}\right)-\left(2n_{1,2}+m_{1,2}\right)m_{2,3}\right\}}{\left(2n_{1,2}+m_{1,2}\right)\left(2n_{2,3}+m_{2,3}\right)+3m_{1,2}m_{2,3}+\left(2n_{2,3}+m_{2,3}\right)^{2}+3m_{2,3}^{2}}.
    \end{align}
    \end{widetext}
    The spatial period of the supermoir\'e pattern is given by 
    $a^{1,2,3} = a^{1,2}\sqrt{n_{1,2}^{2}+m_{1,2}^{2}+n_{1,2}m_{1,2}}= a^{2,3}\sqrt{n_{2,3}^{2}+m_{2,3}^{2}+n_{2,3}m_{2,3}}$.
    Using the condition $\bm{G}_{i}\cdot\bm{a}_{j}=2\pi\delta_{ij}$, the reciprocal lattice vector of supermoir\'e is written as
    \begin{widetext}
    \begin{align}
        \begin{pmatrix}
            G_{1}\\
            G_{2}
        \end{pmatrix}
        &=
        \frac{1}{n_{1,2}^2+n_{1,2}m_{1,2}+m_{1,2}^2}\left(
        \begin{array} {cc}
            n_{1,2}+m_{1,2} & m_{1,2} \\
            -m_{1,2} & n_{1,2}
	\end{array}
        \right)
        \begin{pmatrix}
            G_{1}^{12}\\
            G_{2}^{12}
        \end{pmatrix}
        \nn  \\
        &=
        \frac{1}{n_{2,3}^2+n_{2,3}m_{2,3}+m_{2,3}^2}
        \left(
        \begin{array} {cc}
            n_{2,3}+m_{2,3} & m_{2,3} \\
            -m_{2,3} & n_{2,3}
	\end{array}
        \right)
        \begin{pmatrix}
            G_{1}^{23}\\
            G_{2}^{23}
        \end{pmatrix}.
    \end{align}
    \end{widetext}

In the Fig.~\ref{fig_lattice_relaxed}, we choose $(n_{1,2},m_{1,2},n_{2,3},m_{2,3})=(8,7,8,8)$.
The corresponding twist angle is $(\theta^{1,2},\theta^{2,3},\theta^{3,4})=(2.13,2.27,2.13)$.

\section{Computation method for lattice relaxation}

    We apply the continuum method\cite{PhysRevX.13.041007} to obtain the optimized lattice structure of HTQG.
    We introduce $\bm{s}(\bm{R}^{(\ell)}_{X})$ as the displacement vector of $X$ sublattice atom of layer $\ell$ at position $\bm{R}_{X}$.
    In this model, we assume that the spatial scale of the modulation of lattice relaxation is much larger than the lattice constant of monolayer graphene.
    Under this assumption, the displacement vector can be written as the continuum function of the space $\bm{s}(\bm{R}^{(\ell)}_{X})\to\bm{s}^{(\ell)}(\bm{r})$, where we ignore the sublattice dependence.
    In the continuum method, the optimized lattice structure is obtained by minimizing lattice energy $U=U_{E}+U_{B}^{12}+U_{B}^{23}+U_{B}^{34}$ which is a functional of the displacement vector $\bm{s}^{(\ell)}(\bm{r})$.
    Here, $U_{E}$ is the elastic energy and $U_{B}^{\ell\ell'}$ is the interlayer binding energy between layers $\ell$ and $\ell'$.
    We ignore the binding energy between next nearest negibor layers which could be important only for quite a small angle, for twisted trilayer graphene, it's about $0.3$ degree \cite{park2024tunable}.
    
    The elastic energy of distorted HTQG is written in a standard form \cite{PhysRevB.65.235412,PhysRevB.90.115152} as
    \begin{align}\label{eq:elastic}
        U_E=\sum_{l=1}^{4}\frac{1}{2}\int&\left[\left(\mu+\lambda\right)\left(s_{xx}^{(\ell)}+s_{yy}^{(\ell)}\right)^{2} \right.\notag \\
           &\left. +\mu\left\{\left(s_{xx}^{(\ell)}-s_{yy}^{(\ell)}\right)^{2}+4\left(s_{xy}^{(\ell)}\right)^{2}\right\}\right]\diff^2\bm{r},
    \end{align}
    where $\lambda=325$~eV/nm$^{2}$ and $\mu=957$~eV/nm$^{2}$ are Lam\'e factors of monolayer graphene, and $s_{ij}^{(\ell)}=(\partial_{i}s_{j}^{(\ell)}+\partial_{j}s_{i}^{(\ell)})/2$ is the strain tensor.
    The interlayer binding energy of adjacent layers $(\ell,\ell')=(1,2), (2,3), (3,4)$
    is given by \cite{PhysRevB.96.075311}
     \begin{align}\label{eq:binding}
     U_{B}^{\ell \ell'}&=\int\diff^{2}\bm{r} \sum_{j=1}^{3}2V_{0}\cos\left[\bm{G}_{j}^{\ell \ell'}\cdot\bm{r}+\bm{b}_{j}\cdot\left(\bm{s}^{(\ell')}-\bm{s}^{(\ell)}\right)\right], 
    \end{align}
    where $\bm{b}_{3}=-\bm{b}_{1}-\bm{b}_{2}$, $\bm{G}_{3}^{\ell \ell'}=-\bm{G}_{1}^{\ell \ell'}-\bm{G}_{2}^{\ell \ell'}$, and $V_{0}=0.16$~eV/nm$^{2}$. 

    We assume $\bm{s}^{(\ell)}$'s are periodic in the original supermoir\'e period, and define the Fourier components as
    \begin{equation}
    \label{eq:Fourier_transform_uv}
      \bm{s}^{(\ell)}\left(\br\right) =  \sum_{\bm{G}}\bm{s}^{(\ell)}_{\bm{G}}e^{i\bm{G}\cdot \br}, 
    \end{equation}
    where $\bm{G}=m_{1}\bm{G}_{1}+m_{2}\bm{G}_{2}$ are the supermoir\'e reciprocal lattice vectors.

    The Euler-Lagrange equations $\partial U/\partial s^{(\ell)}_{\mu}$ is written as
    \begin{equation}
    \label{eq:static_sc}
    \begin{aligned}
        \bm{s}^{(1)}_{\bm{G}} &= -2V_{0}\sum_{j=1}^{3}f_{\bm{G},j}^{12} \hat{K}_{\bm{G}}^{-1}\bm{b}_{j}, \notag \\
        \bm{s}^{(2)}_{\bm{G}} &= 2V_{0}\sum_{j=1}^{3}\left(f_{\bm{G},j}^{12}-f_{\bm{G},j}^{23}\right)\hat{K}_{\bm{G}}^{-1}\bm{b}_{j}, \notag \\
        \bm{s}^{(3)}_{\bm{G}} &= 2V_{0}\sum_{j=1}^{3}\left(f_{\bm{G},j}^{23}-f_{\bm{G},j}^{34}\right)\hat{K}_{\bm{G}}^{-1}\bm{b}_{j}, \notag \\
        \bm{s}^{(4)}_{\bm{G}} &= 2V_{0}\sum_{j=1}^{3}f_{\bm{G},j}^{34}\hat{K}_{\bm{G}}^{-1}\bm{b}_{j},
    \end{aligned}
    \end{equation}
    where
        \begin{equation}
\label{eq:def_K}
    \begin{aligned}
        \hat{K}_{\bm{G}} =
	    \left(
			\begin{array} {cc}
			\left(\lambda+2\mu\right)G_{x}^{2}+\mu G_{y}^{2} & \left(\lambda+\mu\right)G_{x}G_{y} \\
			\left(\lambda+\mu\right)G_{x}G_{y} & \left(\lambda+2\mu\right)G_{y}^{2}+\mu G_{x}^{2}
			\end{array}
		\right),
        \end{aligned}
    \end{equation}
    and
        \begin{equation}
    \label{eq:Fourier_transform}
    \begin{aligned}
    &\sin\left[\bm{G}_{j}^{12}\cdot\bm{r}+\bm{b}_{j}\cdot(\bm{s}^{(2)}-\bm{s}^{(1)})\right]
    =  \sum_{\bm{G}}f_{\bm{G},j}^{12}e^{i\bm{G}\cdot\bm{r}},
    \notag\\
    &\sin\left[\bm{G}_{j}^{23}\cdot\bm{r}+\bm{b}_{j}\cdot(\bm{s}^{(3)}-\bm{s}^{(2)})\right]
    =  \sum_{\bm{G}}f_{\bm{G},j}^{23}e^{i\bm{G}\cdot\bm{r}}, \notag \\
    &\sin\left[\bm{G}_{j}^{34}\cdot\bm{r}+\bm{b}_{j}\cdot(\bm{s}^{(4)}-\bm{s}^{(3)})\right]
    =  \sum_{\bm{G}}f_{\bm{G},j}^{34}e^{i\bm{G}\cdot\bm{r}}.
       \end{aligned}
    \end{equation}

    We obtain the optimized $\bm{s}^{(\ell)}_{\bm{G}}$ by solving Eq.~\eqref{eq:Fourier_transform_uv} and \eqref{eq:static_sc} self-consistently.
    We assume that the relaxation preserves the symmetries of the model $\mathcal{C}_{3z}$ and $\mathcal{C}_{2z}$.
    Under this assumption, we take the Fourier components in $0^\circ \leq \phi(\bm{G}) < 60^\circ$, where $\phi(\bm{G})$ is the angle of the vector $\bm{G}$ respected with x-axes.
    Others Fourier components are giving by $\bm{s}^{(\ell)}_{R(\pi/3)\bm{G}}=R(\pi/3)\bm{u}^{(\ell)}_{\bm{G}}$.
    Here, furthermore, We only take a finite number of the Fourier components in $|\bm{G}|<3\max \left(|n|,|m|,|n'|,|m'|\right)$, which is enough to consider the lattice relaxation of the system.
    
    It is important to note that the components of $\bm{G}=0$, which corresponds to the global lateral shift of the layer, cannot be determined by this scheme because of $\hat{K}_{\bm{G}} = 0$ in Eq.~\eqref{eq:static_sc}.
    We can include the effect of the lateral shift by treating $\bm{s}^{(\ell)}_{\bm{G}=0}$ as parameters and solve the above self-consistent equation for each lateral shift.
    In this paper, we take $\bm{s}^{(4)}_{\bm{G}=0}=0$ and $\left(\bm{a}_{1}+\bm{a}_{2}\right)/3$ for $XX$ and $XY$ stack, respectively.

\section{Symmetry}
\label{sec_symmetry_app}
The lattice structure of HTQG exhibits two-fold rotation symmetries about the $x$-, $y$-, and $z$-axes, denoted as $\mathcal{C}_{2x}$, $\mathcal{C}_{2y}$ and $\mathcal{C}_{2z}$, respectively. While $\mathcal{C}_{2x}$ acts within a graphene valley, $\mathcal C_{2y,z}$ exchange valleys but can be combined with time reversal $\T$ to obtain a symmetry within a valley. The Hamiltonian [Eq.\eqref{eq_Hamiltonian}] is symmetric, in the sense that
\begin{equation}
\begin{aligned}
&\mathcal{C}_{2x}H[\boldsymbol{d}_{1,2},\boldsymbol{d}_{3,4}](\br)\mathcal{C}_{2x}^{-1}=H[M_x\boldsymbol{d}_{3,4},M_x\boldsymbol{d}_{1,2}](M_x\br)\\
   &(\mathcal{C}_{2y}\mathcal{T})H[\boldsymbol{d}_{1,2},\boldsymbol{d}_{3,4}](\br)(\mathcal{C}_{2y}\mathcal{T})^{-1}=H[M_y\boldsymbol{d}_{3,4},M_y\boldsymbol{d}_{1,2}](M_y\br)\\
   &(\mathcal{C}_{2z}\mathcal{T})H[\boldsymbol{d}_{1,2},\boldsymbol{d}_{3,4}](\br)(\mathcal{C}_{2z}\mathcal{T})^{-1}=H[-\boldsymbol{d}_{1,2},-\boldsymbol{d}_{3,4}](-\br),
\end{aligned}
\end{equation}
where $M_x(M_y)$ represents mirror reflection across the $x(y)$-axis. The symmetries are explicitly realized as $\mathcal{C}_{2x}=\mu_x\otimes\sigma_x$, $\mathcal{C}_{2y}\mathcal{T}=\mu_x \otimes \mathcal{K}$ and $\mathcal{C}_{2z}\mathcal{T}=\sigma_x\mathcal{K}$, where
\begin{equation}
\mu_x = \left(\begin{array}{cccc} 0 & 0 & 0 & 1 \\ 0 & 0 & 1 & 0 \\  0 & 1 & 0 & 0 \\  1 & 0 & 0 & 0
\end{array}\right)
\end{equation}
acts on the four-layer space, and $\mathcal{K}$ is complex conjugation.

Additionally, when $\boldsymbol{d}_{1,2}$ and $\boldsymbol{d}_{3,4}$ are three-fold rotation symmetric stacking, such as $\alpha\beta\gamma$ and $\alpha\beta\alpha$, the Hamiltonian satisfies
\begin{equation}
\mathcal{C}_{3z}H(\br)\mathcal{C}_{3z}^{-1}=H({C}_{3z}\br)
\end{equation}
The representation of $\mathcal{C}_{3z}$ matrix depends on $\boldsymbol{d}_{1,2}$ and $\boldsymbol{d}_{3,4}$. For example, for $\alpha\beta\gamma$, at $\bm{r}=0$
\begin{equation}
\label{eq_abc_C3z}
\mathcal{C}_{3z} = \left(\begin{array}{cccc} 1 & 0 & 0 & 0 \\ 0 & \omega & 0 & 0 \\  0 & 0 & \omega & 0 \\  0 & 0 & 0 & 1
\end{array}\right)_{\rm layer} \otimes \left(\begin{array}{cc} \omega & 0  \\ 0 & \omega^* 
\end{array}\right)_{\rm sub},
\end{equation}
and for $\alpha\beta\alpha$ at $\bm{r}=0$, the matrix is
\begin{equation}
\label{eq_aba_C3z}
\mathcal{C}_{3z} = \left(\begin{array}{cccc} 1 & 0 & 0 & 0 \\ 0 & \omega^* & 0 & 0 \\  0 & 0 & \omega^* & 0 \\  0 & 0 & 0 & \omega
\end{array}\right)_{\rm layer} \otimes \left(\begin{array}{cc} \omega & 0  \\ 0 & \omega^*
\end{array}\right)_{\rm sub}
\end{equation}
with $\omega=e^{\frac{2\pi i}{3}}$.

The Hamiltonian has translation symmetry by moir\'e lattice vector $\bm{a}_i^M$ such that
$H(\br+\v a_i^M) = T_{\v a_i^M}H(\br)T^{\dagger}_{\v a_i^M}$, where
\begin{equation}
\label{eq_translation}
T_{\bm{a}_i^M} = \left(\begin{array}{cccc} 1 & 0 & 0 & 0 \\ 0 & \omega^* & 0 & 0 \\  0 & 0 & \omega & 0 \\  0 & 0 & 0 & 1
\end{array}\right)_{\rm layer} \otimes \sigma_0.
\end{equation}
The shift of the phase for each layer component are derived from the fact that the layer$\ell$ Dirac cone is misaligned by $\ell\bm{q}_1$ in the momentum space.
Thus, One can obtain the $\mathcal{C}_{3z}$ matrix for $\br = \pm \br_{BA}$ by $T_{\mp\bm{a}_i^M}\mathcal{C}_{3z}$. 

If one sets the rotation of the sublattice Pauli matrix and the momentum-dependent tunneling to zero, the model acquires a particle–hole symmetry. This symmetry takes the form
\begin{equation}
\mathcal{P}H[\boldsymbol{d}_{1,2},\boldsymbol{d}_{3,4}](\br)\mathcal{P}^\dagger=-H[-\boldsymbol{d}_{3,4},-\boldsymbol{d}_{1,2}](\br),
\end{equation}
where
\begin{equation}
\label{eq:particlehole}
\mathcal{P}  = \left(\begin{array}{cccc} 0 & 0 & 0 & 1 \\ 0 & 0 & -1 & 0 \\  0 & 1 & 0 & 0 \\  -1 & 0 & 0 & 0
\end{array}\right)_{\rm layer} \otimes \sigma_x K.
\end{equation}
While $\P$ is weakly broken by the aforementioned terms, it will be useful for characterizing the flat band wavefunctions upon which the particle-hole breaking terms can be projected (to leading order in particle-hole breaking).

Due to $M_{x/y}\boldsymbol{d}_{BA}=\pm\boldsymbol{d}_{BA}$, the Hamitlonian of $\alpha\beta\gamma$ domain and $\alpha\beta\alpha$ domain possesses the different set of symmetries:
\begin{equation}
\begin{aligned}
&\alpha\beta\gamma: \left\{ \mathcal{C}_{3z}, \mathcal{C}_{2y}T, \mathcal{P}\mathcal{C}_{2x}, P\mathcal{C}_{2z}T \right\} \\
&\alpha\beta\alpha: \left\{ \mathcal{C}_{3z}, \mathcal{C}_{2x}, \mathcal{P} \right\}
\end{aligned}
\end{equation}
One can obtain the corresponding inversion partners $\gamma \beta \alpha, \beta \alpha\beta$ by applying $\mathcal{C}_{2z}\mathcal{T}$.

\section{Chiral limit}
In this section, we analyze the electronic structure of Eq. \eqref{eq_Hamiltonian} in the $\kappa = 0$ limit, referred to as the chiral limit\cite{tarnopolskyorigin2019}. It has been established that, in this approximation, the band structure of twisted multilayer graphene can exhibit perfectly flat bands at zero energy at certain discrete twist angles, known as the magic angles\cite{tarnopolskyorigin2019,PhysRevB.103.165113}. Although this phenomenon emerges within an idealized limit, some crucial aspects of the band structure are robust to increasing $\kappa$ to realistic values in practice \cite{ledwith2021168646,PhysRevX.10.031034}.

The presence of chiral symmetry in the system is reflected by the anticommutation of the Hamiltonian with $\Lambda = \sigma_z$. This allows the Hamiltonian to be expressed in a block-off-diagonal form when written in the sublattice basis $|A\rangle$ and $|B\rangle$:
\begin{equation}
\label{eq_Hamiltonian_chiral_general}
H(\bm{r}) = \left(\begin{array}{cc}0 & \mathcal{D}^{\dagger}(\bm{r}) \\ \mathcal{D}(\bm{r}) & 0\end{array}\right).
\end{equation}
At specific magic angles, pairs of zero-energy Bloch eigenstates exist across the entire Brillouin zone (BZ). By chiral symmetry, these zero modes must satisfy
\begin{equation}
\mathcal{D}(\bm{r})\psi_{\bk}(\br)=0, \qquad \mathcal{D}^{\dagger}(\bm{r})\chi_{\bk}(\br)=0,
\end{equation}
where $\psi_{\bk}$ and $\chi_{\bk}$ are zero mode wavefunctions localized on the $A$ and $B$ sublattices, respectively. Since the Hamiltonian is governed by a single dimensionless parameter, $\alpha = w / vk_\theta$, the emergence of magic angles can be understood by tuning $\alpha$.

To construct the zero modes, we adopt an ansatz guided by earlier investigations \cite{tarnopolskyorigin2019,PhysRevResearch.6.L022025,PhysRevB.109.205411}. The combination of chiral symmetry and $\mathcal{C}_{3z}$ symmetry pins the Dirac cones to $E = 0$ at the $\mathcal{C}_{3z}$-symmetric momentum points $k_i = \gamma, \kappa, \kappa^{\prime}$. Since $\mathcal{D}$ involves only antiholomorphic derivatives, one finds that
\begin{equation}
D \left( F(z) \psi_{k_i}(\br)\right)=F(z)D \left(  \psi_{k_i}(\br)\right)=0,
\end{equation}
where $F(z)$ is any holomorphic function in the complex coordinate $z=x+iy$.
This observation naturally leads to the ansatz\cite{tarnopolskyorigin2019,ledwith2021168646,PhysRevResearch.6.L022025,PhysRevB.109.205411,PhysRevResearch.6.013165,PhysRevB.110.195112},
\begin{equation}
\label{eq_chiral_Bloch_wavefunction_general}
\psi_{\bk}(\br)=\sum_{k_i} F_{\bk}^{(k_i)}(z) c_{k_i} \psi_{k_i}(\br).
\end{equation}
By Liouville’s theorem, however, any holomorphic function that remains bounded must be constant, implying that the above form cannot represent a Bloch state at any momentum other than discrete points $k_i$. At first glance, this may suggest that zero modes are possible only at isolated crystal momenta. Nevertheless, there is an alternative: one can allow $F_{\bk}^{(k_i)}(z)$ to be a meromorphic function. Although such functions necessarily exhibit poles within the unit cell, at the magic angle one can arrange $c_{k_i}\psi_{k_i}(\br)$ such that the pole contributions cancel out in the sum.

In what follows, we construct the zero modes for  $\alpha\beta\gamma $ and  $\alpha\beta\alpha $ at the first magic angle(see Appendices \ref{sec_higher_magic_abc} and \ref{sec_higher_magic_aba} for analogous constructions at higher magic angles). We begin by supposing that each meromorphic function $F_{\bk}^{(k_i)}(z) $ develops a pole at some $C_{3z}$ symmetric point $\mathbf{r}_0$. Our task is then to ensure that Eq.\eqref{eq_chiral_Bloch_wavefunction_general} remains nonsingular by identifying coefficients $c_{k_i}$ such that the poles of $F_{k_i}$ are cancelled exactly.
There are two cases of interest 
\\
(1) The entire layer spinor $\psi_{k_i}(\br_0) = 0$ for some $k_i$ so that $\psi_{k_i}(\br) \propto(z - z_0) $ as $z \to z_0$.
\\
(2) No layer spinor vanishes entirely, but one find $c_{k_1} \psi_{k_1}(\br_0) + c_{k_2} \psi_{k_2}(\br_0) = 0$ with $c_{k_{1,2}} \neq 0$ so that $c_{k_1}\psi_{k_1}(\br) + c_{k_2} \psi_{k_2}(\br) \propto (z-z_0)$ for $z \to z_0$.

The scenario(1) is the case of TBG\cite{ledwith2021168646,tarnopolskyorigin2019}. $\Psi_i$ has a single parameter to be zero by tuning $\alpha$. The Bloch wavefunction for the zeromode can be written as
\begin{equation}
\label{eq_chiral_abc_1stMA_psi}
\psi_{\bk}(\br)=\phi_{\bk-k_i^{\prime}}\left(z-z_{0}\right) \psi_{k_i}(\br),
\end{equation}
where 
\begin{equation}
\phi_{\bm{k}}(z)=e^{\frac{i}{2} \bar{k} z} \frac{\sigma(z+i k)}{\sigma(z)}.
\end{equation}
where $k = k_x + i k_y$. Here, $\sigma(z)$ is the Weierstrass $\sigma$-function. $\sigma(z)$ has a pole of the order-one,  and satisfies $\sigma(-z)=-\sigma(z)$ and $\sigma(z+R)=\eta_{\bm{R}} e^{\frac{B}{2} \bar{R}(z+R / 2)} \sigma(z)$, where $\eta_{\bm{R}}=+1$ if $\bm{R} / 2$ is a lattice vector and $-1$ otherwise. The cell-periodic part of the wavefunction is 
\begin{equation}
u_\bk(\br)=e^{-ik\bar{z}}\frac{\sigma(z-z_0+i (k-k_i))}{\sigma(z-z_0)} \psi_{k_i}(\br)
\end{equation}
which is $k$-holomorphic function and host the same quantum geometry as lowest landau level, so called ideal condition. From the $k$-space boundary condition,
\begin{equation}
u_{\bk+\bm{G}_j}=u_{\bk}(\br)e^{-i \bm{G}_j \cdot \br} e^{i f_{\bk,\bm{G}_j}} 
\end{equation}
where $f_{\bk,G_j}=\bar{G}_j(-ik-iG_j/2)+\pi$ and using a formula for computing Chern number
\begin{equation}
C= \frac{1}{2\pi}\left( f_{\bk_0+\bm{G}_2,\bm{G}_1} + f_{\bk_0,\bm{G}_2} - f_{\bk_0,\bm{G}_1} - f_{\bk_0+\bm{G}_1,\bm{G}_2} \right),
\end{equation}
this Bloch wavefunction has $C=1$.

Scenario (2) occurs for the $C=2$ zero mode in HTTG, as shown in Refs. \cite{PhysRevResearch.6.L022025,PhysRevB.109.205411}. In order to satisfy $c_{k_1} \psi_{k_1} + c_{k_2} \psi_{k_2} = 0$, the two spinors are linearly dependent. 
The corresponding Bloch state can be written as
\begin{equation}
\begin{aligned}
\psi_{\bk}(\br)&= c_{k_1} a_{\bk-k_1} \phi_{\bk-k_1}(z-z_0) \psi_{k_1}(\br)\\
&+ c_{k_2}  a_{\bk-k_2}  \phi_{\bk-k_2 }(z-z_0) \psi_{k_2}(\br).
\end{aligned}
\end{equation}
with $a_\bk = \sigma(\bk)$. The boundary conditions lead to a Chern number $C=2$. Alternative constructions of ideal $C=2$ zero mode solutions are given in Refs. \cite{PhysRevLett.128.176404,sarkar2024ideal}.

The above constructions can be generalized, but for the primary magic angles we analyze in this section one of (1),(2) holds. Some generalizations include a linear combination of three terms that vanishes as  $(z - z_0) $, which would naturally lead to $C=3$ for suitable Bloch states $\psi_{k_i}$. For higher magic angles, detailed in the supplement, we obtain a higher order vanishing $\sum_ic_{k_i} \psi_{k_i}(\br_0) \propto (z-z_0)^2$ which leads to increased flat band degeneracy \cite{PhysRevB.108.L081124,httg_devakul}. Note that by replacing $z \to \ov{z}$ we can analogously construct zero modes for $\mathcal{D}^\dag$ that reside on the $B$ sublattice.

\subsection{$\alpha\beta\gamma$}
\label{sec_chiral_abc}

\begin{figure}
\begin{center}
\includegraphics[width=1\hsize]{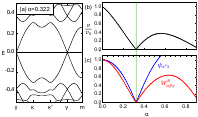}
\caption{(a)The band structure of Eq.\eqref{eq_Hamiltonian_chiral_general} for $\alpha\beta\gamma$ with $\alpha=0.322$. (b)The velocity of Dirac cone at $\kappa$, normalized by the velocity of $\alpha=0$, as a function of $\alpha$. (c) $\psi_{\kappa,3}(-\br_{BA})$ and $W_{\alpha\beta\gamma}^{B}$ [Eq.\eqref{eq_chiral_abc_1stMA_Bsub_W2}] are plotted with blue and red lines, respectively.}
\label{fig_chiral_abc_1stMA}
\end{center}
\end{figure}

The chiral Hamiltonian for the $\alpha\beta\gamma$ is given by
\begin{equation}                                         
D=\left[\begin{array}{cccc}-2 i \bar{\partial} & \alpha U_{0}(\mathbf{r}) & 0 & 0\\
 \alpha U_{-1}(-\mathbf{r}) & -2 i \bar{\partial} & \alpha U_{+1}(\mathbf{r}) & 0 \\
0 &\alpha U_{+1}(-\mathbf{r}) & -2 i \bar{\partial} & \alpha U_{-1}(\mathbf{r})\\
0 & 0 &\alpha U_{0}(-\mathbf{r}) & -2 i \bar{\partial}
\end{array}\right].
\end{equation}
with $\bar{\partial}=\frac{1}{2}(\partial_x+i\partial_y)$. To investigate the dependence of the single particle properties on the parameter $\alpha$, we calculate the velocity of the Dirac cone, show in Fig.~\ref{fig_chiral_abc_1stMA}. 
At $\alpha = 0.322$ the bands become exactly flat, such that there are zero modes of $D$ and $D^\dag$ for all $\bk$. We construct these zero modes below and compute their Chern numbers.

We first consider the eigenstate of $D\psi_{\bk}=0$. 
At $\br=-\br_{BA}$, the $\mathcal{C}_{3z}$ eigenvalue of the $\psi_{\kappa}$ for layer componets are $\rm{diag}(\omega, \omega,1,\omega)$.
By using it and \eqref{eq_translation}, the layer spinor around $-\br_{BA}$ has the form
\begin{equation}
\psi_{\kappa^{\prime}}(\br) = \left[\begin{array}{c}0 \\ 0 \\ \psi_{\kappa^{\prime}, {3}} \\ 0\end{array}\right]+O(z),
\end{equation}
with $\psi_{\kappa^{\prime},3}=\psi_{\kappa^{\prime},3}(-\br_{BA})$. As shown in Fig.~\ref{fig_chiral_abc_1stMA}, at the magic angle, the numerical results indicate $\psi_{\kappa^{\prime}, {3}}=0$, leading to $\psi_{\kappa^{\prime}}(\br) \sim z$. Thus, this is the case for scenario(1). The Bloch wavefunction can be expressed as 
\begin{equation}
\psi_{\bk}(\br)=\phi_{\bk-\kappa^{\prime}}\left(z-z_{0}\right) \psi_{\kappa^{\prime}}(\br).
\end{equation}
The Chern number is $C_A=+1$.

Next, we examine B sublattice polarized wavefunction.  Actually, this case does not match with scenario(1) but scenario (2). 
At $\br=0$, the $\mathcal{C}_{3z}$ eigenvalue of $\chi_{\kappa/\kappa^\prime}$ is given by $\rm{diag}(\omega, 1,1,\omega)$ for layer basis.
By using this $\mathcal{C}_{3z}$ eigenvalues for layers and Eq.\eqref{eq_translation}, we can write the eigenstate as
\begin{equation}
\chi_{\kappa}(\br) = \left[\begin{array}{c}0 \\ \chi_{\kappa, 2} \\ \chi_{\kappa, 3} \\ 0\end{array}\right]+O(\bar{z}).
\end{equation}
where $\chi_{\kappa, 2} = \chi_{\kappa, 2}(0)$ and $\chi_{\kappa, 3}  = \chi_{\kappa, 3}(0)$. The wavefunction at $\kappa^{\prime}$ is obtained from $\chi_{\kappa^{\prime}}(\br)=P\mathcal{C}_{2z}T \chi_{\kappa}(-\br)$ as
\begin{equation}
\chi_{\kappa^{\prime}}(\br) = \left[\begin{array}{c}0 \\ -\chi_{\kappa, 3} \\ \chi_{\kappa, 2} \\ 0\end{array}\right]+O(\bar{z}).
\end{equation}
Here, by using $\mathcal{P}\mathcal{C}_{2x}\chi_{\kappa}(\br)=\chi_{\kappa}, (\mathcal{M}_y\br)^*$, $\chi_{\kappa, 2}$ is real and $\chi_{\kappa, 3}$ is imaginary. 

In the following, numerically, we will verify the linear dependence of the two layer spinor by computing
\begin{equation}
\label{eq_chiral_abc_1stMA_Bsub_W2}
W_{\alpha\beta\gamma}^{B}=\left[\begin{array}{c}\chi_{\kappa, 2} \\ \chi_{\kappa, 3} \end{array}\right] \times
\left[\begin{array}{c}-\chi_{\kappa, 3} \\ \chi_{\kappa, 2} \end{array}\right].
\end{equation}
As shown in Fig.~\ref{fig_chiral_abc_1stMA}, at the magic angle, $W_{\alpha\beta\gamma}^{B}=0$. Thus, we can construct the Bloch wavefunction as
\begin{equation}
\label{eq_chiral_abc_1stMA_chi}
\chi_{\bm{k}}(\bm{r})=\bar{a}_{\bk-\kappa^{\prime}} \bar{\phi}_{\bk-\kappa}(\bar{z}) \chi_{\kappa}(\br) +i \bar{a}_{\bk-\kappa} \bar{\phi}_{\bk-\kappa^{\prime}}(\bar{z}) \chi_{\kappa^{\prime}}(\br)
\end{equation}
where we choose the gauge such as $\chi_{K,2}=+ i\chi_{K,3}$ The Chern number is $C_B=-2$.

At the magic angle, $\psi$ and $\chi$ satisfy the orthogonality relation:
\begin{equation}
\label{eq:chiral_velocity}
v(\br)=\psi_{\bk_1}(\br)\cdot\bar{\chi}_{\bk_2}(\br)=0
\end{equation}
for $\bk_1\neq\bk_2$.\cite{tarnopolskyorigin2019} From Eqs.\eqref{eq_chiral_abc_1stMA_psi} and \eqref{eq_chiral_abc_1stMA_chi}, the total dimension spanned by the layer spinors is three. To span the full layer spinor space, we require another eigenstate. The eigenstate corresponds to the finite velocity Dirac cone at $\Gamma$ point (the detail is presented in appendix.\ref{sec_magic_dirac}).

\subsection{$\alpha\beta\alpha$}
\label{sec_chiral_aba}

\begin{figure}
\begin{center}
\includegraphics[width=1\hsize]{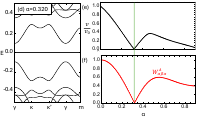}
\caption{(a)The band structure of Eq.\eqref{eq_Hamiltonian_chiral_general} for $\alpha\beta\alpha$ with $\alpha=0.320$. (b)The velocity of Dirac cone at $\kappa$, normalized by the velocity of $\alpha=0$, as a function of $\alpha$. (c) $\psi_{\kappa,3}(-\br_{BA})$ and $W_{\alpha\beta\alpha}^{A}$ [Eq.\eqref{eq_chiral_aba_1stMA_Asub_W2}] is plotted with red line.}
\label{fig_chiral_aba_1sMA}
\end{center}
\end{figure}

The chiral Hamiltonian for $\alpha\beta\alpha$ is written as
\begin{equation}                                         
D=\left[\begin{array}{cccc}-2 i \bar{\partial} & \alpha U_{0}(\mathbf{r}) & 0 & 0\\
 \alpha U_{-1}(-\mathbf{r}) & -2 i \bar{\partial} & \alpha U_{+1}(\mathbf{r}) & 0 \\
0 &\alpha U_{+1}(-\mathbf{r}) & -2 i \bar{\partial} & \alpha U_{-1}(\mathbf{r})\\
0 & 0 &\alpha U_{0}(-\mathbf{r}) & -2 i \bar{\partial}
\end{array}\right].
\end{equation}
As shown in Fig.~\ref{fig_chiral_aba_1sMA}, the velocity of Dirac cones vanishes at $\alpha=0.320$, which closely matches the first magic angle of $\alpha\beta\gamma$. In the following, we explain the origin of the magic angle, noting that the discussion is applicable for other odd magic angle (see appendix\ref{sec_higher_magic_aba} for higher magic angles).

This is the case of the scenario (2).
At $\br=0$, the $\mathcal{C}_{3z}$ eigenvalues of $\psi_{\kappa/\kappa^\prime}$ are given by $\rm{diag}(\omega, 1,1,\omega^*)$ for layer basis.
By using this $\mathcal{C}_{3z}$ eigenvalues, the A sublattice polarized wavefunction for the $\kappa$ and $\kappa^{\prime}$ points are given by
\begin{equation}
\psi_{\kappa/\kappa^\prime}(\br) = \left[\begin{array}{c}  0 \\ \psi_{\kappa/\kappa^\prime,2}  \\ \psi_{\kappa/\kappa^\prime,3} \\ \psi^\prime_{\kappa/\kappa^\prime,4} \bar{z} \end{array}\right]+O(z).
\end{equation}
where $\psi_{\kappa/\kappa^\prime,2}=\psi_{\kappa/\kappa^\prime,2}(0)$ is real, $\psi_{\kappa/\kappa^\prime,3}=\psi_{\kappa/\kappa^\prime,3}(0)$ is pure imaginary and
$\psi^\prime_{\kappa/\kappa^\prime,4}=\bar{\partial}_z\psi_{\kappa/\kappa^\prime,2}(0)$ is real,.
These components are derived using the symmetry relation  $\mathcal{P}\mathcal{C}_{2x}\psi_{\kappa}(\br)=\psi_{\kappa}(\mathcal{M}_y\br)^*$.
Moreover, by using $D^\dagger \chi_{\kappa/\kappa^\prime}=0$, one find $\psi^\prime_{\kappa/\kappa^\prime,4} = -3i\alpha \psi_{\kappa/\kappa^\prime,3} $.

To prove the linear dependence of the two layer spinor at the magic angle, we numerically check
\begin{equation}
\label{eq_chiral_aba_1stMA_Asub_W2}
W_{\alpha\beta\alpha}^{A}=\left[\begin{array}{c} \psi_{\kappa,2}  \\ \psi_{\kappa,3} \end{array}\right] \times \left[\begin{array}{c} \psi_{\kappa^\prime,2}  \\ \psi_{\kappa^\prime,3} \end{array}\right]
\end{equation}
is zero.
As shown in Fig.~\ref{fig_chiral_aba_1sMA}, $W_{\alpha\beta\alpha}^{A}$ is zero at the magic angle. Thus, around $\br=0$
\begin{equation}
c_{\kappa}\psi_{\kappa}(\br)-c_{\kappa^\prime}\psi_{\kappa^{\prime}}(\br)\sim z
\end{equation}
with $c_{\kappa}=\psi_{\kappa^{\prime},2}$ and $c_{\kappa^\prime}=\psi_{\kappa,2}$. The resulting Blcoh wavefunction can be given by
\begin{equation}
\begin{aligned}
\psi_{\bk}(\br)&= c_\kappa a_{\bk-\kappa^{\prime}} \phi_{\bk-\kappa}(z) \psi_{\kappa}(\br)\\
&+ c_{\kappa^\prime} a_{\bk-\kappa}  \phi_{\bk-\kappa^{\prime} }(z) \psi_{\kappa^{\prime}}(\br).
\end{aligned}
\end{equation}
The Chern number for the A sublattice polarized state is $C_A=+2$. The B sublattice polarized state $\chi_\bk$ is obtained from the $\mathcal{C}_{2x}$ operation to $\psi_\bk$, resulting in $C_B=-2$.

\subsection{geometrical origin of a finite velocity Dirac cone}
\label{sec_magic_dirac}
At the magic angle, the wavefunction of the Dirac cone at $\gamma$ can be explicitly constructed using the geometric relationship between $\psi$ and $\chi$.
$\chi$ and $\psi$ satisfies the orthogonality relation.~\ref{eq:chiral_velocity}
In addition, since $\psi_\bk = C \psi_{\kappa^{\prime}}$ and $\chi_\bk = C_1 \chi_{\kappa} + C_2 \chi_{\kappa^{\prime}}$, 
the dimensions of the layer spinor are one for $\psi_\bk$ and two for $\chi_\bk$, respectively.
This allows us to generate an additional zeromode that satisfies
\begin{equation} 
\psi_{-\bk_1-\bk_2-\bk_3}=\bar{\phi}_{\bk_1} \wedge \bar{\chi}_{\bk_2} \wedge \bar{\chi}_{\bk_3}
\end{equation}
where $\wedge$ denotes the wedge product.
For the construction of the Dirac cone at $\gamma$, we choose $(\bk_1,\bk_2,\bk_3)=(\gamma,\kappa,\kappa^{\prime})$.
Since $D\psi_{\bk}=0$, we find
\begin{equation}
D^T[\bar{\phi}_{\gamma}] \wedge \bar{\chi}_{\kappa} \wedge \bar{\chi}_{\kappa^\prime}=0.
\end{equation}
Here, we used $D^\dagger\chi_\bk=0$.
This indicates that $D^T\bar{\phi}_{\gamma}$ is on the plane spanned by $\bar{\chi}_{\kappa}$ and $\bar{\chi}_{\kappa^{\prime}}$ such as
\begin{equation}
D^T\bar{\phi}_{\gamma} = C_1(\br) \bar{\chi}_{\kappa} + C_2(\br) \bar{\chi}_{\kappa^{\prime}}.
\end{equation}
with $C_1(\br+\bm{R}) = e^{i\kappa \cdot \bm{R}}C_1(\br)$ and $C_2(\br+\bm{R}) = e^{i\kappa^\prime \cdot \bm{R}}C_2(\br)$.
By introducing a shift $\bar{\phi}_{\gamma} \rightarrow \bar{\phi}_{\gamma} - B_1(\br)\bar{\chi}_{\kappa}-B_2(\br)\bar{\chi}_{\kappa^{\prime}}$ such that $\bar{\partial} B_i(\br)=C_i(\br)$, this mode satisfies $D^\dagger\phi_{\gamma}=0$.
In the context of HTTG, the construction of the Dirac cones at the magic angle has been previously discussed in detail by \cite{PhysRevB.109.205411, PhysRevResearch.5.043079}.

\subsection{higher magic angles for $\alpha\beta\gamma$}
\label{sec_higher_magic_abc}
\begin{table}[]
\begin{tabular}{|c|c|c|c|c|c|c|c|c|}
\hline
& $\alpha_1$ & $\alpha_2$ & $\alpha_3$ & $\alpha_4$ & $\alpha_5$ & $\alpha_6$ & $\alpha_7$ & $\alpha_8$ \\ \hline
$\alpha\beta\gamma$ & 0.322                    & 0.931                    & 1.30                     & 2.12                     & 2.39                     & 3.09                     & 3.45     & 4.18                \\ \hline
$\alpha\beta\alpha$ & 0.320                    & 0.945                    & 1.50                     & 2.40                     & 2.62                     & 3.69                     & 4.08    & 4.78                   \\ \hline
\end{tabular}
\caption{}
\label{table_chiral_magic_angles}
\end{table}

\begin{figure}
\begin{center}
\includegraphics[width=1\hsize]{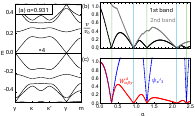}
\caption{(a)The band structure of Eq.\eqref{eq_Hamiltonian_chiral_general} for $\alpha\beta\gamma$ with $\alpha=0.945$. (b)The black line reprensents the velocity of Dirac cone at $\kappa^\prime$ (black line), normalized by the velocity of $\alpha=0$. The gray line denotes the velocity of the second band.(c) $\psi_{\kappa^\prime,3}(-\br_{BA})$ and $W_{\alpha\beta\gamma}^{A}$ [Eq.\eqref{eq_chiral_abc_2ndMA_Asub_W2}] are plotted with blue and red lines, respectively.}
\label{fig_chiral_abc_magic}
\end{center}
\end{figure}

In Section~\ref{sec_chiral_abc}, the analysis was about the first magic angle, where two flat bands emerge at charge neutrality. In this section, we explain about the origin of the higher magic angles. As illustrated in Fig.~\ref{fig_chiral_abc_magic}, the velocity of Dirac cone (shown in black) vanishes at each magic angle, while the first remote bands (shown in gray) also become flat at the even magic angles. 
For instance, as shown in Fig.~\ref{fig_chiral_abc_magic}, at the second magic angle, 
$\alpha=0.931$, four flat bands appear at zero energy.
From this perspective, the magic angles can be classified into two categories:
\begin{itemize}
    \item Odd magic angles $\alpha_{2n-1}$ exhibit two flat bands 
    \item Even magic angles $\alpha_{2n}$ exhibit four flat bands
\end{itemize}
Table~\ref{table_chiral_magic_angles} lists the computed values of the magic angles.
The approximate constancy of the difference $\alpha_{2n+2}-\alpha_{2n}\approx \alpha_{2n+1}-\alpha_{2n-1}\approx 1.05$ implies a regular spacing between successive angles, in agreement with prior numerical studies(see Ref.\cite{PhysRevB.110.195112} for the emergence of the first three magic angles).

For the odd magic angle, the same scenario in Sec.\ref{sec_chiral_abc} is applicable. For example, for A sublattice, $\psi_{\kappa^\prime,3}(\br_0)=0$ (see Fig.~\ref{fig_chiral_abc_magic}) leads to the Bloch wavefunction [Eq.\eqref{eq_chiral_abc_1stMA_psi}].
For B sublattice, it is readily to show that $W_{\alpha\beta\gamma}^{B}=0$ allows us to construct the Bloch wavefunction [Eq.\eqref{eq_chiral_abc_1stMA_chi}].

At the even magic angle, we will see that $D$ has two zeromodes solution in the entire BZ. Here, we explain how we construct the two Bloch wavefunctions.
In the first magic angle, there are two cases such that $\psi_{k_i}(\br) \propto (z-z_0)$ as $z \rightarrow z_0$ for some $k_i$ and $c_{k_1}\psi_{k_1}(\br) + c_{k_2}\psi_{k_2}(\br)=(z-z_0)$ with $c_{k_{1,2}}\neq0$ as $z \rightarrow z_0$.
They can cancel a pole originated from $F_\bk^{(k_i)}(z)$.
On the other hand, for the second magic angle of $\alpha\beta\gamma$, we will see that the cases of interest are
\\
(1) The entire layer spinor $\psi_{k_i}(\br_0) = 0$ for some $k_i$ so that $\psi_{k_i}(\br) \propto(z - z_0)^2 $ as $z \to z_0$,
\\
(2) No layer spinor vanishes entirely, but one finds $c_{k_1} \psi_{k_1}(\br_0) + c_{k_2} \psi_{k_2}(\br_0) = 0$ with $c_{k_{1,2}} \neq 0$ so that $c_{k_1}\psi_{k_1}(\br) + c_{k_2} \psi_{k_2}(\br) \propto (z-z_0)^2$ for $z \to z_0$.
In these case, we can cancel the pole up to order-two.

The scenario (1) also appears in the second magic angle HTTG.\cite{PhysRevB.109.205411, PhysRevB.108.L081124,PhysRevResearch.5.043079}
Mathematically, $\psi_{k_i}(\br) \propto(z - z_0)^2 $ as $z \to z_0$ enables to host the following analytical zero-energy solution\cite{PhysRevB.109.205411}
\begin{equation}
\label{eq_higher_magic_abc_C=1}
\psi_{\bk}(\br)=\phi_{-\bk^{\prime}}\left(z+z_{0}\right) \phi_{\bk+\bk^{\prime}-k_i}\left(z+z_{0}\right) \psi_{k_i}(\br),
\end{equation}
where $\bk^\prime$ is an arbitrary wave vector and most two distinct $\bk^\prime$ create different Bloch wavefunction.
For instance, we choose $\bk^\prime=0$, leading
\begin{equation}
\psi_{\bk}^{(1)}(\br)= \phi_{\bk-k_i}(z+z_0)\psi_{k_i}(\br),
\end{equation}
and $\bk^\prime=k_i$ gives
\begin{equation}
\psi_{\bk}^{(2)}(\br)= \phi_{\bk}(z+z_0) \phi_{-k_i}(z+z_0) \psi_{k_i}(\br).
\end{equation}

For scenario (2), the analytical zero-energy solution is given by
\begin{equation}
\label{eq_higher_magic_abc_C=2}
\begin{aligned}
\psi_{\bk}(\br)&= c_1a_{\bk-k_2 + \bk^\prime} \phi_{-\bk^\prime}(z) \phi_{\bk-k_1 +\bk^\prime}(z) \psi_{k_1}(\br)\\
&+ c_2 a_{\bk-k_1 + \bk^\prime} \phi_{-\bk^\prime}(z) \phi_{\bk-k_2 +\bk^\prime}(z) \psi_{k_2}(\br).
\end{aligned}
\end{equation}
where we choose a gauge such as $\psi_{\kappa,3}=+ i\psi_{\kappa,2}$.
$\bk^\prime$ is arbitrary vector and at most two distinct $\bk^\prime$ create different Bloch wavefunctions.
Here, for example, we take $\bk^\prime=0$
\begin{equation}
\begin{aligned}
\psi_{\bk}^{(1)}(\br)&= c_1 a_{\bk-k_2} \phi_{\bk-k_1 }(z) \psi_{k_1}(\br)\\
&+ c_2 a_{\bk-k_1 } \phi_{\bk-k_2 }(z) \psi_{k_2}(\br).
\end{aligned}
\end{equation}
and $\bk^\prime=k_2$ 
\begin{equation}
\begin{aligned}
\psi_{\bk}^{(2)}(\br)&= c_1 a_{\bk} \phi_{-k_2}(z) \phi_{\bk}(z) \psi_{k_1}(\br)\\
&+ c_2 a_{\bk-k_2} \phi_{-k_2}(z) \phi_{\bk}(z) \psi_{k_2}(\br).
\end{aligned}
\end{equation}

The procedure to identify which scenario the zero-energy solution belongs to is the same as the main text. Combining with the symmetric analysis and the solution of $D\psi_{k_i}=0$ around $\mathcal{C}_{3z}$ points in real space, we obtain the wavefunction by $O(z^2)$. Then, numerically, we check if the wavefunction itself or the linear combination for $\kappa$ and $\kappa^\prime$ is $\propto (z-z_0)^2$.

We begin with A sublattice polarized state.
This belongs to scenario (1).
By solving $D \psi_{\kappa}=0$ around $\br=0$ with the $\mathcal{C}_{3z}$ eigenvalue for $\psi_{\kappa}$, $\mathrm{diag}(\omega^*,1,1,\omega^*)$, the zeromode is given by
\begin{equation}
\psi_{\kappa}(\br) = \frac{4i}{3\alpha}\left[\begin{array}{c}  0 \\ \psi_{\kappa,2}  \\ \psi_{\kappa,3} \\ 0 \end{array}\right]+\left[\begin{array}{c} 2 \psi_{\kappa,2} \bar{z} \\ \psi_{\kappa,3} z\bar{z}  \\ -\psi_{\kappa,2} z\bar{z} \\ 2 \psi_{\kappa,3} \bar{z}  \end{array}\right] + O(z^2)
\end{equation}
where $\psi_{\kappa, 2} = \psi_{\kappa, 2}(0)$ and $\psi_{\kappa, 3}  = \psi_{\kappa, 3}(0)$.
By using $\mathcal{P}\mathcal{C}_{2x}\psi_{\kappa}(\br)=\psi_{\kappa}, (\mathcal{M}_y\br)^*$, we show $\psi_{\kappa, 2}$, $\psi_{\kappa, 2/3}$ is real and pure imaginary, respectively.
The wavefunction at $\kappa^{\prime}$ is obtained from $\psi_{\kappa^{\prime}}(\br)=P\mathcal{C}_{2z}T \psi_{\kappa}(-\br)$ as
\begin{equation}
\psi_{\kappa}(\br) = \frac{4i}{3\alpha}\left[\begin{array}{c}  0 \\ -\psi_{\kappa,3}  \\ \psi_{\kappa,2} \\ 0 \end{array}\right]+\left[\begin{array}{c} -2 \psi_{\kappa,3} \bar{z} \\ \psi_{\kappa,2} z\bar{z}  \\ \psi_{\kappa,3} z\bar{z} \\ 2 \psi_{\kappa,2} \bar{z}  \end{array}\right] + O(z^2)
\end{equation}
The geometric relationship between the two spinors is analyzed by computing
\begin{equation}
\label{eq_chiral_abc_2ndMA_Asub_W2}
	W_{\alpha\beta\gamma}^{A}= \left[\begin{array}{c}  \psi_{\kappa,2}  \\ \psi_{\kappa,3}  \end{array}\right] \times \left[\begin{array}{c}  -\psi_{\kappa,3}  \\ \psi_{\kappa,2}  \end{array}\right].
\end{equation}
At the magic angle, $W_{\alpha\beta\gamma}^{A}=0$ as shown in Fig.~\ref{fig_chiral_abc_magic} (blue line).
It implies that the two spinors are linearly dependent, with $\psi_{\kappa,3}=\pm i\psi_{\kappa,2}$.
This gives the linear combination of $\psi_{\kappa}$ and $\psi_{\kappa^\prime}$ $ \propto z^2$, thus the resulting Bloch wavefunction is Eq.\eqref{eq_higher_magic_abc_C=2} with $c_1 =1$ and $c_2=i$, where we choose a gauge such as $\psi_{\kappa,3}=+ i\psi_{\kappa,2}$.
There are two-fold degenerate flat bands for A sublattice, with each band hosting  $C_A=+2$.

Let us consider the zeromode of B sublattice.
This belongs to scenario (2).
By solve $D^\dagger \chi_{\kappa^{\prime}}(\br)=0$ around $-\br_{BA}$ and the $C_{3z}$ eigenvalue of $\chi_{\kappa^{\prime}}(-\br_{BA})$, $\mathrm{diag}(\omega,\omega,1,\omega)$, we obtain
\begin{equation}
\chi_{\kappa^{\prime}}(\br) = \left[\begin{array}{c} 0 \\ 0  \\ \chi_{\kappa^{\prime},3} \\0 \end{array}\right]+\frac{4i}{3\alpha}\left[\begin{array}{c} 0 \\ \chi_{\kappa^{\prime},3}z  \\ 0 \\ \chi_{\kappa^{\prime},3} z \end{array}\right]+O(\bar{z}^2)
\end{equation}
with $\chi_{\kappa^{\prime},3}=\chi_{\kappa^{\prime},3}(-\br_{BA})$.
As shown in Fig.~\ref{fig_chiral_abc_magic}, at the magic angle, the numerical result indicates $\chi_{\kappa^{\prime},3}=0$, resulting in $\chi(\br)\propto \bar{z}^2$.
Therefore, the Bloch wavefunction is Eq.\eqref{eq_higher_magic_abc_C=1}.
Thus, at the even magic angles, $D^\dagger$ has two flat bands and the Chern number for each band is $C_B=-1$.

At the even magic angles, $\gamma$ point has a Dirac cone with a finite velocity which no symmetry protects.
Since $\psi_{\bk}^{(i)}$ lies on the plane spanned by $\psi_\kappa \wedge \psi_\kappa^\prime$ and $\chi_{\bk}^{(i)} \parallel \chi_\kappa^\prime$, the total dimension of the layer spinors is three.
Therefore, following the discussion in appendix~\ref{sec_magic_dirac}, we find an additional dimension to create a Dirac point at $E=0$.

\subsection{higher magic angles for $\alpha\beta\alpha$}
\label{sec_higher_magic_aba}
\begin{figure}
\begin{center}
\includegraphics[width=1\hsize]{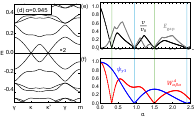}
\caption{(a)The band structure of Eq.\eqref{eq_Hamiltonian_chiral_general} for $\alpha\beta\alpha$ with $\alpha=0.945$. (b)The velocity of Dirac cone at $\kappa$, normalized by the velocity of $\alpha=0$, as a function of $\alpha$. (c) $\psi_{\gamma,1}(0)$ and $W_{\alpha\beta\alpha}^A$ [Eq.\eqref{eq_chiral_aba_1stMA_Asub_W2}] are plotted with blue and red lines, respectively.}
\label{fig_chiral_aba_magic}
\end{center}
\end{figure}

In Section~\ref{sec_chiral_aba}, we focused on the first magic angle, where $C_{A/B}=\pm2$ flat bands emerge for $\alpha\beta\alpha$.
In this section, we will see the electronic structure at higher magic angles.
At these higher magic angles, distinct features in the band structure appear.
For instance, at the second magic angle, $\alpha=0.945$, the band structure exhibits two flat bands and four-fold degeneracy at $\kappa$ and $\kappa^\prime$ at zero energy (see Figure~\ref{fig_chiral_aba_magic}).
Figure~\ref{fig_chiral_aba_magic} also plots the bandwidth of the flat bands and the gap between the flat bands and the remote bands. The plot shows that we can classify the series of magic angles into odd and even as follow.
\begin{itemize}
    \item odd magic angle $\alpha_{2n-1}$ has two isolated flat bands
    \item even magic angle $\alpha_{2n}$ has tow isolated flat bands with four-fold degeneracy at $\kappa$ and $\kappa^\prime$
\end{itemize}
Table~\ref{table_chiral_magic_angles} lists the computed values of these angles. 
The intervals between magic angles are approximately given by $\alpha_{2n+2}-\alpha_{2n}\approx \alpha_{2n+1}-\alpha_{2n-1}\approx 1.26$.

The origin of flat bands at odd magic angles can be explained using the same scenario as Sec.\ref{sec_chiral_aba}.
The red line in Fig.~\ref{fig_chiral_aba_magic} represents $W_{\alpha\beta\alpha}^{A}$.
At the odd magic angle, $W_{\alpha\beta\alpha}^{A}=0$, indicating that the two spinors $\chi_{\kappa}$ and $\chi_{\kappa^\prime}$ are linearly dependent.
The resulting Bloch wavefunction has Chern number $C_A=2$.

In the following, we delve into the origin of flat bands at even magic angles.
As we will see the zero energy mode belong to the scenario (1) in Appendix~\ref{sec_chiral_abc}.
We examine the zeromode at $\gamma$ point, which can be obtained from solving $D\psi_\gamma=0$ around $\br=0$ and the $C_{3z}$ eigenvalue of $\psi_{\gamma}(0)$, $\mathrm{diag}(1,\omega^*,\omega^*,\omega)$:
\begin{equation}
\psi_{\gamma}(\br) = \frac{1}{4A}\left[\begin{array}{c} 4A\psi_{\Gamma,1} \\ \psi_{\Gamma,1} \bar{z}\\0 \\ 0 \end{array}\right]+ O(z)
\end{equation}
with $A=-3i\alpha/4$.
At the magic angle, Fig.~\ref{fig_chiral_aba_magic} shows that $\psi_{\Gamma,1}=0$, resulting in $\psi_{\gamma}(\br)\sim z $.
Thus, we can construct the Bloch wavefunction
\begin{equation} 
\psi_{\bk}(\br)= \phi_{\bk}(z) \psi_{\gamma}(\br).
\end{equation}
For B sublattice, $\chi_\bk$ is obtained from $\mathcal{C}_{2x}$ operation to $\psi_\bk$.
Consequently, the Bloch wavefunction for A and B sublattice $C_{A/B}= \pm 1$, respectively.
$\psi_{\bk}(\br)$ and $\chi_{\bk}(\br)$ are spanned by one-dimensional layer spinor, respectively so that there are two additional dimension which are not spanned by the two wavefunction. By repeating the similar approach as appendix~\ref{sec_magic_dirac}, one can construct the two additional zeromodes, which appear as the Dirac cone at $\kappa$ and at $\kappa^\prime$.

\section{Breaking of symmetries}
\label{sec_breaking_symmetry}

\begin{figure}
\begin{center}
\includegraphics[width=1\hsize]{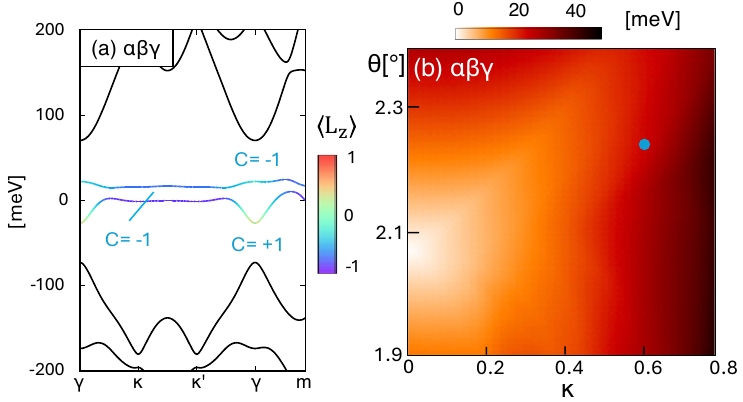}
\caption{(a)The band structure for $\alpha\beta\gamma$ for $\theta=2.25^\circ$. The gap Chern numbers in $\alpha\beta\gamma$ are $\pm1$ for gaps. (b)The bandwidth as a function of $\kappa$ and $\theta$ for $\alpha\beta\gamma$}
\label{fig_abc_band_width}
\end{center}
\end{figure}

Figure~\ref{fig_htqg_summary}(d) shows the band structure obtained for the $\alpha\beta\alpha$ domain. Because the $\mathcal{C}_{3z}$ and $\mathcal{P}\mathcal{C}_{2x}$ protects Dirac cones at $\kappa$ and $\kappa^\prime$ points, the particle-hole symmetry breaking gaps these Dirac cones, while the Dirac cone at the $\gamma$ point remains protected by the $\mathcal{C}_{2x}$ symmetry.

In Fig.~\ref{fig_abc_band_width}, we plot the corresponding (a) band structure for the $\alpha\beta\gamma$ and (b)the bandwidth for $\kappa$ and $\theta$. Here, again, because the $\mathcal{C}_{3z}$ and $\mathcal{P}\mathcal{C}_{2x}$ protects Dirac cones at $\kappa$ and $\kappa^\prime$ points, the particle-hole symmetry breaking opens gaps at the $\kappa$ and $\kappa^\prime$ Dirac cones. On the other hand, the mechanism of the opening gap for the four-fold dgeneracy at $\gamma$ point is different. Originally, the combination of $\mathcal{C}_{3z}$, $\mathcal{C}_{2y}T$ and the chiral symmetry protects the four-fold degeneracy. This arises because $\mathcal{C}_{2y}T$ imposes different angular momentum constraints on A(B) sublattice of layer 1 and B(A) sublattice of layer 4, as Eq.\eqref{eq_abc_C3z} shows. Thus, $\kappa>0$ strongly mixes the Dirac cone at $\gamma$ with the nearby flat bands. One can see this mixing explicitly by examining the expectation value of $ L_z = \mathcal{P}_{\rm Dirac} - \mathcal{P}_{\rm flat}$, where $\mathcal{P}_{\rm Dirac/flat}$ project onto the subspaces spanned by the Dirac and flat-band states of the chiral limit, respectively. Near the $\gamma$ point, $\langle L_z\rangle \approx 0$, indicating that the Dirac cone and the flat band subspaces become almost maximally entangled.

It is worth to comment on the effect of the displacement field.
In the $\alpha\beta\alpha$, the displacement field breaks the $\mathcal{C}_{2x}$ symmetry, thereby gaping out the Dirac cone at $\gamma$. However, the displacement field only weakly affects the low-energy electronic structure in $\alpha\beta\alpha$. $\alpha\beta\gamma$ shows the similar behavior such that the displacement field does not change the electronic structure much.

\bibliography{htqg.bib}

\begin{thebibliography}{152}%
\makeatletter
\providecommand \@ifxundefined [1]{%
 \@ifx{#1\undefined}
}%
\providecommand \@ifnum [1]{%
 \ifnum #1\expandafter \@firstoftwo
 \else \expandafter \@secondoftwo
 \fi
}%
\providecommand \@ifx [1]{%
 \ifx #1\expandafter \@firstoftwo
 \else \expandafter \@secondoftwo
 \fi
}%
\providecommand \natexlab [1]{#1}%
\providecommand \enquote  [1]{``#1''}%
\providecommand \bibnamefont  [1]{#1}%
\providecommand \bibfnamefont [1]{#1}%
\providecommand \citenamefont [1]{#1}%
\providecommand \href@noop [0]{\@secondoftwo}%
\providecommand \href [0]{\begingroup \@sanitize@url \@href}%
\providecommand \@href[1]{\@@startlink{#1}\@@href}%
\providecommand \@@href[1]{\endgroup#1\@@endlink}%
\providecommand \@sanitize@url [0]{\catcode `\\12\catcode `\$12\catcode `\&12\catcode `\#12\catcode `\^12\catcode `\_12\catcode `\%12\relax}%
\providecommand \@@startlink[1]{}%
\providecommand \@@endlink[0]{}%
\providecommand \url  [0]{\begingroup\@sanitize@url \@url }%
\providecommand \@url [1]{\endgroup\@href {#1}{\urlprefix }}%
\providecommand \urlprefix  [0]{URL }%
\providecommand \Eprint [0]{\href }%
\providecommand \doibase [0]{https://doi.org/}%
\providecommand \selectlanguage [0]{\@gobble}%
\providecommand \bibinfo  [0]{\@secondoftwo}%
\providecommand \bibfield  [0]{\@secondoftwo}%
\providecommand \translation [1]{[#1]}%
\providecommand \BibitemOpen [0]{}%
\providecommand \bibitemStop [0]{}%
\providecommand \bibitemNoStop [0]{.\EOS\space}%
\providecommand \EOS [0]{\spacefactor3000\relax}%
\providecommand \BibitemShut  [1]{\csname bibitem#1\endcsname}%
\let\auto@bib@innerbib\@empty
\bibitem [{\citenamefont {Cao}\ \emph {et~al.}(2018{\natexlab{a}})\citenamefont {Cao}, \citenamefont {Fatemi}, \citenamefont {Fang}, \citenamefont {Watanabe}, \citenamefont {Taniguchi}, \citenamefont {Kaxiras},\ and\ \citenamefont {Jarillo-Herrero}}]{cao2018unconventional}%
  \BibitemOpen
  \bibfield  {author} {\bibinfo {author} {\bibfnamefont {Y.}~\bibnamefont {Cao}}, \bibinfo {author} {\bibfnamefont {V.}~\bibnamefont {Fatemi}}, \bibinfo {author} {\bibfnamefont {S.}~\bibnamefont {Fang}}, \bibinfo {author} {\bibfnamefont {K.}~\bibnamefont {Watanabe}}, \bibinfo {author} {\bibfnamefont {T.}~\bibnamefont {Taniguchi}}, \bibinfo {author} {\bibfnamefont {E.}~\bibnamefont {Kaxiras}},\ and\ \bibinfo {author} {\bibfnamefont {P.}~\bibnamefont {Jarillo-Herrero}},\ }\bibfield  {title} {\bibinfo {title} {Unconventional superconductivity in magic-angle graphene superlattices},\ }\href {https://doi.org/10.1038/nature26160} {\bibfield  {journal} {\bibinfo  {journal} {Nature}\ }\textbf {\bibinfo {volume} {556}},\ \bibinfo {pages} {43} (\bibinfo {year} {2018}{\natexlab{a}})}\BibitemShut {NoStop}%
\bibitem [{\citenamefont {Cao}\ \emph {et~al.}(2018{\natexlab{b}})\citenamefont {Cao}, \citenamefont {Fatemi}, \citenamefont {Demir}, \citenamefont {Fang}, \citenamefont {Tomarken}, \citenamefont {Luo}, \citenamefont {Sanchez-Yamagishi}, \citenamefont {Watanabe}, \citenamefont {Taniguchi}, \citenamefont {Kaxiras} \emph {et~al.}}]{cao2018correlated}%
  \BibitemOpen
  \bibfield  {author} {\bibinfo {author} {\bibfnamefont {Y.}~\bibnamefont {Cao}}, \bibinfo {author} {\bibfnamefont {V.}~\bibnamefont {Fatemi}}, \bibinfo {author} {\bibfnamefont {A.}~\bibnamefont {Demir}}, \bibinfo {author} {\bibfnamefont {S.}~\bibnamefont {Fang}}, \bibinfo {author} {\bibfnamefont {S.~L.}\ \bibnamefont {Tomarken}}, \bibinfo {author} {\bibfnamefont {J.~Y.}\ \bibnamefont {Luo}}, \bibinfo {author} {\bibfnamefont {J.~D.}\ \bibnamefont {Sanchez-Yamagishi}}, \bibinfo {author} {\bibfnamefont {K.}~\bibnamefont {Watanabe}}, \bibinfo {author} {\bibfnamefont {T.}~\bibnamefont {Taniguchi}}, \bibinfo {author} {\bibfnamefont {E.}~\bibnamefont {Kaxiras}}, \emph {et~al.},\ }\bibfield  {title} {\bibinfo {title} {Correlated insulator behaviour at half-filling in magic-angle graphene superlattices},\ }\href {https://doi.org/10.1038/nature26154} {\bibfield  {journal} {\bibinfo  {journal} {Nature}\ }\textbf {\bibinfo {volume} {556}},\ \bibinfo {pages} {80} (\bibinfo {year} {2018}{\natexlab{b}})}\BibitemShut
  {NoStop}%
\bibitem [{\citenamefont {Yankowitz}\ \emph {et~al.}(2019)\citenamefont {Yankowitz}, \citenamefont {Chen}, \citenamefont {Polshyn}, \citenamefont {Zhang}, \citenamefont {Watanabe}, \citenamefont {Taniguchi}, \citenamefont {Graf}, \citenamefont {Young},\ and\ \citenamefont {Dean}}]{yankowitz2019tuning}%
  \BibitemOpen
  \bibfield  {author} {\bibinfo {author} {\bibfnamefont {M.}~\bibnamefont {Yankowitz}}, \bibinfo {author} {\bibfnamefont {S.}~\bibnamefont {Chen}}, \bibinfo {author} {\bibfnamefont {H.}~\bibnamefont {Polshyn}}, \bibinfo {author} {\bibfnamefont {Y.}~\bibnamefont {Zhang}}, \bibinfo {author} {\bibfnamefont {K.}~\bibnamefont {Watanabe}}, \bibinfo {author} {\bibfnamefont {T.}~\bibnamefont {Taniguchi}}, \bibinfo {author} {\bibfnamefont {D.}~\bibnamefont {Graf}}, \bibinfo {author} {\bibfnamefont {A.~F.}\ \bibnamefont {Young}},\ and\ \bibinfo {author} {\bibfnamefont {C.~R.}\ \bibnamefont {Dean}},\ }\bibfield  {title} {\bibinfo {title} {Tuning superconductivity in twisted bilayer graphene},\ }\href {https://doi.org/10.1126/science.aav1910} {\bibfield  {journal} {\bibinfo  {journal} {Science}\ }\textbf {\bibinfo {volume} {363}},\ \bibinfo {pages} {1059} (\bibinfo {year} {2019})}\BibitemShut {NoStop}%
\bibitem [{\citenamefont {Lu}\ \emph {et~al.}(2019)\citenamefont {Lu}, \citenamefont {Stepanov}, \citenamefont {Yang}, \citenamefont {Xie}, \citenamefont {Aamir}, \citenamefont {Das}, \citenamefont {Urgell}, \citenamefont {Watanabe}, \citenamefont {Taniguchi}, \citenamefont {Zhang} \emph {et~al.}}]{lu2019superconductors}%
  \BibitemOpen
  \bibfield  {author} {\bibinfo {author} {\bibfnamefont {X.}~\bibnamefont {Lu}}, \bibinfo {author} {\bibfnamefont {P.}~\bibnamefont {Stepanov}}, \bibinfo {author} {\bibfnamefont {W.}~\bibnamefont {Yang}}, \bibinfo {author} {\bibfnamefont {M.}~\bibnamefont {Xie}}, \bibinfo {author} {\bibfnamefont {M.~A.}\ \bibnamefont {Aamir}}, \bibinfo {author} {\bibfnamefont {I.}~\bibnamefont {Das}}, \bibinfo {author} {\bibfnamefont {C.}~\bibnamefont {Urgell}}, \bibinfo {author} {\bibfnamefont {K.}~\bibnamefont {Watanabe}}, \bibinfo {author} {\bibfnamefont {T.}~\bibnamefont {Taniguchi}}, \bibinfo {author} {\bibfnamefont {G.}~\bibnamefont {Zhang}}, \emph {et~al.},\ }\bibfield  {title} {\bibinfo {title} {Superconductors, orbital magnets and correlated states in magic-angle bilayer graphene},\ }\href {https://doi.org/10.1038/s41586-019-1695-0} {\bibfield  {journal} {\bibinfo  {journal} {Nature}\ }\textbf {\bibinfo {volume} {574}},\ \bibinfo {pages} {653} (\bibinfo {year} {2019})}\BibitemShut {NoStop}%
\bibitem [{\citenamefont {Stepanov}\ \emph {et~al.}(2020)\citenamefont {Stepanov}, \citenamefont {Das}, \citenamefont {Lu}, \citenamefont {Fahimniya}, \citenamefont {Watanabe}, \citenamefont {Taniguchi}, \citenamefont {Koppens}, \citenamefont {Lischner}, \citenamefont {Levitov},\ and\ \citenamefont {Efetov}}]{stepanov2020untying}%
  \BibitemOpen
  \bibfield  {author} {\bibinfo {author} {\bibfnamefont {P.}~\bibnamefont {Stepanov}}, \bibinfo {author} {\bibfnamefont {I.}~\bibnamefont {Das}}, \bibinfo {author} {\bibfnamefont {X.}~\bibnamefont {Lu}}, \bibinfo {author} {\bibfnamefont {A.}~\bibnamefont {Fahimniya}}, \bibinfo {author} {\bibfnamefont {K.}~\bibnamefont {Watanabe}}, \bibinfo {author} {\bibfnamefont {T.}~\bibnamefont {Taniguchi}}, \bibinfo {author} {\bibfnamefont {F.~H.}\ \bibnamefont {Koppens}}, \bibinfo {author} {\bibfnamefont {J.}~\bibnamefont {Lischner}}, \bibinfo {author} {\bibfnamefont {L.}~\bibnamefont {Levitov}},\ and\ \bibinfo {author} {\bibfnamefont {D.~K.}\ \bibnamefont {Efetov}},\ }\bibfield  {title} {\bibinfo {title} {Untying the insulating and superconducting orders in magic-angle graphene},\ }\href {https://doi.org/10.1038/s41586-020-2459-6} {\bibfield  {journal} {\bibinfo  {journal} {Nature}\ }\textbf {\bibinfo {volume} {583}},\ \bibinfo {pages} {375} (\bibinfo {year} {2020})}\BibitemShut {NoStop}%
\bibitem [{\citenamefont {Saito}\ \emph {et~al.}(2020)\citenamefont {Saito}, \citenamefont {Ge}, \citenamefont {Watanabe}, \citenamefont {Taniguchi},\ and\ \citenamefont {Young}}]{saito2020independent}%
  \BibitemOpen
  \bibfield  {author} {\bibinfo {author} {\bibfnamefont {Y.}~\bibnamefont {Saito}}, \bibinfo {author} {\bibfnamefont {J.}~\bibnamefont {Ge}}, \bibinfo {author} {\bibfnamefont {K.}~\bibnamefont {Watanabe}}, \bibinfo {author} {\bibfnamefont {T.}~\bibnamefont {Taniguchi}},\ and\ \bibinfo {author} {\bibfnamefont {A.~F.}\ \bibnamefont {Young}},\ }\bibfield  {title} {\bibinfo {title} {Independent superconductors and correlated insulators in twisted bilayer graphene},\ }\href {https://doi.org/10.1038/s41567-020-0928-3} {\bibfield  {journal} {\bibinfo  {journal} {Nature Physics}\ }\textbf {\bibinfo {volume} {16}},\ \bibinfo {pages} {926} (\bibinfo {year} {2020})}\BibitemShut {NoStop}%
\bibitem [{\citenamefont {Sharpe}\ \emph {et~al.}(2019)\citenamefont {Sharpe}, \citenamefont {Fox}, \citenamefont {Barnard}, \citenamefont {Finney}, \citenamefont {Watanabe}, \citenamefont {Taniguchi}, \citenamefont {Kastner},\ and\ \citenamefont {Goldhaber-Gordon}}]{sharpe2019emergent}%
  \BibitemOpen
  \bibfield  {author} {\bibinfo {author} {\bibfnamefont {A.~L.}\ \bibnamefont {Sharpe}}, \bibinfo {author} {\bibfnamefont {E.~J.}\ \bibnamefont {Fox}}, \bibinfo {author} {\bibfnamefont {A.~W.}\ \bibnamefont {Barnard}}, \bibinfo {author} {\bibfnamefont {J.}~\bibnamefont {Finney}}, \bibinfo {author} {\bibfnamefont {K.}~\bibnamefont {Watanabe}}, \bibinfo {author} {\bibfnamefont {T.}~\bibnamefont {Taniguchi}}, \bibinfo {author} {\bibfnamefont {M.~A.}\ \bibnamefont {Kastner}},\ and\ \bibinfo {author} {\bibfnamefont {D.}~\bibnamefont {Goldhaber-Gordon}},\ }\bibfield  {title} {\bibinfo {title} {Emergent ferromagnetism near three-quarters filling in twisted bilayer graphene},\ }\href {https://doi.org/10.1126/science.aaw3780} {\bibfield  {journal} {\bibinfo  {journal} {Science}\ }\textbf {\bibinfo {volume} {365}},\ \bibinfo {pages} {605} (\bibinfo {year} {2019})}\BibitemShut {NoStop}%
\bibitem [{\citenamefont {Serlin}\ \emph {et~al.}(2020)\citenamefont {Serlin}, \citenamefont {Tschirhart}, \citenamefont {Polshyn}, \citenamefont {Zhang}, \citenamefont {Zhu}, \citenamefont {Watanabe}, \citenamefont {Taniguchi}, \citenamefont {Balents},\ and\ \citenamefont {Young}}]{serlin2020intrinsic}%
  \BibitemOpen
  \bibfield  {author} {\bibinfo {author} {\bibfnamefont {M.}~\bibnamefont {Serlin}}, \bibinfo {author} {\bibfnamefont {C.~L.}\ \bibnamefont {Tschirhart}}, \bibinfo {author} {\bibfnamefont {H.}~\bibnamefont {Polshyn}}, \bibinfo {author} {\bibfnamefont {Y.}~\bibnamefont {Zhang}}, \bibinfo {author} {\bibfnamefont {J.}~\bibnamefont {Zhu}}, \bibinfo {author} {\bibfnamefont {K.}~\bibnamefont {Watanabe}}, \bibinfo {author} {\bibfnamefont {T.}~\bibnamefont {Taniguchi}}, \bibinfo {author} {\bibfnamefont {L.}~\bibnamefont {Balents}},\ and\ \bibinfo {author} {\bibfnamefont {A.~F.}\ \bibnamefont {Young}},\ }\bibfield  {title} {\bibinfo {title} {Intrinsic quantized anomalous hall effect in a moiré heterostructure},\ }\href {https://doi.org/10.1126/science.aay5533} {\bibfield  {journal} {\bibinfo  {journal} {Science}\ }\textbf {\bibinfo {volume} {367}},\ \bibinfo {pages} {900} (\bibinfo {year} {2020})}\BibitemShut {NoStop}%
\bibitem [{\citenamefont {Nuckolls}\ \emph {et~al.}(2020)\citenamefont {Nuckolls}, \citenamefont {Oh}, \citenamefont {Wong}, \citenamefont {Lian}, \citenamefont {Watanabe}, \citenamefont {Taniguchi}, \citenamefont {Bernevig},\ and\ \citenamefont {Yazdani}}]{nuckolls2020strongly}%
  \BibitemOpen
  \bibfield  {author} {\bibinfo {author} {\bibfnamefont {K.~P.}\ \bibnamefont {Nuckolls}}, \bibinfo {author} {\bibfnamefont {M.}~\bibnamefont {Oh}}, \bibinfo {author} {\bibfnamefont {D.}~\bibnamefont {Wong}}, \bibinfo {author} {\bibfnamefont {B.}~\bibnamefont {Lian}}, \bibinfo {author} {\bibfnamefont {K.}~\bibnamefont {Watanabe}}, \bibinfo {author} {\bibfnamefont {T.}~\bibnamefont {Taniguchi}}, \bibinfo {author} {\bibfnamefont {B.~A.}\ \bibnamefont {Bernevig}},\ and\ \bibinfo {author} {\bibfnamefont {A.}~\bibnamefont {Yazdani}},\ }\bibfield  {title} {\bibinfo {title} {Strongly correlated chern insulators in magic-angle twisted bilayer graphene},\ }\href {https://doi.org/10.1038/s41586-020-3028-8} {\bibfield  {journal} {\bibinfo  {journal} {Nature}\ }\textbf {\bibinfo {volume} {588}},\ \bibinfo {pages} {610} (\bibinfo {year} {2020})}\BibitemShut {NoStop}%
\bibitem [{\citenamefont {Saito}\ \emph {et~al.}(2021)\citenamefont {Saito}, \citenamefont {Ge}, \citenamefont {Rademaker}, \citenamefont {Watanabe}, \citenamefont {Taniguchi}, \citenamefont {Abanin},\ and\ \citenamefont {Young}}]{saito2021hofstadter}%
  \BibitemOpen
  \bibfield  {author} {\bibinfo {author} {\bibfnamefont {Y.}~\bibnamefont {Saito}}, \bibinfo {author} {\bibfnamefont {J.}~\bibnamefont {Ge}}, \bibinfo {author} {\bibfnamefont {L.}~\bibnamefont {Rademaker}}, \bibinfo {author} {\bibfnamefont {K.}~\bibnamefont {Watanabe}}, \bibinfo {author} {\bibfnamefont {T.}~\bibnamefont {Taniguchi}}, \bibinfo {author} {\bibfnamefont {D.~A.}\ \bibnamefont {Abanin}},\ and\ \bibinfo {author} {\bibfnamefont {A.~F.}\ \bibnamefont {Young}},\ }\bibfield  {title} {\bibinfo {title} {Hofstadter subband ferromagnetism and symmetry-broken chern insulators in twisted bilayer graphene},\ }\href {https://doi.org/10.1038/s41567-020-01129-4} {\bibfield  {journal} {\bibinfo  {journal} {Nature Physics}\ }\textbf {\bibinfo {volume} {17}},\ \bibinfo {pages} {478} (\bibinfo {year} {2021})}\BibitemShut {NoStop}%
\bibitem [{\citenamefont {Choi}\ \emph {et~al.}(2021)\citenamefont {Choi}, \citenamefont {Kim}, \citenamefont {Peng}, \citenamefont {Thomson}, \citenamefont {Lewandowski}, \citenamefont {Polski}, \citenamefont {Zhang}, \citenamefont {Arora}, \citenamefont {Watanabe}, \citenamefont {Taniguchi} \emph {et~al.}}]{choi2021correlation}%
  \BibitemOpen
  \bibfield  {author} {\bibinfo {author} {\bibfnamefont {Y.}~\bibnamefont {Choi}}, \bibinfo {author} {\bibfnamefont {H.}~\bibnamefont {Kim}}, \bibinfo {author} {\bibfnamefont {Y.}~\bibnamefont {Peng}}, \bibinfo {author} {\bibfnamefont {A.}~\bibnamefont {Thomson}}, \bibinfo {author} {\bibfnamefont {C.}~\bibnamefont {Lewandowski}}, \bibinfo {author} {\bibfnamefont {R.}~\bibnamefont {Polski}}, \bibinfo {author} {\bibfnamefont {Y.}~\bibnamefont {Zhang}}, \bibinfo {author} {\bibfnamefont {H.~S.}\ \bibnamefont {Arora}}, \bibinfo {author} {\bibfnamefont {K.}~\bibnamefont {Watanabe}}, \bibinfo {author} {\bibfnamefont {T.}~\bibnamefont {Taniguchi}}, \emph {et~al.},\ }\bibfield  {title} {\bibinfo {title} {Correlation-driven topological phases in magic-angle twisted bilayer graphene},\ }\href {https://doi.org/10.1038/s41586-020-03159-7} {\bibfield  {journal} {\bibinfo  {journal} {Nature}\ }\textbf {\bibinfo {volume} {589}},\ \bibinfo {pages} {536} (\bibinfo {year} {2021})}\BibitemShut {NoStop}%
\bibitem [{\citenamefont {Wu}\ \emph {et~al.}(2021)\citenamefont {Wu}, \citenamefont {Zhang}, \citenamefont {Watanabe}, \citenamefont {Taniguchi},\ and\ \citenamefont {Andrei}}]{wu2021chern}%
  \BibitemOpen
  \bibfield  {author} {\bibinfo {author} {\bibfnamefont {S.}~\bibnamefont {Wu}}, \bibinfo {author} {\bibfnamefont {Z.}~\bibnamefont {Zhang}}, \bibinfo {author} {\bibfnamefont {K.}~\bibnamefont {Watanabe}}, \bibinfo {author} {\bibfnamefont {T.}~\bibnamefont {Taniguchi}},\ and\ \bibinfo {author} {\bibfnamefont {E.~Y.}\ \bibnamefont {Andrei}},\ }\bibfield  {title} {\bibinfo {title} {Chern insulators, van hove singularities and topological flat bands in magic-angle twisted bilayer graphene},\ }\href {https://doi.org/10.1038/s41563-020-00911-2} {\bibfield  {journal} {\bibinfo  {journal} {Nature materials}\ }\textbf {\bibinfo {volume} {20}},\ \bibinfo {pages} {488} (\bibinfo {year} {2021})}\BibitemShut {NoStop}%
\bibitem [{\citenamefont {Pierce}\ \emph {et~al.}(2021)\citenamefont {Pierce}, \citenamefont {Xie}, \citenamefont {Park}, \citenamefont {Khalaf}, \citenamefont {Lee}, \citenamefont {Cao}, \citenamefont {Parker}, \citenamefont {Forrester}, \citenamefont {Chen}, \citenamefont {Watanabe} \emph {et~al.}}]{pierce2021unconventional}%
  \BibitemOpen
  \bibfield  {author} {\bibinfo {author} {\bibfnamefont {A.~T.}\ \bibnamefont {Pierce}}, \bibinfo {author} {\bibfnamefont {Y.}~\bibnamefont {Xie}}, \bibinfo {author} {\bibfnamefont {J.~M.}\ \bibnamefont {Park}}, \bibinfo {author} {\bibfnamefont {E.}~\bibnamefont {Khalaf}}, \bibinfo {author} {\bibfnamefont {S.~H.}\ \bibnamefont {Lee}}, \bibinfo {author} {\bibfnamefont {Y.}~\bibnamefont {Cao}}, \bibinfo {author} {\bibfnamefont {D.~E.}\ \bibnamefont {Parker}}, \bibinfo {author} {\bibfnamefont {P.~R.}\ \bibnamefont {Forrester}}, \bibinfo {author} {\bibfnamefont {S.}~\bibnamefont {Chen}}, \bibinfo {author} {\bibfnamefont {K.}~\bibnamefont {Watanabe}}, \emph {et~al.},\ }\bibfield  {title} {\bibinfo {title} {Unconventional sequence of correlated chern insulators in magic-angle twisted bilayer graphene},\ }\href {https://doi.org/10.1038/s41567-021-01347-4} {\bibfield  {journal} {\bibinfo  {journal} {Nature Physics}\ }\textbf {\bibinfo {volume} {17}},\ \bibinfo {pages} {1210} (\bibinfo {year} {2021})}\BibitemShut
  {NoStop}%
\bibitem [{\citenamefont {Xie}\ \emph {et~al.}(2021)\citenamefont {Xie}, \citenamefont {Pierce}, \citenamefont {Park}, \citenamefont {Parker}, \citenamefont {Khalaf}, \citenamefont {Ledwith}, \citenamefont {Cao}, \citenamefont {Lee}, \citenamefont {Chen}, \citenamefont {Forrester} \emph {et~al.}}]{xie2021fractional}%
  \BibitemOpen
  \bibfield  {author} {\bibinfo {author} {\bibfnamefont {Y.}~\bibnamefont {Xie}}, \bibinfo {author} {\bibfnamefont {A.~T.}\ \bibnamefont {Pierce}}, \bibinfo {author} {\bibfnamefont {J.~M.}\ \bibnamefont {Park}}, \bibinfo {author} {\bibfnamefont {D.~E.}\ \bibnamefont {Parker}}, \bibinfo {author} {\bibfnamefont {E.}~\bibnamefont {Khalaf}}, \bibinfo {author} {\bibfnamefont {P.}~\bibnamefont {Ledwith}}, \bibinfo {author} {\bibfnamefont {Y.}~\bibnamefont {Cao}}, \bibinfo {author} {\bibfnamefont {S.~H.}\ \bibnamefont {Lee}}, \bibinfo {author} {\bibfnamefont {S.}~\bibnamefont {Chen}}, \bibinfo {author} {\bibfnamefont {P.~R.}\ \bibnamefont {Forrester}}, \emph {et~al.},\ }\bibfield  {title} {\bibinfo {title} {Fractional chern insulators in magic-angle twisted bilayer graphene},\ }\href {https://doi.org/10.1038/s41586-021-04002-3} {\bibfield  {journal} {\bibinfo  {journal} {Nature}\ }\textbf {\bibinfo {volume} {600}},\ \bibinfo {pages} {439} (\bibinfo {year} {2021})}\BibitemShut {NoStop}%
\bibitem [{\citenamefont {Finney}\ \emph {et~al.}(2025)\citenamefont {Finney}, \citenamefont {Sharpe}, \citenamefont {Rodenbach}, \citenamefont {Kang}, \citenamefont {Wang}, \citenamefont {Watanabe}, \citenamefont {Taniguchi}, \citenamefont {Kastner}, \citenamefont {Vafek},\ and\ \citenamefont {{Goldhaber-Gordon}}}]{finneyExtendedFractionalChern2025}%
  \BibitemOpen
  \bibfield  {author} {\bibinfo {author} {\bibfnamefont {J.}~\bibnamefont {Finney}}, \bibinfo {author} {\bibfnamefont {A.~L.}\ \bibnamefont {Sharpe}}, \bibinfo {author} {\bibfnamefont {L.~K.}\ \bibnamefont {Rodenbach}}, \bibinfo {author} {\bibfnamefont {J.}~\bibnamefont {Kang}}, \bibinfo {author} {\bibfnamefont {X.}~\bibnamefont {Wang}}, \bibinfo {author} {\bibfnamefont {K.}~\bibnamefont {Watanabe}}, \bibinfo {author} {\bibfnamefont {T.}~\bibnamefont {Taniguchi}}, \bibinfo {author} {\bibfnamefont {M.~A.}\ \bibnamefont {Kastner}}, \bibinfo {author} {\bibfnamefont {O.}~\bibnamefont {Vafek}},\ and\ \bibinfo {author} {\bibfnamefont {D.}~\bibnamefont {{Goldhaber-Gordon}}},\ }\href {https://doi.org/10.48550/arXiv.2503.12819} {\bibinfo {title} {Extended {{Fractional Chern Insulators Near Half Flux}} in {{Twisted Bilayer Graphene Above}} the {{Magic Angle}}}} (\bibinfo {year} {2025}),\ \Eprint {https://arxiv.org/abs/2503.12819} {arXiv:2503.12819 [cond-mat]} \BibitemShut {NoStop}%
\bibitem [{\citenamefont {Guinea}\ and\ \citenamefont {Walet}(2018)}]{Guineaelectrostatic2018}%
  \BibitemOpen
  \bibfield  {author} {\bibinfo {author} {\bibfnamefont {F.}~\bibnamefont {Guinea}}\ and\ \bibinfo {author} {\bibfnamefont {N.~R.}\ \bibnamefont {Walet}},\ }\bibfield  {title} {\bibinfo {title} {Electrostatic effects, band distortions, and superconductivity in twisted graphene bilayers},\ }\href {https://doi.org/10.1073/pnas.1810947115} {\bibfield  {journal} {\bibinfo  {journal} {Proceedings of the National Academy of Sciences}\ }\textbf {\bibinfo {volume} {115}},\ \bibinfo {pages} {13174} (\bibinfo {year} {2018})}\BibitemShut {NoStop}%
\bibitem [{\citenamefont {Rademaker}\ \emph {et~al.}(2019)\citenamefont {Rademaker}, \citenamefont {Abanin},\ and\ \citenamefont {Mellado}}]{ardemakercharge2019}%
  \BibitemOpen
  \bibfield  {author} {\bibinfo {author} {\bibfnamefont {L.}~\bibnamefont {Rademaker}}, \bibinfo {author} {\bibfnamefont {D.~A.}\ \bibnamefont {Abanin}},\ and\ \bibinfo {author} {\bibfnamefont {P.}~\bibnamefont {Mellado}},\ }\bibfield  {title} {\bibinfo {title} {Charge smoothening and band flattening due to hartree corrections in twisted bilayer graphene},\ }\href {https://doi.org/10.1103/PhysRevB.100.205114} {\bibfield  {journal} {\bibinfo  {journal} {Phys. Rev. B}\ }\textbf {\bibinfo {volume} {100}},\ \bibinfo {pages} {205114} (\bibinfo {year} {2019})}\BibitemShut {NoStop}%
\bibitem [{\citenamefont {Cea}\ \emph {et~al.}(2019)\citenamefont {Cea}, \citenamefont {Walet},\ and\ \citenamefont {Guinea}}]{ceapinning2019}%
  \BibitemOpen
  \bibfield  {author} {\bibinfo {author} {\bibfnamefont {T.}~\bibnamefont {Cea}}, \bibinfo {author} {\bibfnamefont {N.~R.}\ \bibnamefont {Walet}},\ and\ \bibinfo {author} {\bibfnamefont {F.}~\bibnamefont {Guinea}},\ }\bibfield  {title} {\bibinfo {title} {Electronic band structure and pinning of fermi energy to van hove singularities in twisted bilayer graphene: A self-consistent approach},\ }\href {https://doi.org/10.1103/PhysRevB.100.205113} {\bibfield  {journal} {\bibinfo  {journal} {Phys. Rev. B}\ }\textbf {\bibinfo {volume} {100}},\ \bibinfo {pages} {205113} (\bibinfo {year} {2019})}\BibitemShut {NoStop}%
\bibitem [{\citenamefont {Goodwin}\ \emph {et~al.}(2020)\citenamefont {Goodwin}, \citenamefont {Vitale}, \citenamefont {Liang}, \citenamefont {Mostofi},\ and\ \citenamefont {Lischner}}]{goodwin2020hartree}%
  \BibitemOpen
  \bibfield  {author} {\bibinfo {author} {\bibfnamefont {Z.~A.}\ \bibnamefont {Goodwin}}, \bibinfo {author} {\bibfnamefont {V.}~\bibnamefont {Vitale}}, \bibinfo {author} {\bibfnamefont {X.}~\bibnamefont {Liang}}, \bibinfo {author} {\bibfnamefont {A.~A.}\ \bibnamefont {Mostofi}},\ and\ \bibinfo {author} {\bibfnamefont {J.}~\bibnamefont {Lischner}},\ }\bibfield  {title} {\bibinfo {title} {Hartree theory calculations of quasiparticle properties in twisted bilayer graphene},\ }\href@noop {} {\bibfield  {journal} {\bibinfo  {journal} {Electronic Structure}\ }\textbf {\bibinfo {volume} {2}},\ \bibinfo {pages} {034001} (\bibinfo {year} {2020})}\BibitemShut {NoStop}%
\bibitem [{\citenamefont {Kang}\ \emph {et~al.}(2021)\citenamefont {Kang}, \citenamefont {Bernevig},\ and\ \citenamefont {Vafek}}]{Kangcascades2021}%
  \BibitemOpen
  \bibfield  {author} {\bibinfo {author} {\bibfnamefont {J.}~\bibnamefont {Kang}}, \bibinfo {author} {\bibfnamefont {B.~A.}\ \bibnamefont {Bernevig}},\ and\ \bibinfo {author} {\bibfnamefont {O.}~\bibnamefont {Vafek}},\ }\bibfield  {title} {\bibinfo {title} {Cascades between light and heavy fermions in the normal state of magic-angle twisted bilayer graphene},\ }\href {https://doi.org/10.1103/PhysRevLett.127.266402} {\bibfield  {journal} {\bibinfo  {journal} {Phys. Rev. Lett.}\ }\textbf {\bibinfo {volume} {127}},\ \bibinfo {pages} {266402} (\bibinfo {year} {2021})}\BibitemShut {NoStop}%
\bibitem [{\citenamefont {Parker}\ \emph {et~al.}(2021{\natexlab{a}})\citenamefont {Parker}, \citenamefont {Ledwith}, \citenamefont {Khalaf}, \citenamefont {Soejima}, \citenamefont {Hauschild}, \citenamefont {Xie}, \citenamefont {Pierce}, \citenamefont {Zaletel}, \citenamefont {Yacoby},\ and\ \citenamefont {Vishwanath}}]{parker2021field}%
  \BibitemOpen
  \bibfield  {author} {\bibinfo {author} {\bibfnamefont {D.}~\bibnamefont {Parker}}, \bibinfo {author} {\bibfnamefont {P.}~\bibnamefont {Ledwith}}, \bibinfo {author} {\bibfnamefont {E.}~\bibnamefont {Khalaf}}, \bibinfo {author} {\bibfnamefont {T.}~\bibnamefont {Soejima}}, \bibinfo {author} {\bibfnamefont {J.}~\bibnamefont {Hauschild}}, \bibinfo {author} {\bibfnamefont {Y.}~\bibnamefont {Xie}}, \bibinfo {author} {\bibfnamefont {A.}~\bibnamefont {Pierce}}, \bibinfo {author} {\bibfnamefont {M.~P.}\ \bibnamefont {Zaletel}}, \bibinfo {author} {\bibfnamefont {A.}~\bibnamefont {Yacoby}},\ and\ \bibinfo {author} {\bibfnamefont {A.}~\bibnamefont {Vishwanath}},\ }\bibfield  {title} {\bibinfo {title} {Field-tuned and zero-field fractional chern insulators in magic angle graphene},\ }\href {https://arxiv.org/abs/2112.13837} {\bibfield  {journal} {\bibinfo  {journal} {arXiv preprint arXiv:2112.13837}\ } (\bibinfo {year} {2021}{\natexlab{a}})}\BibitemShut {NoStop}%
\bibitem [{\citenamefont {Liu}\ \emph {et~al.}(2021{\natexlab{a}})\citenamefont {Liu}, \citenamefont {Khalaf}, \citenamefont {Lee},\ and\ \citenamefont {Vishwanath}}]{PhysRevResearch.3.013033}%
  \BibitemOpen
  \bibfield  {author} {\bibinfo {author} {\bibfnamefont {S.}~\bibnamefont {Liu}}, \bibinfo {author} {\bibfnamefont {E.}~\bibnamefont {Khalaf}}, \bibinfo {author} {\bibfnamefont {J.~Y.}\ \bibnamefont {Lee}},\ and\ \bibinfo {author} {\bibfnamefont {A.}~\bibnamefont {Vishwanath}},\ }\bibfield  {title} {\bibinfo {title} {Nematic topological semimetal and insulator in magic-angle bilayer graphene at charge neutrality},\ }\href {https://doi.org/10.1103/PhysRevResearch.3.013033} {\bibfield  {journal} {\bibinfo  {journal} {Phys. Rev. Res.}\ }\textbf {\bibinfo {volume} {3}},\ \bibinfo {pages} {013033} (\bibinfo {year} {2021}{\natexlab{a}})}\BibitemShut {NoStop}%
\bibitem [{\citenamefont {Parker}\ \emph {et~al.}(2021{\natexlab{b}})\citenamefont {Parker}, \citenamefont {Soejima}, \citenamefont {Hauschild}, \citenamefont {Zaletel},\ and\ \citenamefont {Bultinck}}]{PhysRevLett.127.027601}%
  \BibitemOpen
  \bibfield  {author} {\bibinfo {author} {\bibfnamefont {D.~E.}\ \bibnamefont {Parker}}, \bibinfo {author} {\bibfnamefont {T.}~\bibnamefont {Soejima}}, \bibinfo {author} {\bibfnamefont {J.}~\bibnamefont {Hauschild}}, \bibinfo {author} {\bibfnamefont {M.~P.}\ \bibnamefont {Zaletel}},\ and\ \bibinfo {author} {\bibfnamefont {N.}~\bibnamefont {Bultinck}},\ }\bibfield  {title} {\bibinfo {title} {Strain-induced quantum phase transitions in magic-angle graphene},\ }\href {https://doi.org/10.1103/PhysRevLett.127.027601} {\bibfield  {journal} {\bibinfo  {journal} {Phys. Rev. Lett.}\ }\textbf {\bibinfo {volume} {127}},\ \bibinfo {pages} {027601} (\bibinfo {year} {2021}{\natexlab{b}})}\BibitemShut {NoStop}%
\bibitem [{\citenamefont {Soejima}\ \emph {et~al.}(2020)\citenamefont {Soejima}, \citenamefont {Parker}, \citenamefont {Bultinck}, \citenamefont {Hauschild},\ and\ \citenamefont {Zaletel}}]{PhysRevB.102.205111}%
  \BibitemOpen
  \bibfield  {author} {\bibinfo {author} {\bibfnamefont {T.}~\bibnamefont {Soejima}}, \bibinfo {author} {\bibfnamefont {D.~E.}\ \bibnamefont {Parker}}, \bibinfo {author} {\bibfnamefont {N.}~\bibnamefont {Bultinck}}, \bibinfo {author} {\bibfnamefont {J.}~\bibnamefont {Hauschild}},\ and\ \bibinfo {author} {\bibfnamefont {M.~P.}\ \bibnamefont {Zaletel}},\ }\bibfield  {title} {\bibinfo {title} {Efficient simulation of moir\'e materials using the density matrix renormalization group},\ }\href {https://doi.org/10.1103/PhysRevB.102.205111} {\bibfield  {journal} {\bibinfo  {journal} {Phys. Rev. B}\ }\textbf {\bibinfo {volume} {102}},\ \bibinfo {pages} {205111} (\bibinfo {year} {2020})}\BibitemShut {NoStop}%
\bibitem [{\citenamefont {Bocarsly}\ \emph {et~al.}(2025)\citenamefont {Bocarsly}, \citenamefont {Roy}, \citenamefont {Bhardwaj}, \citenamefont {Uzan}, \citenamefont {Ledwith}, \citenamefont {Shavit}, \citenamefont {Banu}, \citenamefont {Zhou}, \citenamefont {Myasoedov}, \citenamefont {Watanabe}, \citenamefont {Taniguchi}, \citenamefont {Oreg}, \citenamefont {Parker}, \citenamefont {Ronen},\ and\ \citenamefont {Zeldov}}]{bocarslyCoulombInteractionsMigrating2025}%
  \BibitemOpen
  \bibfield  {author} {\bibinfo {author} {\bibfnamefont {M.}~\bibnamefont {Bocarsly}}, \bibinfo {author} {\bibfnamefont {I.}~\bibnamefont {Roy}}, \bibinfo {author} {\bibfnamefont {V.}~\bibnamefont {Bhardwaj}}, \bibinfo {author} {\bibfnamefont {M.}~\bibnamefont {Uzan}}, \bibinfo {author} {\bibfnamefont {P.}~\bibnamefont {Ledwith}}, \bibinfo {author} {\bibfnamefont {G.}~\bibnamefont {Shavit}}, \bibinfo {author} {\bibfnamefont {N.}~\bibnamefont {Banu}}, \bibinfo {author} {\bibfnamefont {Y.}~\bibnamefont {Zhou}}, \bibinfo {author} {\bibfnamefont {Y.}~\bibnamefont {Myasoedov}}, \bibinfo {author} {\bibfnamefont {K.}~\bibnamefont {Watanabe}}, \bibinfo {author} {\bibfnamefont {T.}~\bibnamefont {Taniguchi}}, \bibinfo {author} {\bibfnamefont {Y.}~\bibnamefont {Oreg}}, \bibinfo {author} {\bibfnamefont {D.~E.}\ \bibnamefont {Parker}}, \bibinfo {author} {\bibfnamefont {Y.}~\bibnamefont {Ronen}},\ and\ \bibinfo {author} {\bibfnamefont {E.}~\bibnamefont {Zeldov}},\ }\bibfield  {title} {\bibinfo {title} {Coulomb interactions
  and migrating {{Dirac}} cones imaged by local quantum oscillations in twisted graphene},\ }\href {https://doi.org/10.1038/s41567-025-02786-z} {\bibfield  {journal} {\bibinfo  {journal} {Nature Physics}\ }\textbf {\bibinfo {volume} {21}},\ \bibinfo {pages} {421} (\bibinfo {year} {2025})}\BibitemShut {NoStop}%
\bibitem [{\citenamefont {Kwan}\ \emph {et~al.}(2021)\citenamefont {Kwan}, \citenamefont {Wagner}, \citenamefont {Soejima}, \citenamefont {Zaletel}, \citenamefont {Simon}, \citenamefont {Parameswaran},\ and\ \citenamefont {Bultinck}}]{PhysRevX.11.041063}%
  \BibitemOpen
  \bibfield  {author} {\bibinfo {author} {\bibfnamefont {Y.~H.}\ \bibnamefont {Kwan}}, \bibinfo {author} {\bibfnamefont {G.}~\bibnamefont {Wagner}}, \bibinfo {author} {\bibfnamefont {T.}~\bibnamefont {Soejima}}, \bibinfo {author} {\bibfnamefont {M.~P.}\ \bibnamefont {Zaletel}}, \bibinfo {author} {\bibfnamefont {S.~H.}\ \bibnamefont {Simon}}, \bibinfo {author} {\bibfnamefont {S.~A.}\ \bibnamefont {Parameswaran}},\ and\ \bibinfo {author} {\bibfnamefont {N.}~\bibnamefont {Bultinck}},\ }\bibfield  {title} {\bibinfo {title} {Kekul\'e spiral order at all nonzero integer fillings in twisted bilayer graphene},\ }\href {https://doi.org/10.1103/PhysRevX.11.041063} {\bibfield  {journal} {\bibinfo  {journal} {Phys. Rev. X}\ }\textbf {\bibinfo {volume} {11}},\ \bibinfo {pages} {041063} (\bibinfo {year} {2021})}\BibitemShut {NoStop}%
\bibitem [{\citenamefont {Wang}\ \emph {et~al.}(2023{\natexlab{a}})\citenamefont {Wang}, \citenamefont {Parker}, \citenamefont {Soejima}, \citenamefont {Hauschild}, \citenamefont {Anand}, \citenamefont {Bultinck},\ and\ \citenamefont {Zaletel}}]{PhysRevB.108.235128}%
  \BibitemOpen
  \bibfield  {author} {\bibinfo {author} {\bibfnamefont {T.}~\bibnamefont {Wang}}, \bibinfo {author} {\bibfnamefont {D.~E.}\ \bibnamefont {Parker}}, \bibinfo {author} {\bibfnamefont {T.}~\bibnamefont {Soejima}}, \bibinfo {author} {\bibfnamefont {J.}~\bibnamefont {Hauschild}}, \bibinfo {author} {\bibfnamefont {S.}~\bibnamefont {Anand}}, \bibinfo {author} {\bibfnamefont {N.}~\bibnamefont {Bultinck}},\ and\ \bibinfo {author} {\bibfnamefont {M.~P.}\ \bibnamefont {Zaletel}},\ }\bibfield  {title} {\bibinfo {title} {Ground-state order in magic-angle graphene at filling $\ensuremath{\nu}=\ensuremath{-}3$: A full-scale density matrix renormalization group study},\ }\href {https://doi.org/10.1103/PhysRevB.108.235128} {\bibfield  {journal} {\bibinfo  {journal} {Phys. Rev. B}\ }\textbf {\bibinfo {volume} {108}},\ \bibinfo {pages} {235128} (\bibinfo {year} {2023}{\natexlab{a}})}\BibitemShut {NoStop}%
\bibitem [{\citenamefont {Nuckolls}\ \emph {et~al.}(2023)\citenamefont {Nuckolls}, \citenamefont {Lee}, \citenamefont {Oh}, \citenamefont {Wong}, \citenamefont {Soejima}, \citenamefont {Hong}, \citenamefont {C{\u{a}}lug{\u{a}}ru}, \citenamefont {Herzog-Arbeitman}, \citenamefont {Bernevig}, \citenamefont {Watanabe} \emph {et~al.}}]{nuckolls2023quantum}%
  \BibitemOpen
  \bibfield  {author} {\bibinfo {author} {\bibfnamefont {K.~P.}\ \bibnamefont {Nuckolls}}, \bibinfo {author} {\bibfnamefont {R.~L.}\ \bibnamefont {Lee}}, \bibinfo {author} {\bibfnamefont {M.}~\bibnamefont {Oh}}, \bibinfo {author} {\bibfnamefont {D.}~\bibnamefont {Wong}}, \bibinfo {author} {\bibfnamefont {T.}~\bibnamefont {Soejima}}, \bibinfo {author} {\bibfnamefont {J.~P.}\ \bibnamefont {Hong}}, \bibinfo {author} {\bibfnamefont {D.}~\bibnamefont {C{\u{a}}lug{\u{a}}ru}}, \bibinfo {author} {\bibfnamefont {J.}~\bibnamefont {Herzog-Arbeitman}}, \bibinfo {author} {\bibfnamefont {B.~A.}\ \bibnamefont {Bernevig}}, \bibinfo {author} {\bibfnamefont {K.}~\bibnamefont {Watanabe}}, \emph {et~al.},\ }\bibfield  {title} {\bibinfo {title} {Quantum textures of the many-body wavefunctions in magic-angle graphene},\ }\href@noop {} {\bibfield  {journal} {\bibinfo  {journal} {Nature}\ }\textbf {\bibinfo {volume} {620}},\ \bibinfo {pages} {525} (\bibinfo {year} {2023})}\BibitemShut {NoStop}%
\bibitem [{\citenamefont {Kim}\ \emph {et~al.}(2023)\citenamefont {Kim}, \citenamefont {Choi}, \citenamefont {{Lantagne-Hurtubise}}, \citenamefont {Lewandowski}, \citenamefont {Thomson}, \citenamefont {Kong}, \citenamefont {Zhou}, \citenamefont {Baum}, \citenamefont {Zhang}, \citenamefont {Holleis}, \citenamefont {Watanabe}, \citenamefont {Taniguchi}, \citenamefont {Young}, \citenamefont {Alicea},\ and\ \citenamefont {{Nadj-Perge}}}]{kimImagingIntervalleyCoherent2023}%
  \BibitemOpen
  \bibfield  {author} {\bibinfo {author} {\bibfnamefont {H.}~\bibnamefont {Kim}}, \bibinfo {author} {\bibfnamefont {Y.}~\bibnamefont {Choi}}, \bibinfo {author} {\bibfnamefont {{\'E}.}~\bibnamefont {{Lantagne-Hurtubise}}}, \bibinfo {author} {\bibfnamefont {C.}~\bibnamefont {Lewandowski}}, \bibinfo {author} {\bibfnamefont {A.}~\bibnamefont {Thomson}}, \bibinfo {author} {\bibfnamefont {L.}~\bibnamefont {Kong}}, \bibinfo {author} {\bibfnamefont {H.}~\bibnamefont {Zhou}}, \bibinfo {author} {\bibfnamefont {E.}~\bibnamefont {Baum}}, \bibinfo {author} {\bibfnamefont {Y.}~\bibnamefont {Zhang}}, \bibinfo {author} {\bibfnamefont {L.}~\bibnamefont {Holleis}}, \bibinfo {author} {\bibfnamefont {K.}~\bibnamefont {Watanabe}}, \bibinfo {author} {\bibfnamefont {T.}~\bibnamefont {Taniguchi}}, \bibinfo {author} {\bibfnamefont {A.~F.}\ \bibnamefont {Young}}, \bibinfo {author} {\bibfnamefont {J.}~\bibnamefont {Alicea}},\ and\ \bibinfo {author} {\bibfnamefont {S.}~\bibnamefont {{Nadj-Perge}}},\ }\bibfield  {title} {\bibinfo {title}
  {Imaging inter-valley coherent order in magic-angle twisted trilayer graphene},\ }\href {https://doi.org/10.1038/s41586-023-06663-8} {\bibfield  {journal} {\bibinfo  {journal} {Nature}\ }\textbf {\bibinfo {volume} {623}},\ \bibinfo {pages} {942} (\bibinfo {year} {2023})}\BibitemShut {NoStop}%
\bibitem [{\citenamefont {Devakul}\ \emph {et~al.}(2023)\citenamefont {Devakul}, \citenamefont {Ledwith}, \citenamefont {Xia}, \citenamefont {Uri}, \citenamefont {de~la Barrera}, \citenamefont {Jarillo-Herrero},\ and\ \citenamefont {Fu}}]{httg_devakul}%
  \BibitemOpen
  \bibfield  {author} {\bibinfo {author} {\bibfnamefont {T.}~\bibnamefont {Devakul}}, \bibinfo {author} {\bibfnamefont {P.~J.}\ \bibnamefont {Ledwith}}, \bibinfo {author} {\bibfnamefont {L.-Q.}\ \bibnamefont {Xia}}, \bibinfo {author} {\bibfnamefont {A.}~\bibnamefont {Uri}}, \bibinfo {author} {\bibfnamefont {S.~C.}\ \bibnamefont {de~la Barrera}}, \bibinfo {author} {\bibfnamefont {P.}~\bibnamefont {Jarillo-Herrero}},\ and\ \bibinfo {author} {\bibfnamefont {L.}~\bibnamefont {Fu}},\ }\bibfield  {title} {\bibinfo {title} {Magic-angle helical trilayer graphene},\ }\href {https://doi.org/10.1126/sciadv.adi6063} {\bibfield  {journal} {\bibinfo  {journal} {Science Advances}\ }\textbf {\bibinfo {volume} {9}},\ \bibinfo {pages} {eadi6063} (\bibinfo {year} {2023})}\BibitemShut {NoStop}%
\bibitem [{\citenamefont {Mora}\ \emph {et~al.}(2019)\citenamefont {Mora}, \citenamefont {Regnault},\ and\ \citenamefont {Bernevig}}]{PhysRevLett.123.026402}%
  \BibitemOpen
  \bibfield  {author} {\bibinfo {author} {\bibfnamefont {C.}~\bibnamefont {Mora}}, \bibinfo {author} {\bibfnamefont {N.}~\bibnamefont {Regnault}},\ and\ \bibinfo {author} {\bibfnamefont {B.~A.}\ \bibnamefont {Bernevig}},\ }\bibfield  {title} {\bibinfo {title} {Flatbands and perfect metal in trilayer moir\'e graphene},\ }\href {https://doi.org/10.1103/PhysRevLett.123.026402} {\bibfield  {journal} {\bibinfo  {journal} {Phys. Rev. Lett.}\ }\textbf {\bibinfo {volume} {123}},\ \bibinfo {pages} {026402} (\bibinfo {year} {2019})}\BibitemShut {NoStop}%
\bibitem [{\citenamefont {Zhu}\ \emph {et~al.}(2020{\natexlab{a}})\citenamefont {Zhu}, \citenamefont {Cazeaux}, \citenamefont {Luskin},\ and\ \citenamefont {Kaxiras}}]{PhysRevB.101.224107}%
  \BibitemOpen
  \bibfield  {author} {\bibinfo {author} {\bibfnamefont {Z.}~\bibnamefont {Zhu}}, \bibinfo {author} {\bibfnamefont {P.}~\bibnamefont {Cazeaux}}, \bibinfo {author} {\bibfnamefont {M.}~\bibnamefont {Luskin}},\ and\ \bibinfo {author} {\bibfnamefont {E.}~\bibnamefont {Kaxiras}},\ }\bibfield  {title} {\bibinfo {title} {Modeling mechanical relaxation in incommensurate trilayer van der waals heterostructures},\ }\href {https://doi.org/10.1103/PhysRevB.101.224107} {\bibfield  {journal} {\bibinfo  {journal} {Phys. Rev. B}\ }\textbf {\bibinfo {volume} {101}},\ \bibinfo {pages} {224107} (\bibinfo {year} {2020}{\natexlab{a}})}\BibitemShut {NoStop}%
\bibitem [{\citenamefont {Zhu}\ \emph {et~al.}(2020{\natexlab{b}})\citenamefont {Zhu}, \citenamefont {Carr}, \citenamefont {Massatt}, \citenamefont {Luskin},\ and\ \citenamefont {Kaxiras}}]{PhysRevLett.125.116404}%
  \BibitemOpen
  \bibfield  {author} {\bibinfo {author} {\bibfnamefont {Z.}~\bibnamefont {Zhu}}, \bibinfo {author} {\bibfnamefont {S.}~\bibnamefont {Carr}}, \bibinfo {author} {\bibfnamefont {D.}~\bibnamefont {Massatt}}, \bibinfo {author} {\bibfnamefont {M.}~\bibnamefont {Luskin}},\ and\ \bibinfo {author} {\bibfnamefont {E.}~\bibnamefont {Kaxiras}},\ }\bibfield  {title} {\bibinfo {title} {Twisted trilayer graphene: A precisely tunable platform for correlated electrons},\ }\href {https://doi.org/10.1103/PhysRevLett.125.116404} {\bibfield  {journal} {\bibinfo  {journal} {Phys. Rev. Lett.}\ }\textbf {\bibinfo {volume} {125}},\ \bibinfo {pages} {116404} (\bibinfo {year} {2020}{\natexlab{b}})}\BibitemShut {NoStop}%
\bibitem [{\citenamefont {Zhang}\ \emph {et~al.}(2021)\citenamefont {Zhang}, \citenamefont {Tsai}, \citenamefont {Zhu}, \citenamefont {Ren}, \citenamefont {Luo}, \citenamefont {Carr}, \citenamefont {Luskin}, \citenamefont {Kaxiras},\ and\ \citenamefont {Wang}}]{PhysRevLett.127.166802}%
  \BibitemOpen
  \bibfield  {author} {\bibinfo {author} {\bibfnamefont {X.}~\bibnamefont {Zhang}}, \bibinfo {author} {\bibfnamefont {K.-T.}\ \bibnamefont {Tsai}}, \bibinfo {author} {\bibfnamefont {Z.}~\bibnamefont {Zhu}}, \bibinfo {author} {\bibfnamefont {W.}~\bibnamefont {Ren}}, \bibinfo {author} {\bibfnamefont {Y.}~\bibnamefont {Luo}}, \bibinfo {author} {\bibfnamefont {S.}~\bibnamefont {Carr}}, \bibinfo {author} {\bibfnamefont {M.}~\bibnamefont {Luskin}}, \bibinfo {author} {\bibfnamefont {E.}~\bibnamefont {Kaxiras}},\ and\ \bibinfo {author} {\bibfnamefont {K.}~\bibnamefont {Wang}},\ }\bibfield  {title} {\bibinfo {title} {Correlated insulating states and transport signature of superconductivity in twisted trilayer graphene superlattices},\ }\href {https://doi.org/10.1103/PhysRevLett.127.166802} {\bibfield  {journal} {\bibinfo  {journal} {Phys. Rev. Lett.}\ }\textbf {\bibinfo {volume} {127}},\ \bibinfo {pages} {166802} (\bibinfo {year} {2021})}\BibitemShut {NoStop}%
\bibitem [{\citenamefont {Mao}\ \emph {et~al.}(2023)\citenamefont {Mao}, \citenamefont {Guerci},\ and\ \citenamefont {Mora}}]{PhysRevB.107.125423}%
  \BibitemOpen
  \bibfield  {author} {\bibinfo {author} {\bibfnamefont {Y.}~\bibnamefont {Mao}}, \bibinfo {author} {\bibfnamefont {D.}~\bibnamefont {Guerci}},\ and\ \bibinfo {author} {\bibfnamefont {C.}~\bibnamefont {Mora}},\ }\bibfield  {title} {\bibinfo {title} {Supermoir\'e low-energy effective theory of twisted trilayer graphene},\ }\href {https://doi.org/10.1103/PhysRevB.107.125423} {\bibfield  {journal} {\bibinfo  {journal} {Phys. Rev. B}\ }\textbf {\bibinfo {volume} {107}},\ \bibinfo {pages} {125423} (\bibinfo {year} {2023})}\BibitemShut {NoStop}%
\bibitem [{\citenamefont {Dunbrack}\ and\ \citenamefont {Cano}(2023)}]{PhysRevB.107.235425}%
  \BibitemOpen
  \bibfield  {author} {\bibinfo {author} {\bibfnamefont {A.}~\bibnamefont {Dunbrack}}\ and\ \bibinfo {author} {\bibfnamefont {J.}~\bibnamefont {Cano}},\ }\bibfield  {title} {\bibinfo {title} {Intrinsically multilayer moir\'e heterostructures},\ }\href {https://doi.org/10.1103/PhysRevB.107.235425} {\bibfield  {journal} {\bibinfo  {journal} {Phys. Rev. B}\ }\textbf {\bibinfo {volume} {107}},\ \bibinfo {pages} {235425} (\bibinfo {year} {2023})}\BibitemShut {NoStop}%
\bibitem [{\citenamefont {Uri}\ \emph {et~al.}(2023)\citenamefont {Uri}, \citenamefont {de~la Barrera}, \citenamefont {Randeria}, \citenamefont {Rodan-Legrain}, \citenamefont {Devakul}, \citenamefont {Crowley}, \citenamefont {Paul}, \citenamefont {Watanabe}, \citenamefont {Taniguchi}, \citenamefont {Lifshitz} \emph {et~al.}}]{uri2023superconductivity}%
  \BibitemOpen
  \bibfield  {author} {\bibinfo {author} {\bibfnamefont {A.}~\bibnamefont {Uri}}, \bibinfo {author} {\bibfnamefont {S.~C.}\ \bibnamefont {de~la Barrera}}, \bibinfo {author} {\bibfnamefont {M.~T.}\ \bibnamefont {Randeria}}, \bibinfo {author} {\bibfnamefont {D.}~\bibnamefont {Rodan-Legrain}}, \bibinfo {author} {\bibfnamefont {T.}~\bibnamefont {Devakul}}, \bibinfo {author} {\bibfnamefont {P.~J.}\ \bibnamefont {Crowley}}, \bibinfo {author} {\bibfnamefont {N.}~\bibnamefont {Paul}}, \bibinfo {author} {\bibfnamefont {K.}~\bibnamefont {Watanabe}}, \bibinfo {author} {\bibfnamefont {T.}~\bibnamefont {Taniguchi}}, \bibinfo {author} {\bibfnamefont {R.}~\bibnamefont {Lifshitz}}, \emph {et~al.},\ }\bibfield  {title} {\bibinfo {title} {Superconductivity and strong interactions in a tunable moir{\'e} quasicrystal},\ }\href {https://doi.org/10.1038/s41586-023-06294-z} {\bibfield  {journal} {\bibinfo  {journal} {Nature}\ }\textbf {\bibinfo {volume} {620}},\ \bibinfo {pages} {762} (\bibinfo {year} {2023})}\BibitemShut {NoStop}%
\bibitem [{\citenamefont {Becker}\ \emph {et~al.}(2023)\citenamefont {Becker}, \citenamefont {Humbert}, \citenamefont {Wittsten},\ and\ \citenamefont {Yang}}]{becker2023chiral}%
  \BibitemOpen
  \bibfield  {author} {\bibinfo {author} {\bibfnamefont {S.}~\bibnamefont {Becker}}, \bibinfo {author} {\bibfnamefont {T.}~\bibnamefont {Humbert}}, \bibinfo {author} {\bibfnamefont {J.}~\bibnamefont {Wittsten}},\ and\ \bibinfo {author} {\bibfnamefont {M.}~\bibnamefont {Yang}},\ }\bibfield  {title} {\bibinfo {title} {Chiral limit of twisted trilayer graphene},\ }\href {https://doi.org/10.48550/arXiv.2308.10859} {\bibfield  {journal} {\bibinfo  {journal} {arXiv preprint arXiv:2308.10859}\ } (\bibinfo {year} {2023})}\BibitemShut {NoStop}%
\bibitem [{\citenamefont {Popov}\ and\ \citenamefont {Tarnopolsky}(2023{\natexlab{a}})}]{PhysRevB.108.L081124}%
  \BibitemOpen
  \bibfield  {author} {\bibinfo {author} {\bibfnamefont {F.~K.}\ \bibnamefont {Popov}}\ and\ \bibinfo {author} {\bibfnamefont {G.}~\bibnamefont {Tarnopolsky}},\ }\bibfield  {title} {\bibinfo {title} {Magic angles in equal-twist trilayer graphene},\ }\href {https://doi.org/10.1103/PhysRevB.108.L081124} {\bibfield  {journal} {\bibinfo  {journal} {Phys. Rev. B}\ }\textbf {\bibinfo {volume} {108}},\ \bibinfo {pages} {L081124} (\bibinfo {year} {2023}{\natexlab{a}})}\BibitemShut {NoStop}%
\bibitem [{\citenamefont {Nakatsuji}\ \emph {et~al.}(2023)\citenamefont {Nakatsuji}, \citenamefont {Kawakami},\ and\ \citenamefont {Koshino}}]{PhysRevX.13.041007}%
  \BibitemOpen
  \bibfield  {author} {\bibinfo {author} {\bibfnamefont {N.}~\bibnamefont {Nakatsuji}}, \bibinfo {author} {\bibfnamefont {T.}~\bibnamefont {Kawakami}},\ and\ \bibinfo {author} {\bibfnamefont {M.}~\bibnamefont {Koshino}},\ }\bibfield  {title} {\bibinfo {title} {Multiscale lattice relaxation in general twisted trilayer graphenes},\ }\href {https://doi.org/10.1103/PhysRevX.13.041007} {\bibfield  {journal} {\bibinfo  {journal} {Phys. Rev. X}\ }\textbf {\bibinfo {volume} {13}},\ \bibinfo {pages} {041007} (\bibinfo {year} {2023})}\BibitemShut {NoStop}%
\bibitem [{\citenamefont {Popov}\ and\ \citenamefont {Tarnopolsky}(2023{\natexlab{b}})}]{PhysRevResearch.5.043079}%
  \BibitemOpen
  \bibfield  {author} {\bibinfo {author} {\bibfnamefont {F.~K.}\ \bibnamefont {Popov}}\ and\ \bibinfo {author} {\bibfnamefont {G.}~\bibnamefont {Tarnopolsky}},\ }\bibfield  {title} {\bibinfo {title} {Magic angle butterfly in twisted trilayer graphene},\ }\href {https://doi.org/10.1103/PhysRevResearch.5.043079} {\bibfield  {journal} {\bibinfo  {journal} {Phys. Rev. Res.}\ }\textbf {\bibinfo {volume} {5}},\ \bibinfo {pages} {043079} (\bibinfo {year} {2023}{\natexlab{b}})}\BibitemShut {NoStop}%
\bibitem [{\citenamefont {Hao}\ \emph {et~al.}(2024)\citenamefont {Hao}, \citenamefont {Zhan}, \citenamefont {Pantale{\'o}n}, \citenamefont {He}, \citenamefont {Zhao}, \citenamefont {Watanabe}, \citenamefont {Taniguchi}, \citenamefont {Guinea},\ and\ \citenamefont {He}}]{hao2024robust}%
  \BibitemOpen
  \bibfield  {author} {\bibinfo {author} {\bibfnamefont {C.-Y.}\ \bibnamefont {Hao}}, \bibinfo {author} {\bibfnamefont {Z.}~\bibnamefont {Zhan}}, \bibinfo {author} {\bibfnamefont {P.~A.}\ \bibnamefont {Pantale{\'o}n}}, \bibinfo {author} {\bibfnamefont {J.-Q.}\ \bibnamefont {He}}, \bibinfo {author} {\bibfnamefont {Y.-X.}\ \bibnamefont {Zhao}}, \bibinfo {author} {\bibfnamefont {K.}~\bibnamefont {Watanabe}}, \bibinfo {author} {\bibfnamefont {T.}~\bibnamefont {Taniguchi}}, \bibinfo {author} {\bibfnamefont {F.}~\bibnamefont {Guinea}},\ and\ \bibinfo {author} {\bibfnamefont {L.}~\bibnamefont {He}},\ }\bibfield  {title} {\bibinfo {title} {Robust flat bands in twisted trilayer graphene quasicrystals},\ }\href {https://doi.org/10.48550/arXiv.2401.09010} {\bibfield  {journal} {\bibinfo  {journal} {arXiv preprint arXiv:2401.09010}\ } (\bibinfo {year} {2024})}\BibitemShut {NoStop}%
\bibitem [{\citenamefont {Foo}\ \emph {et~al.}(2024)\citenamefont {Foo}, \citenamefont {Zhan}, \citenamefont {Al~Ezzi}, \citenamefont {Peng}, \citenamefont {Adam},\ and\ \citenamefont {Guinea}}]{PhysRevResearch.6.013165}%
  \BibitemOpen
  \bibfield  {author} {\bibinfo {author} {\bibfnamefont {D.~C.~W.}\ \bibnamefont {Foo}}, \bibinfo {author} {\bibfnamefont {Z.}~\bibnamefont {Zhan}}, \bibinfo {author} {\bibfnamefont {M.~M.}\ \bibnamefont {Al~Ezzi}}, \bibinfo {author} {\bibfnamefont {L.}~\bibnamefont {Peng}}, \bibinfo {author} {\bibfnamefont {S.}~\bibnamefont {Adam}},\ and\ \bibinfo {author} {\bibfnamefont {F.}~\bibnamefont {Guinea}},\ }\bibfield  {title} {\bibinfo {title} {Extended magic phase in twisted graphene multilayers},\ }\href {https://doi.org/10.1103/PhysRevResearch.6.013165} {\bibfield  {journal} {\bibinfo  {journal} {Phys. Rev. Res.}\ }\textbf {\bibinfo {volume} {6}},\ \bibinfo {pages} {013165} (\bibinfo {year} {2024})}\BibitemShut {NoStop}%
\bibitem [{\citenamefont {Makov}\ \emph {et~al.}(2024)\citenamefont {Makov}, \citenamefont {Guinea},\ and\ \citenamefont {Stern}}]{PhysRevB.110.195112}%
  \BibitemOpen
  \bibfield  {author} {\bibinfo {author} {\bibfnamefont {R.}~\bibnamefont {Makov}}, \bibinfo {author} {\bibfnamefont {F.}~\bibnamefont {Guinea}},\ and\ \bibinfo {author} {\bibfnamefont {A.}~\bibnamefont {Stern}},\ }\bibfield  {title} {\bibinfo {title} {Flat bands in chiral multilayer graphene},\ }\href {https://doi.org/10.1103/PhysRevB.110.195112} {\bibfield  {journal} {\bibinfo  {journal} {Phys. Rev. B}\ }\textbf {\bibinfo {volume} {110}},\ \bibinfo {pages} {195112} (\bibinfo {year} {2024})}\BibitemShut {NoStop}%
\bibitem [{\citenamefont {Park}\ \emph {et~al.}(2024)\citenamefont {Park}, \citenamefont {Park}, \citenamefont {Ko}, \citenamefont {Yananose}, \citenamefont {Engelke}, \citenamefont {Zhang}, \citenamefont {Davydov}, \citenamefont {Green}, \citenamefont {Park}, \citenamefont {Lee} \emph {et~al.}}]{park2024tunable}%
  \BibitemOpen
  \bibfield  {author} {\bibinfo {author} {\bibfnamefont {D.}~\bibnamefont {Park}}, \bibinfo {author} {\bibfnamefont {C.}~\bibnamefont {Park}}, \bibinfo {author} {\bibfnamefont {E.}~\bibnamefont {Ko}}, \bibinfo {author} {\bibfnamefont {K.}~\bibnamefont {Yananose}}, \bibinfo {author} {\bibfnamefont {R.}~\bibnamefont {Engelke}}, \bibinfo {author} {\bibfnamefont {X.}~\bibnamefont {Zhang}}, \bibinfo {author} {\bibfnamefont {K.}~\bibnamefont {Davydov}}, \bibinfo {author} {\bibfnamefont {M.}~\bibnamefont {Green}}, \bibinfo {author} {\bibfnamefont {S.~H.}\ \bibnamefont {Park}}, \bibinfo {author} {\bibfnamefont {J.~H.}\ \bibnamefont {Lee}}, \emph {et~al.},\ }\bibfield  {title} {\bibinfo {title} {Tunable incommensurability and spontaneous symmetry breaking in the reconstructed moir$\backslash$'e-of-moir$\backslash$'e lattices},\ }\href {https://arxiv.org/abs/2402.15760} {\bibfield  {journal} {\bibinfo  {journal} {arXiv preprint arXiv:2402.15760}\ } (\bibinfo {year} {2024})}\BibitemShut {NoStop}%
\bibitem [{\citenamefont {Kwan}\ \emph {et~al.}(2024{\natexlab{a}})\citenamefont {Kwan}, \citenamefont {Ledwith}, \citenamefont {Lo},\ and\ \citenamefont {Devakul}}]{PhysRevB.109.125141}%
  \BibitemOpen
  \bibfield  {author} {\bibinfo {author} {\bibfnamefont {Y.~H.}\ \bibnamefont {Kwan}}, \bibinfo {author} {\bibfnamefont {P.~J.}\ \bibnamefont {Ledwith}}, \bibinfo {author} {\bibfnamefont {C.~F.~B.}\ \bibnamefont {Lo}},\ and\ \bibinfo {author} {\bibfnamefont {T.}~\bibnamefont {Devakul}},\ }\bibfield  {title} {\bibinfo {title} {Strong-coupling topological states and phase transitions in helical trilayer graphene},\ }\href {https://doi.org/10.1103/PhysRevB.109.125141} {\bibfield  {journal} {\bibinfo  {journal} {Phys. Rev. B}\ }\textbf {\bibinfo {volume} {109}},\ \bibinfo {pages} {125141} (\bibinfo {year} {2024}{\natexlab{a}})}\BibitemShut {NoStop}%
\bibitem [{\citenamefont {Craig}\ \emph {et~al.}(2024)\citenamefont {Craig}, \citenamefont {Van~Winkle}, \citenamefont {Groschner}, \citenamefont {Zhang}, \citenamefont {Dowlatshahi}, \citenamefont {Zhu}, \citenamefont {Taniguchi}, \citenamefont {Watanabe}, \citenamefont {Griffin},\ and\ \citenamefont {Bediako}}]{craig2024local}%
  \BibitemOpen
  \bibfield  {author} {\bibinfo {author} {\bibfnamefont {I.~M.}\ \bibnamefont {Craig}}, \bibinfo {author} {\bibfnamefont {M.}~\bibnamefont {Van~Winkle}}, \bibinfo {author} {\bibfnamefont {C.}~\bibnamefont {Groschner}}, \bibinfo {author} {\bibfnamefont {K.}~\bibnamefont {Zhang}}, \bibinfo {author} {\bibfnamefont {N.}~\bibnamefont {Dowlatshahi}}, \bibinfo {author} {\bibfnamefont {Z.}~\bibnamefont {Zhu}}, \bibinfo {author} {\bibfnamefont {T.}~\bibnamefont {Taniguchi}}, \bibinfo {author} {\bibfnamefont {K.}~\bibnamefont {Watanabe}}, \bibinfo {author} {\bibfnamefont {S.~M.}\ \bibnamefont {Griffin}},\ and\ \bibinfo {author} {\bibfnamefont {D.~K.}\ \bibnamefont {Bediako}},\ }\bibfield  {title} {\bibinfo {title} {Local atomic stacking and symmetry in twisted graphene trilayers},\ }\href {https://doi.org/10.1038/s41563-023-01783-y} {\bibfield  {journal} {\bibinfo  {journal} {Nature Materials}\ }\textbf {\bibinfo {volume} {23}},\ \bibinfo {pages} {323} (\bibinfo {year} {2024})}\BibitemShut {NoStop}%
\bibitem [{\citenamefont {Guerci}\ \emph {et~al.}(2024{\natexlab{a}})\citenamefont {Guerci}, \citenamefont {Mao},\ and\ \citenamefont {Mora}}]{PhysRevResearch.6.L022025}%
  \BibitemOpen
  \bibfield  {author} {\bibinfo {author} {\bibfnamefont {D.}~\bibnamefont {Guerci}}, \bibinfo {author} {\bibfnamefont {Y.}~\bibnamefont {Mao}},\ and\ \bibinfo {author} {\bibfnamefont {C.}~\bibnamefont {Mora}},\ }\bibfield  {title} {\bibinfo {title} {Chern mosaic and ideal flat bands in equal-twist trilayer graphene},\ }\href {https://doi.org/10.1103/PhysRevResearch.6.L022025} {\bibfield  {journal} {\bibinfo  {journal} {Phys. Rev. Res.}\ }\textbf {\bibinfo {volume} {6}},\ \bibinfo {pages} {L022025} (\bibinfo {year} {2024}{\natexlab{a}})}\BibitemShut {NoStop}%
\bibitem [{\citenamefont {Guerci}\ \emph {et~al.}(2024{\natexlab{b}})\citenamefont {Guerci}, \citenamefont {Mao},\ and\ \citenamefont {Mora}}]{PhysRevB.109.205411}%
  \BibitemOpen
  \bibfield  {author} {\bibinfo {author} {\bibfnamefont {D.}~\bibnamefont {Guerci}}, \bibinfo {author} {\bibfnamefont {Y.}~\bibnamefont {Mao}},\ and\ \bibinfo {author} {\bibfnamefont {C.}~\bibnamefont {Mora}},\ }\bibfield  {title} {\bibinfo {title} {Nature of even and odd magic angles in helical twisted trilayer graphene},\ }\href {https://doi.org/10.1103/PhysRevB.109.205411} {\bibfield  {journal} {\bibinfo  {journal} {Phys. Rev. B}\ }\textbf {\bibinfo {volume} {109}},\ \bibinfo {pages} {205411} (\bibinfo {year} {2024}{\natexlab{b}})}\BibitemShut {NoStop}%
\bibitem [{\citenamefont {Datta}\ \emph {et~al.}(2024)\citenamefont {Datta}, \citenamefont {Guerci}, \citenamefont {Goerbig},\ and\ \citenamefont {Mora}}]{PhysRevB.110.075417}%
  \BibitemOpen
  \bibfield  {author} {\bibinfo {author} {\bibfnamefont {A.}~\bibnamefont {Datta}}, \bibinfo {author} {\bibfnamefont {D.}~\bibnamefont {Guerci}}, \bibinfo {author} {\bibfnamefont {M.~O.}\ \bibnamefont {Goerbig}},\ and\ \bibinfo {author} {\bibfnamefont {C.}~\bibnamefont {Mora}},\ }\bibfield  {title} {\bibinfo {title} {Helical trilayer graphene in magnetic field: Chern mosaic and higher chern number ideal flat bands},\ }\href {https://doi.org/10.1103/PhysRevB.110.075417} {\bibfield  {journal} {\bibinfo  {journal} {Phys. Rev. B}\ }\textbf {\bibinfo {volume} {110}},\ \bibinfo {pages} {075417} (\bibinfo {year} {2024})}\BibitemShut {NoStop}%
\bibitem [{\citenamefont {Long}\ \emph {et~al.}(2024)\citenamefont {Long}, \citenamefont {Jimeno-Pozo}, \citenamefont {Sainz-Cruz}, \citenamefont {Pantaleón},\ and\ \citenamefont {Guinea}}]{min_Evolution}%
  \BibitemOpen
  \bibfield  {author} {\bibinfo {author} {\bibfnamefont {M.}~\bibnamefont {Long}}, \bibinfo {author} {\bibfnamefont {A.}~\bibnamefont {Jimeno-Pozo}}, \bibinfo {author} {\bibfnamefont {H.}~\bibnamefont {Sainz-Cruz}}, \bibinfo {author} {\bibfnamefont {P.~A.}\ \bibnamefont {Pantaleón}},\ and\ \bibinfo {author} {\bibfnamefont {F.}~\bibnamefont {Guinea}},\ }\bibfield  {title} {\bibinfo {title} {Evolution of superconductivity in twisted graphene multilayers},\ }\href {https://doi.org/10.1073/pnas.2405259121} {\bibfield  {journal} {\bibinfo  {journal} {Proceedings of the National Academy of Sciences}\ }\textbf {\bibinfo {volume} {121}},\ \bibinfo {pages} {e2405259121} (\bibinfo {year} {2024})}\BibitemShut {NoStop}%
\bibitem [{\citenamefont {Ren}\ \emph {et~al.}(2024)\citenamefont {Ren}, \citenamefont {Davydov}, \citenamefont {Zhu}, \citenamefont {Ma}, \citenamefont {Watanabe}, \citenamefont {Taniguchi}, \citenamefont {Kaxiras}, \citenamefont {Luskin},\ and\ \citenamefont {Wang}}]{PhysRevB.110.115404}%
  \BibitemOpen
  \bibfield  {author} {\bibinfo {author} {\bibfnamefont {W.}~\bibnamefont {Ren}}, \bibinfo {author} {\bibfnamefont {K.}~\bibnamefont {Davydov}}, \bibinfo {author} {\bibfnamefont {Z.}~\bibnamefont {Zhu}}, \bibinfo {author} {\bibfnamefont {J.}~\bibnamefont {Ma}}, \bibinfo {author} {\bibfnamefont {K.}~\bibnamefont {Watanabe}}, \bibinfo {author} {\bibfnamefont {T.}~\bibnamefont {Taniguchi}}, \bibinfo {author} {\bibfnamefont {E.}~\bibnamefont {Kaxiras}}, \bibinfo {author} {\bibfnamefont {M.}~\bibnamefont {Luskin}},\ and\ \bibinfo {author} {\bibfnamefont {K.}~\bibnamefont {Wang}},\ }\bibfield  {title} {\bibinfo {title} {Tunable inter-moir\'e physics in consecutively twisted trilayer graphene},\ }\href {https://doi.org/10.1103/PhysRevB.110.115404} {\bibfield  {journal} {\bibinfo  {journal} {Phys. Rev. B}\ }\textbf {\bibinfo {volume} {110}},\ \bibinfo {pages} {115404} (\bibinfo {year} {2024})}\BibitemShut {NoStop}%
\bibitem [{\citenamefont {Yang}\ \emph {et~al.}(2024)\citenamefont {Yang}, \citenamefont {May-Mann}, \citenamefont {Zhu},\ and\ \citenamefont {Devakul}}]{PhysRevB.110.115434}%
  \BibitemOpen
  \bibfield  {author} {\bibinfo {author} {\bibfnamefont {C.}~\bibnamefont {Yang}}, \bibinfo {author} {\bibfnamefont {J.}~\bibnamefont {May-Mann}}, \bibinfo {author} {\bibfnamefont {Z.}~\bibnamefont {Zhu}},\ and\ \bibinfo {author} {\bibfnamefont {T.}~\bibnamefont {Devakul}},\ }\bibfield  {title} {\bibinfo {title} {Multi-moir\'e trilayer graphene: Lattice relaxation, electronic structure, and magic angles},\ }\href {https://doi.org/10.1103/PhysRevB.110.115434} {\bibfield  {journal} {\bibinfo  {journal} {Phys. Rev. B}\ }\textbf {\bibinfo {volume} {110}},\ \bibinfo {pages} {115434} (\bibinfo {year} {2024})}\BibitemShut {NoStop}%
\bibitem [{\citenamefont {Kwan}\ \emph {et~al.}(2024{\natexlab{b}})\citenamefont {Kwan}, \citenamefont {Tan},\ and\ \citenamefont {Devakul}}]{kwan2024fractional}%
  \BibitemOpen
  \bibfield  {author} {\bibinfo {author} {\bibfnamefont {Y.~H.}\ \bibnamefont {Kwan}}, \bibinfo {author} {\bibfnamefont {T.}~\bibnamefont {Tan}},\ and\ \bibinfo {author} {\bibfnamefont {T.}~\bibnamefont {Devakul}},\ }\bibfield  {title} {\bibinfo {title} {Fractional chern mosaic in supermoir$\backslash$'e graphene},\ }\href {https://doi.org/10.48550/arXiv.2411.08880} {\bibfield  {journal} {\bibinfo  {journal} {arXiv preprint arXiv:2411.08880}\ } (\bibinfo {year} {2024}{\natexlab{b}})}\BibitemShut {NoStop}%
\bibitem [{\citenamefont {Niu}\ \emph {et~al.}(2025)\citenamefont {Niu}, \citenamefont {Alicea}, \citenamefont {Sheng},\ and\ \citenamefont {Peng}}]{niu2025quantumanomaloushalleffects}%
  \BibitemOpen
  \bibfield  {author} {\bibinfo {author} {\bibfnamefont {S.}~\bibnamefont {Niu}}, \bibinfo {author} {\bibfnamefont {J.}~\bibnamefont {Alicea}}, \bibinfo {author} {\bibfnamefont {D.~N.}\ \bibnamefont {Sheng}},\ and\ \bibinfo {author} {\bibfnamefont {Y.}~\bibnamefont {Peng}},\ }\href {https://arxiv.org/abs/2505.24146} {\bibinfo {title} {Quantum anomalous hall effects and emergent $\rm{SU}(2)$ hall ferromagnets at fractional filling of helical trilayer graphene}} (\bibinfo {year} {2025}),\ \Eprint {https://arxiv.org/abs/2505.24146} {arXiv:2505.24146 [cond-mat.str-el]} \BibitemShut {NoStop}%
\bibitem [{\citenamefont {Hoke}\ \emph {et~al.}(2024)\citenamefont {Hoke}, \citenamefont {Li}, \citenamefont {Hu}, \citenamefont {May-Mann}, \citenamefont {Watanabe}, \citenamefont {Taniguchi}, \citenamefont {Devakul},\ and\ \citenamefont {Feldman}}]{hoke2024imaging}%
  \BibitemOpen
  \bibfield  {author} {\bibinfo {author} {\bibfnamefont {J.~C.}\ \bibnamefont {Hoke}}, \bibinfo {author} {\bibfnamefont {Y.}~\bibnamefont {Li}}, \bibinfo {author} {\bibfnamefont {Y.}~\bibnamefont {Hu}}, \bibinfo {author} {\bibfnamefont {J.}~\bibnamefont {May-Mann}}, \bibinfo {author} {\bibfnamefont {K.}~\bibnamefont {Watanabe}}, \bibinfo {author} {\bibfnamefont {T.}~\bibnamefont {Taniguchi}}, \bibinfo {author} {\bibfnamefont {T.}~\bibnamefont {Devakul}},\ and\ \bibinfo {author} {\bibfnamefont {B.~E.}\ \bibnamefont {Feldman}},\ }\bibfield  {title} {\bibinfo {title} {Imaging supermoire relaxation and conductive domain walls in helical trilayer graphene},\ }\href {https://doi.org/10.48550/arXiv.2410.16269} {\bibfield  {journal} {\bibinfo  {journal} {arXiv preprint arXiv:2410.16269}\ } (\bibinfo {year} {2024})}\BibitemShut {NoStop}%
\bibitem [{\citenamefont {Xia}\ \emph {et~al.}(2025)\citenamefont {Xia}, \citenamefont {de~la Barrera}, \citenamefont {Uri}, \citenamefont {Sharpe}, \citenamefont {Kwan}, \citenamefont {Zhu}, \citenamefont {Watanabe}, \citenamefont {Taniguchi}, \citenamefont {Goldhaber-Gordon}, \citenamefont {Fu} \emph {et~al.}}]{xia2025topological}%
  \BibitemOpen
  \bibfield  {author} {\bibinfo {author} {\bibfnamefont {L.-Q.}\ \bibnamefont {Xia}}, \bibinfo {author} {\bibfnamefont {S.~C.}\ \bibnamefont {de~la Barrera}}, \bibinfo {author} {\bibfnamefont {A.}~\bibnamefont {Uri}}, \bibinfo {author} {\bibfnamefont {A.}~\bibnamefont {Sharpe}}, \bibinfo {author} {\bibfnamefont {Y.~H.}\ \bibnamefont {Kwan}}, \bibinfo {author} {\bibfnamefont {Z.}~\bibnamefont {Zhu}}, \bibinfo {author} {\bibfnamefont {K.}~\bibnamefont {Watanabe}}, \bibinfo {author} {\bibfnamefont {T.}~\bibnamefont {Taniguchi}}, \bibinfo {author} {\bibfnamefont {D.}~\bibnamefont {Goldhaber-Gordon}}, \bibinfo {author} {\bibfnamefont {L.}~\bibnamefont {Fu}}, \emph {et~al.},\ }\bibfield  {title} {\bibinfo {title} {Topological bands and correlated states in helical trilayer graphene},\ }\href {https://doi.org/10.1038/s41567-024-02731-6} {\bibfield  {journal} {\bibinfo  {journal} {Nature Physics}\ ,\ \bibinfo {pages} {1}} (\bibinfo {year} {2025})}\BibitemShut {NoStop}%
\bibitem [{\citenamefont {Polshyn}\ \emph {et~al.}(2020)\citenamefont {Polshyn}, \citenamefont {Zhu}, \citenamefont {Kumar}, \citenamefont {Zhang}, \citenamefont {Yang}, \citenamefont {Tschirhart}, \citenamefont {Serlin}, \citenamefont {Watanabe}, \citenamefont {Taniguchi}, \citenamefont {MacDonald} \emph {et~al.}}]{polshyn2020electrical}%
  \BibitemOpen
  \bibfield  {author} {\bibinfo {author} {\bibfnamefont {H.}~\bibnamefont {Polshyn}}, \bibinfo {author} {\bibfnamefont {J.}~\bibnamefont {Zhu}}, \bibinfo {author} {\bibfnamefont {M.~A.}\ \bibnamefont {Kumar}}, \bibinfo {author} {\bibfnamefont {Y.}~\bibnamefont {Zhang}}, \bibinfo {author} {\bibfnamefont {F.}~\bibnamefont {Yang}}, \bibinfo {author} {\bibfnamefont {C.~L.}\ \bibnamefont {Tschirhart}}, \bibinfo {author} {\bibfnamefont {M.}~\bibnamefont {Serlin}}, \bibinfo {author} {\bibfnamefont {K.}~\bibnamefont {Watanabe}}, \bibinfo {author} {\bibfnamefont {T.}~\bibnamefont {Taniguchi}}, \bibinfo {author} {\bibfnamefont {A.~H.}\ \bibnamefont {MacDonald}}, \emph {et~al.},\ }\bibfield  {title} {\bibinfo {title} {Electrical switching of magnetic order in an orbital chern insulator},\ }\href {https://doi.org/10.1038/s41586-020-2963-8} {\bibfield  {journal} {\bibinfo  {journal} {Nature}\ }\textbf {\bibinfo {volume} {588}},\ \bibinfo {pages} {66} (\bibinfo {year} {2020})}\BibitemShut {NoStop}%
\bibitem [{\citenamefont {Chen}\ \emph {et~al.}(2021)\citenamefont {Chen}, \citenamefont {He}, \citenamefont {Zhang}, \citenamefont {Hsieh}, \citenamefont {Fei}, \citenamefont {Watanabe}, \citenamefont {Taniguchi}, \citenamefont {Cobden}, \citenamefont {Xu}, \citenamefont {Dean} \emph {et~al.}}]{chen2021electrically}%
  \BibitemOpen
  \bibfield  {author} {\bibinfo {author} {\bibfnamefont {S.}~\bibnamefont {Chen}}, \bibinfo {author} {\bibfnamefont {M.}~\bibnamefont {He}}, \bibinfo {author} {\bibfnamefont {Y.-H.}\ \bibnamefont {Zhang}}, \bibinfo {author} {\bibfnamefont {V.}~\bibnamefont {Hsieh}}, \bibinfo {author} {\bibfnamefont {Z.}~\bibnamefont {Fei}}, \bibinfo {author} {\bibfnamefont {K.}~\bibnamefont {Watanabe}}, \bibinfo {author} {\bibfnamefont {T.}~\bibnamefont {Taniguchi}}, \bibinfo {author} {\bibfnamefont {D.~H.}\ \bibnamefont {Cobden}}, \bibinfo {author} {\bibfnamefont {X.}~\bibnamefont {Xu}}, \bibinfo {author} {\bibfnamefont {C.~R.}\ \bibnamefont {Dean}}, \emph {et~al.},\ }\bibfield  {title} {\bibinfo {title} {Electrically tunable correlated and topological states in twisted monolayer--bilayer graphene},\ }\href {https://doi.org/10.1038/s41567-020-01062-6} {\bibfield  {journal} {\bibinfo  {journal} {Nature Physics}\ }\textbf {\bibinfo {volume} {17}},\ \bibinfo {pages} {374} (\bibinfo {year} {2021})}\BibitemShut {NoStop}%
\bibitem [{\citenamefont {He}\ \emph {et~al.}(2021)\citenamefont {He}, \citenamefont {Zhang}, \citenamefont {Li}, \citenamefont {Fei}, \citenamefont {Watanabe}, \citenamefont {Taniguchi}, \citenamefont {Xu},\ and\ \citenamefont {Yankowitz}}]{he2021competing}%
  \BibitemOpen
  \bibfield  {author} {\bibinfo {author} {\bibfnamefont {M.}~\bibnamefont {He}}, \bibinfo {author} {\bibfnamefont {Y.-H.}\ \bibnamefont {Zhang}}, \bibinfo {author} {\bibfnamefont {Y.}~\bibnamefont {Li}}, \bibinfo {author} {\bibfnamefont {Z.}~\bibnamefont {Fei}}, \bibinfo {author} {\bibfnamefont {K.}~\bibnamefont {Watanabe}}, \bibinfo {author} {\bibfnamefont {T.}~\bibnamefont {Taniguchi}}, \bibinfo {author} {\bibfnamefont {X.}~\bibnamefont {Xu}},\ and\ \bibinfo {author} {\bibfnamefont {M.}~\bibnamefont {Yankowitz}},\ }\bibfield  {title} {\bibinfo {title} {Competing correlated states and abundant orbital magnetism in twisted monolayer-bilayer graphene},\ }\href {https://doi.org/10.1038/s41467-021-25044-1} {\bibfield  {journal} {\bibinfo  {journal} {Nature Communications}\ }\textbf {\bibinfo {volume} {12}},\ \bibinfo {pages} {4727} (\bibinfo {year} {2021})}\BibitemShut {NoStop}%
\bibitem [{\citenamefont {Zhang}\ \emph {et~al.}(2023)\citenamefont {Zhang}, \citenamefont {Zhu}, \citenamefont {Soejima}, \citenamefont {Kahn}, \citenamefont {Watanabe}, \citenamefont {Taniguchi}, \citenamefont {Zettl}, \citenamefont {Wang}, \citenamefont {Zaletel},\ and\ \citenamefont {Crommie}}]{zhang2023local}%
  \BibitemOpen
  \bibfield  {author} {\bibinfo {author} {\bibfnamefont {C.}~\bibnamefont {Zhang}}, \bibinfo {author} {\bibfnamefont {T.}~\bibnamefont {Zhu}}, \bibinfo {author} {\bibfnamefont {T.}~\bibnamefont {Soejima}}, \bibinfo {author} {\bibfnamefont {S.}~\bibnamefont {Kahn}}, \bibinfo {author} {\bibfnamefont {K.}~\bibnamefont {Watanabe}}, \bibinfo {author} {\bibfnamefont {T.}~\bibnamefont {Taniguchi}}, \bibinfo {author} {\bibfnamefont {A.}~\bibnamefont {Zettl}}, \bibinfo {author} {\bibfnamefont {F.}~\bibnamefont {Wang}}, \bibinfo {author} {\bibfnamefont {M.~P.}\ \bibnamefont {Zaletel}},\ and\ \bibinfo {author} {\bibfnamefont {M.~F.}\ \bibnamefont {Crommie}},\ }\bibfield  {title} {\bibinfo {title} {Local spectroscopy of a gate-switchable moir{\'e} quantum anomalous hall insulator},\ }\href {https://doi.org/10.1038/s41467-023-39110-3} {\bibfield  {journal} {\bibinfo  {journal} {Nature Communications}\ }\textbf {\bibinfo {volume} {14}},\ \bibinfo {pages} {3595} (\bibinfo {year} {2023})}\BibitemShut {NoStop}%
\bibitem [{\citenamefont {Peng}\ \emph {et~al.}(2024)\citenamefont {Peng}, \citenamefont {Zhong}, \citenamefont {Feng}, \citenamefont {Hu}, \citenamefont {Li}, \citenamefont {Zhang}, \citenamefont {Mao}, \citenamefont {Duan},\ and\ \citenamefont {Yao}}]{peng2024abundant}%
  \BibitemOpen
  \bibfield  {author} {\bibinfo {author} {\bibfnamefont {H.}~\bibnamefont {Peng}}, \bibinfo {author} {\bibfnamefont {J.}~\bibnamefont {Zhong}}, \bibinfo {author} {\bibfnamefont {Q.}~\bibnamefont {Feng}}, \bibinfo {author} {\bibfnamefont {Y.}~\bibnamefont {Hu}}, \bibinfo {author} {\bibfnamefont {Q.}~\bibnamefont {Li}}, \bibinfo {author} {\bibfnamefont {S.}~\bibnamefont {Zhang}}, \bibinfo {author} {\bibfnamefont {J.}~\bibnamefont {Mao}}, \bibinfo {author} {\bibfnamefont {J.}~\bibnamefont {Duan}},\ and\ \bibinfo {author} {\bibfnamefont {Y.}~\bibnamefont {Yao}},\ }\bibfield  {title} {\bibinfo {title} {Abundant electric-field tunable symmetry-broken states in twisted monolayer-bilayer graphene},\ }\href {https://doi.org/10.1038/s42005-024-01722-6} {\bibfield  {journal} {\bibinfo  {journal} {Communications Physics}\ }\textbf {\bibinfo {volume} {7}},\ \bibinfo {pages} {240} (\bibinfo {year} {2024})}\BibitemShut {NoStop}%
\bibitem [{\citenamefont {Kuiri}\ \emph {et~al.}(2022)\citenamefont {Kuiri}, \citenamefont {Coleman}, \citenamefont {Gao}, \citenamefont {Vishnuradhan}, \citenamefont {Watanabe}, \citenamefont {Taniguchi}, \citenamefont {Zhu}, \citenamefont {MacDonald},\ and\ \citenamefont {Folk}}]{kuiri2022spontaneous}%
  \BibitemOpen
  \bibfield  {author} {\bibinfo {author} {\bibfnamefont {M.}~\bibnamefont {Kuiri}}, \bibinfo {author} {\bibfnamefont {C.}~\bibnamefont {Coleman}}, \bibinfo {author} {\bibfnamefont {Z.}~\bibnamefont {Gao}}, \bibinfo {author} {\bibfnamefont {A.}~\bibnamefont {Vishnuradhan}}, \bibinfo {author} {\bibfnamefont {K.}~\bibnamefont {Watanabe}}, \bibinfo {author} {\bibfnamefont {T.}~\bibnamefont {Taniguchi}}, \bibinfo {author} {\bibfnamefont {J.}~\bibnamefont {Zhu}}, \bibinfo {author} {\bibfnamefont {A.~H.}\ \bibnamefont {MacDonald}},\ and\ \bibinfo {author} {\bibfnamefont {J.}~\bibnamefont {Folk}},\ }\bibfield  {title} {\bibinfo {title} {Spontaneous time-reversal symmetry breaking in twisted double bilayer graphene},\ }\href {https://doi.org/10.1038/s41467-022-34192-x} {\bibfield  {journal} {\bibinfo  {journal} {Nature Communications}\ }\textbf {\bibinfo {volume} {13}},\ \bibinfo {pages} {6468} (\bibinfo {year} {2022})}\BibitemShut {NoStop}%
\bibitem [{\citenamefont {He}\ \emph {et~al.}(2023)\citenamefont {He}, \citenamefont {Cai}, \citenamefont {Zhang}, \citenamefont {Liu}, \citenamefont {Li}, \citenamefont {Taniguchi}, \citenamefont {Watanabe}, \citenamefont {Cobden}, \citenamefont {Yankowitz},\ and\ \citenamefont {Xu}}]{he2023symmetry}%
  \BibitemOpen
  \bibfield  {author} {\bibinfo {author} {\bibfnamefont {M.}~\bibnamefont {He}}, \bibinfo {author} {\bibfnamefont {J.}~\bibnamefont {Cai}}, \bibinfo {author} {\bibfnamefont {Y.-H.}\ \bibnamefont {Zhang}}, \bibinfo {author} {\bibfnamefont {Y.}~\bibnamefont {Liu}}, \bibinfo {author} {\bibfnamefont {Y.}~\bibnamefont {Li}}, \bibinfo {author} {\bibfnamefont {T.}~\bibnamefont {Taniguchi}}, \bibinfo {author} {\bibfnamefont {K.}~\bibnamefont {Watanabe}}, \bibinfo {author} {\bibfnamefont {D.~H.}\ \bibnamefont {Cobden}}, \bibinfo {author} {\bibfnamefont {M.}~\bibnamefont {Yankowitz}},\ and\ \bibinfo {author} {\bibfnamefont {X.}~\bibnamefont {Xu}},\ }\bibfield  {title} {\bibinfo {title} {Symmetry-broken chern insulators in twisted double bilayer graphene},\ }\href {https://doi.org/10.1021/acs.nanolett.3c03414} {\bibfield  {journal} {\bibinfo  {journal} {Nano Letters}\ }\textbf {\bibinfo {volume} {23}},\ \bibinfo {pages} {11066} (\bibinfo {year} {2023})}\BibitemShut {NoStop}%
\bibitem [{\citenamefont {Khalaf}\ \emph {et~al.}(2019)\citenamefont {Khalaf}, \citenamefont {Kruchkov}, \citenamefont {Tarnopolsky},\ and\ \citenamefont {Vishwanath}}]{khalafMagicAngleHierarchy2019}%
  \BibitemOpen
  \bibfield  {author} {\bibinfo {author} {\bibfnamefont {E.}~\bibnamefont {Khalaf}}, \bibinfo {author} {\bibfnamefont {A.~J.}\ \bibnamefont {Kruchkov}}, \bibinfo {author} {\bibfnamefont {G.}~\bibnamefont {Tarnopolsky}},\ and\ \bibinfo {author} {\bibfnamefont {A.}~\bibnamefont {Vishwanath}},\ }\bibfield  {title} {\bibinfo {title} {Magic angle hierarchy in twisted graphene multilayers},\ }\href {https://doi.org/10.1103/PhysRevB.100.085109} {\bibfield  {journal} {\bibinfo  {journal} {Physical Review B}\ }\textbf {\bibinfo {volume} {100}},\ \bibinfo {pages} {085109} (\bibinfo {year} {2019})}\BibitemShut {NoStop}%
\bibitem [{\citenamefont {Park}\ \emph {et~al.}(2021)\citenamefont {Park}, \citenamefont {Cao}, \citenamefont {Watanabe}, \citenamefont {Taniguchi},\ and\ \citenamefont {Jarillo-Herrero}}]{park2021tunable}%
  \BibitemOpen
  \bibfield  {author} {\bibinfo {author} {\bibfnamefont {J.~M.}\ \bibnamefont {Park}}, \bibinfo {author} {\bibfnamefont {Y.}~\bibnamefont {Cao}}, \bibinfo {author} {\bibfnamefont {K.}~\bibnamefont {Watanabe}}, \bibinfo {author} {\bibfnamefont {T.}~\bibnamefont {Taniguchi}},\ and\ \bibinfo {author} {\bibfnamefont {P.}~\bibnamefont {Jarillo-Herrero}},\ }\bibfield  {title} {\bibinfo {title} {Tunable strongly coupled superconductivity in magic-angle twisted trilayer graphene},\ }\href@noop {} {\bibfield  {journal} {\bibinfo  {journal} {Nature}\ }\textbf {\bibinfo {volume} {590}},\ \bibinfo {pages} {249} (\bibinfo {year} {2021})}\BibitemShut {NoStop}%
\bibitem [{\citenamefont {Hao}\ \emph {et~al.}(2021)\citenamefont {Hao}, \citenamefont {Zimmerman}, \citenamefont {Ledwith}, \citenamefont {Khalaf}, \citenamefont {Najafabadi}, \citenamefont {Watanabe}, \citenamefont {Taniguchi}, \citenamefont {Vishwanath},\ and\ \citenamefont {Kim}}]{hao2021electric}%
  \BibitemOpen
  \bibfield  {author} {\bibinfo {author} {\bibfnamefont {Z.}~\bibnamefont {Hao}}, \bibinfo {author} {\bibfnamefont {A.}~\bibnamefont {Zimmerman}}, \bibinfo {author} {\bibfnamefont {P.}~\bibnamefont {Ledwith}}, \bibinfo {author} {\bibfnamefont {E.}~\bibnamefont {Khalaf}}, \bibinfo {author} {\bibfnamefont {D.~H.}\ \bibnamefont {Najafabadi}}, \bibinfo {author} {\bibfnamefont {K.}~\bibnamefont {Watanabe}}, \bibinfo {author} {\bibfnamefont {T.}~\bibnamefont {Taniguchi}}, \bibinfo {author} {\bibfnamefont {A.}~\bibnamefont {Vishwanath}},\ and\ \bibinfo {author} {\bibfnamefont {P.}~\bibnamefont {Kim}},\ }\bibfield  {title} {\bibinfo {title} {Electric field--tunable superconductivity in alternating-twist magic-angle trilayer graphene},\ }\href@noop {} {\bibfield  {journal} {\bibinfo  {journal} {Science}\ }\textbf {\bibinfo {volume} {371}},\ \bibinfo {pages} {1133} (\bibinfo {year} {2021})}\BibitemShut {NoStop}%
\bibitem [{\citenamefont {Cao}\ \emph {et~al.}(2021)\citenamefont {Cao}, \citenamefont {Park}, \citenamefont {Watanabe}, \citenamefont {Taniguchi},\ and\ \citenamefont {Jarillo-Herrero}}]{cao2021pauli}%
  \BibitemOpen
  \bibfield  {author} {\bibinfo {author} {\bibfnamefont {Y.}~\bibnamefont {Cao}}, \bibinfo {author} {\bibfnamefont {J.~M.}\ \bibnamefont {Park}}, \bibinfo {author} {\bibfnamefont {K.}~\bibnamefont {Watanabe}}, \bibinfo {author} {\bibfnamefont {T.}~\bibnamefont {Taniguchi}},\ and\ \bibinfo {author} {\bibfnamefont {P.}~\bibnamefont {Jarillo-Herrero}},\ }\bibfield  {title} {\bibinfo {title} {Pauli-limit violation and re-entrant superconductivity in moir{\'e} graphene},\ }\href@noop {} {\bibfield  {journal} {\bibinfo  {journal} {Nature}\ }\textbf {\bibinfo {volume} {595}},\ \bibinfo {pages} {526} (\bibinfo {year} {2021})}\BibitemShut {NoStop}%
\bibitem [{\citenamefont {Park}\ \emph {et~al.}(2022)\citenamefont {Park}, \citenamefont {Cao}, \citenamefont {Xia}, \citenamefont {Sun}, \citenamefont {Watanabe}, \citenamefont {Taniguchi},\ and\ \citenamefont {Jarillo-Herrero}}]{park2022robust}%
  \BibitemOpen
  \bibfield  {author} {\bibinfo {author} {\bibfnamefont {J.~M.}\ \bibnamefont {Park}}, \bibinfo {author} {\bibfnamefont {Y.}~\bibnamefont {Cao}}, \bibinfo {author} {\bibfnamefont {L.-Q.}\ \bibnamefont {Xia}}, \bibinfo {author} {\bibfnamefont {S.}~\bibnamefont {Sun}}, \bibinfo {author} {\bibfnamefont {K.}~\bibnamefont {Watanabe}}, \bibinfo {author} {\bibfnamefont {T.}~\bibnamefont {Taniguchi}},\ and\ \bibinfo {author} {\bibfnamefont {P.}~\bibnamefont {Jarillo-Herrero}},\ }\bibfield  {title} {\bibinfo {title} {Robust superconductivity in magic-angle multilayer graphene family},\ }\href@noop {} {\bibfield  {journal} {\bibinfo  {journal} {Nature Materials}\ }\textbf {\bibinfo {volume} {21}},\ \bibinfo {pages} {877} (\bibinfo {year} {2022})}\BibitemShut {NoStop}%
\bibitem [{\citenamefont {Zhang}\ \emph {et~al.}(2022)\citenamefont {Zhang}, \citenamefont {Polski}, \citenamefont {Lewandowski}, \citenamefont {Thomson}, \citenamefont {Peng}, \citenamefont {Choi}, \citenamefont {Kim}, \citenamefont {Watanabe}, \citenamefont {Taniguchi}, \citenamefont {Alicea} \emph {et~al.}}]{zhang2022promotion}%
  \BibitemOpen
  \bibfield  {author} {\bibinfo {author} {\bibfnamefont {Y.}~\bibnamefont {Zhang}}, \bibinfo {author} {\bibfnamefont {R.}~\bibnamefont {Polski}}, \bibinfo {author} {\bibfnamefont {C.}~\bibnamefont {Lewandowski}}, \bibinfo {author} {\bibfnamefont {A.}~\bibnamefont {Thomson}}, \bibinfo {author} {\bibfnamefont {Y.}~\bibnamefont {Peng}}, \bibinfo {author} {\bibfnamefont {Y.}~\bibnamefont {Choi}}, \bibinfo {author} {\bibfnamefont {H.}~\bibnamefont {Kim}}, \bibinfo {author} {\bibfnamefont {K.}~\bibnamefont {Watanabe}}, \bibinfo {author} {\bibfnamefont {T.}~\bibnamefont {Taniguchi}}, \bibinfo {author} {\bibfnamefont {J.}~\bibnamefont {Alicea}}, \emph {et~al.},\ }\bibfield  {title} {\bibinfo {title} {Promotion of superconductivity in magic-angle graphene multilayers},\ }\href@noop {} {\bibfield  {journal} {\bibinfo  {journal} {Science}\ }\textbf {\bibinfo {volume} {377}},\ \bibinfo {pages} {1538} (\bibinfo {year} {2022})}\BibitemShut {NoStop}%
\bibitem [{\citenamefont {Nam}\ and\ \citenamefont {Koshino}(2017)}]{PhysRevB.96.075311}%
  \BibitemOpen
  \bibfield  {author} {\bibinfo {author} {\bibfnamefont {N.~N.~T.}\ \bibnamefont {Nam}}\ and\ \bibinfo {author} {\bibfnamefont {M.}~\bibnamefont {Koshino}},\ }\bibfield  {title} {\bibinfo {title} {Lattice relaxation and energy band modulation in twisted bilayer graphene},\ }\href {https://doi.org/10.1103/PhysRevB.96.075311} {\bibfield  {journal} {\bibinfo  {journal} {Phys. Rev. B}\ }\textbf {\bibinfo {volume} {96}},\ \bibinfo {pages} {075311} (\bibinfo {year} {2017})}\BibitemShut {NoStop}%
\bibitem [{\citenamefont {Koshino}\ and\ \citenamefont {Son}(2019)}]{PhysRevB.100.075416}%
  \BibitemOpen
  \bibfield  {author} {\bibinfo {author} {\bibfnamefont {M.}~\bibnamefont {Koshino}}\ and\ \bibinfo {author} {\bibfnamefont {Y.-W.}\ \bibnamefont {Son}},\ }\bibfield  {title} {\bibinfo {title} {Moir\'e phonons in twisted bilayer graphene},\ }\href {https://doi.org/10.1103/PhysRevB.100.075416} {\bibfield  {journal} {\bibinfo  {journal} {Phys. Rev. B}\ }\textbf {\bibinfo {volume} {100}},\ \bibinfo {pages} {075416} (\bibinfo {year} {2019})}\BibitemShut {NoStop}%
\bibitem [{\citenamefont {Krisna}\ and\ \citenamefont {Koshino}(2023)}]{PhysRevB.107.115301}%
  \BibitemOpen
  \bibfield  {author} {\bibinfo {author} {\bibfnamefont {L.~P.~A.}\ \bibnamefont {Krisna}}\ and\ \bibinfo {author} {\bibfnamefont {M.}~\bibnamefont {Koshino}},\ }\bibfield  {title} {\bibinfo {title} {Moir\'e phonons in graphene/hexagonal boron nitride moir\'e superlattice},\ }\href {https://doi.org/10.1103/PhysRevB.107.115301} {\bibfield  {journal} {\bibinfo  {journal} {Phys. Rev. B}\ }\textbf {\bibinfo {volume} {107}},\ \bibinfo {pages} {115301} (\bibinfo {year} {2023})}\BibitemShut {NoStop}%
\bibitem [{\citenamefont {Guinea}\ and\ \citenamefont {Walet}(2019)}]{PhysRevB.99.205134}%
  \BibitemOpen
  \bibfield  {author} {\bibinfo {author} {\bibfnamefont {F.}~\bibnamefont {Guinea}}\ and\ \bibinfo {author} {\bibfnamefont {N.~R.}\ \bibnamefont {Walet}},\ }\bibfield  {title} {\bibinfo {title} {Continuum models for twisted bilayer graphene: Effect of lattice deformation and hopping parameters},\ }\href {https://doi.org/10.1103/PhysRevB.99.205134} {\bibfield  {journal} {\bibinfo  {journal} {Phys. Rev. B}\ }\textbf {\bibinfo {volume} {99}},\ \bibinfo {pages} {205134} (\bibinfo {year} {2019})}\BibitemShut {NoStop}%
\bibitem [{\citenamefont {Fang}\ \emph {et~al.}(2019)\citenamefont {Fang}, \citenamefont {Carr}, \citenamefont {Zhu}, \citenamefont {Massatt},\ and\ \citenamefont {Kaxiras}}]{fang2019angle}%
  \BibitemOpen
  \bibfield  {author} {\bibinfo {author} {\bibfnamefont {S.}~\bibnamefont {Fang}}, \bibinfo {author} {\bibfnamefont {S.}~\bibnamefont {Carr}}, \bibinfo {author} {\bibfnamefont {Z.}~\bibnamefont {Zhu}}, \bibinfo {author} {\bibfnamefont {D.}~\bibnamefont {Massatt}},\ and\ \bibinfo {author} {\bibfnamefont {E.}~\bibnamefont {Kaxiras}},\ }\bibfield  {title} {\bibinfo {title} {Angle-dependent $\{$$\backslash$it Ab initio$\}$ low-energy hamiltonians for a relaxed twisted bilayer graphene heterostructure},\ }\href {https://doi.org/10.48550/arXiv.1908.00058} {\bibfield  {journal} {\bibinfo  {journal} {arXiv preprint arXiv:1908.00058}\ } (\bibinfo {year} {2019})}\BibitemShut {NoStop}%
\bibitem [{\citenamefont {Carr}\ \emph {et~al.}(2019)\citenamefont {Carr}, \citenamefont {Fang}, \citenamefont {Zhu},\ and\ \citenamefont {Kaxiras}}]{PhysRevResearch.1.013001}%
  \BibitemOpen
  \bibfield  {author} {\bibinfo {author} {\bibfnamefont {S.}~\bibnamefont {Carr}}, \bibinfo {author} {\bibfnamefont {S.}~\bibnamefont {Fang}}, \bibinfo {author} {\bibfnamefont {Z.}~\bibnamefont {Zhu}},\ and\ \bibinfo {author} {\bibfnamefont {E.}~\bibnamefont {Kaxiras}},\ }\bibfield  {title} {\bibinfo {title} {Exact continuum model for low-energy electronic states of twisted bilayer graphene},\ }\href {https://doi.org/10.1103/PhysRevResearch.1.013001} {\bibfield  {journal} {\bibinfo  {journal} {Phys. Rev. Res.}\ }\textbf {\bibinfo {volume} {1}},\ \bibinfo {pages} {013001} (\bibinfo {year} {2019})}\BibitemShut {NoStop}%
\bibitem [{\citenamefont {Koshino}\ and\ \citenamefont {Nam}(2020)}]{PhysRevB.101.195425}%
  \BibitemOpen
  \bibfield  {author} {\bibinfo {author} {\bibfnamefont {M.}~\bibnamefont {Koshino}}\ and\ \bibinfo {author} {\bibfnamefont {N.~N.~T.}\ \bibnamefont {Nam}},\ }\bibfield  {title} {\bibinfo {title} {Effective continuum model for relaxed twisted bilayer graphene and moir\'e electron-phonon interaction},\ }\href {https://doi.org/10.1103/PhysRevB.101.195425} {\bibfield  {journal} {\bibinfo  {journal} {Phys. Rev. B}\ }\textbf {\bibinfo {volume} {101}},\ \bibinfo {pages} {195425} (\bibinfo {year} {2020})}\BibitemShut {NoStop}%
\bibitem [{\citenamefont {Carr}\ \emph {et~al.}(2018)\citenamefont {Carr}, \citenamefont {Massatt}, \citenamefont {Torrisi}, \citenamefont {Cazeaux}, \citenamefont {Luskin},\ and\ \citenamefont {Kaxiras}}]{PhysRevB.98.224102}%
  \BibitemOpen
  \bibfield  {author} {\bibinfo {author} {\bibfnamefont {S.}~\bibnamefont {Carr}}, \bibinfo {author} {\bibfnamefont {D.}~\bibnamefont {Massatt}}, \bibinfo {author} {\bibfnamefont {S.~B.}\ \bibnamefont {Torrisi}}, \bibinfo {author} {\bibfnamefont {P.}~\bibnamefont {Cazeaux}}, \bibinfo {author} {\bibfnamefont {M.}~\bibnamefont {Luskin}},\ and\ \bibinfo {author} {\bibfnamefont {E.}~\bibnamefont {Kaxiras}},\ }\bibfield  {title} {\bibinfo {title} {Relaxation and domain formation in incommensurate two-dimensional heterostructures},\ }\href {https://doi.org/10.1103/PhysRevB.98.224102} {\bibfield  {journal} {\bibinfo  {journal} {Phys. Rev. B}\ }\textbf {\bibinfo {volume} {98}},\ \bibinfo {pages} {224102} (\bibinfo {year} {2018})}\BibitemShut {NoStop}%
\bibitem [{\citenamefont {Ledwith}\ \emph {et~al.}(2021{\natexlab{a}})\citenamefont {Ledwith}, \citenamefont {Khalaf}, \citenamefont {Zhu}, \citenamefont {Carr}, \citenamefont {Kaxiras},\ and\ \citenamefont {Vishwanath}}]{ledwith2021tb}%
  \BibitemOpen
  \bibfield  {author} {\bibinfo {author} {\bibfnamefont {P.~J.}\ \bibnamefont {Ledwith}}, \bibinfo {author} {\bibfnamefont {E.}~\bibnamefont {Khalaf}}, \bibinfo {author} {\bibfnamefont {Z.}~\bibnamefont {Zhu}}, \bibinfo {author} {\bibfnamefont {S.}~\bibnamefont {Carr}}, \bibinfo {author} {\bibfnamefont {E.}~\bibnamefont {Kaxiras}},\ and\ \bibinfo {author} {\bibfnamefont {A.}~\bibnamefont {Vishwanath}},\ }\bibfield  {title} {\bibinfo {title} {Tb or not tb? contrasting properties of twisted bilayer graphene and the alternating twist $ n $-layer structures ($ n= 3, 4, 5,$¥backslash$dots $)},\ }\href {https://doi.org/10.48550/arXiv.2111.11060} {\bibfield  {journal} {\bibinfo  {journal} {arXiv preprint arXiv:2111.11060}\ } (\bibinfo {year} {2021}{\natexlab{a}})}\BibitemShut {NoStop}%
\bibitem [{\citenamefont {Koshino}\ \emph {et~al.}(2018)\citenamefont {Koshino}, \citenamefont {Yuan}, \citenamefont {Koretsune}, \citenamefont {Ochi}, \citenamefont {Kuroki},\ and\ \citenamefont {Fu}}]{PhysRevX.8.031087}%
  \BibitemOpen
  \bibfield  {author} {\bibinfo {author} {\bibfnamefont {M.}~\bibnamefont {Koshino}}, \bibinfo {author} {\bibfnamefont {N.~F.~Q.}\ \bibnamefont {Yuan}}, \bibinfo {author} {\bibfnamefont {T.}~\bibnamefont {Koretsune}}, \bibinfo {author} {\bibfnamefont {M.}~\bibnamefont {Ochi}}, \bibinfo {author} {\bibfnamefont {K.}~\bibnamefont {Kuroki}},\ and\ \bibinfo {author} {\bibfnamefont {L.}~\bibnamefont {Fu}},\ }\bibfield  {title} {\bibinfo {title} {Maximally localized wannier orbitals and the extended hubbard model for twisted bilayer graphene},\ }\href {https://doi.org/10.1103/PhysRevX.8.031087} {\bibfield  {journal} {\bibinfo  {journal} {Phys. Rev. X}\ }\textbf {\bibinfo {volume} {8}},\ \bibinfo {pages} {031087} (\bibinfo {year} {2018})}\BibitemShut {NoStop}%
\bibitem [{\citenamefont {Das}\ \emph {et~al.}(2021)\citenamefont {Das}, \citenamefont {Lu}, \citenamefont {Herzog-Arbeitman}, \citenamefont {Song}, \citenamefont {Watanabe}, \citenamefont {Taniguchi}, \citenamefont {Bernevig},\ and\ \citenamefont {Efetov}}]{das2021symmetry}%
  \BibitemOpen
  \bibfield  {author} {\bibinfo {author} {\bibfnamefont {I.}~\bibnamefont {Das}}, \bibinfo {author} {\bibfnamefont {X.}~\bibnamefont {Lu}}, \bibinfo {author} {\bibfnamefont {J.}~\bibnamefont {Herzog-Arbeitman}}, \bibinfo {author} {\bibfnamefont {Z.-D.}\ \bibnamefont {Song}}, \bibinfo {author} {\bibfnamefont {K.}~\bibnamefont {Watanabe}}, \bibinfo {author} {\bibfnamefont {T.}~\bibnamefont {Taniguchi}}, \bibinfo {author} {\bibfnamefont {B.~A.}\ \bibnamefont {Bernevig}},\ and\ \bibinfo {author} {\bibfnamefont {D.~K.}\ \bibnamefont {Efetov}},\ }\bibfield  {title} {\bibinfo {title} {Symmetry-broken chern insulators and rashba-like landau-level crossings in magic-angle bilayer graphene},\ }\href {https://doi.org/10.1038/s41567-021-01186-3} {\bibfield  {journal} {\bibinfo  {journal} {Nature Physics}\ }\textbf {\bibinfo {volume} {17}},\ \bibinfo {pages} {710} (\bibinfo {year} {2021})}\BibitemShut {NoStop}%
\bibitem [{\citenamefont {Vafek}\ and\ \citenamefont {Kang}(2020)}]{PhysRevLett.125.257602}%
  \BibitemOpen
  \bibfield  {author} {\bibinfo {author} {\bibfnamefont {O.}~\bibnamefont {Vafek}}\ and\ \bibinfo {author} {\bibfnamefont {J.}~\bibnamefont {Kang}},\ }\bibfield  {title} {\bibinfo {title} {Renormalization group study of hidden symmetry in twisted bilayer graphene with coulomb interactions},\ }\href {https://doi.org/10.1103/PhysRevLett.125.257602} {\bibfield  {journal} {\bibinfo  {journal} {Phys. Rev. Lett.}\ }\textbf {\bibinfo {volume} {125}},\ \bibinfo {pages} {257602} (\bibinfo {year} {2020})}\BibitemShut {NoStop}%
\bibitem [{\citenamefont {Liu}\ \emph {et~al.}(2019)\citenamefont {Liu}, \citenamefont {Liu},\ and\ \citenamefont {Dai}}]{PhysRevB.99.155415}%
  \BibitemOpen
  \bibfield  {author} {\bibinfo {author} {\bibfnamefont {J.}~\bibnamefont {Liu}}, \bibinfo {author} {\bibfnamefont {J.}~\bibnamefont {Liu}},\ and\ \bibinfo {author} {\bibfnamefont {X.}~\bibnamefont {Dai}},\ }\bibfield  {title} {\bibinfo {title} {Pseudo landau level representation of twisted bilayer graphene: Band topology and implications on the correlated insulating phase},\ }\href {https://doi.org/10.1103/PhysRevB.99.155415} {\bibfield  {journal} {\bibinfo  {journal} {Phys. Rev. B}\ }\textbf {\bibinfo {volume} {99}},\ \bibinfo {pages} {155415} (\bibinfo {year} {2019})}\BibitemShut {NoStop}%
\bibitem [{\citenamefont {Bultinck}\ \emph {et~al.}(2020)\citenamefont {Bultinck}, \citenamefont {Khalaf}, \citenamefont {Liu}, \citenamefont {Chatterjee}, \citenamefont {Vishwanath},\ and\ \citenamefont {Zaletel}}]{PhysRevX.10.031034}%
  \BibitemOpen
  \bibfield  {author} {\bibinfo {author} {\bibfnamefont {N.}~\bibnamefont {Bultinck}}, \bibinfo {author} {\bibfnamefont {E.}~\bibnamefont {Khalaf}}, \bibinfo {author} {\bibfnamefont {S.}~\bibnamefont {Liu}}, \bibinfo {author} {\bibfnamefont {S.}~\bibnamefont {Chatterjee}}, \bibinfo {author} {\bibfnamefont {A.}~\bibnamefont {Vishwanath}},\ and\ \bibinfo {author} {\bibfnamefont {M.~P.}\ \bibnamefont {Zaletel}},\ }\bibfield  {title} {\bibinfo {title} {Ground state and hidden symmetry of magic-angle graphene at even integer filling},\ }\href {https://doi.org/10.1103/PhysRevX.10.031034} {\bibfield  {journal} {\bibinfo  {journal} {Phys. Rev. X}\ }\textbf {\bibinfo {volume} {10}},\ \bibinfo {pages} {031034} (\bibinfo {year} {2020})}\BibitemShut {NoStop}%
\bibitem [{\citenamefont {Bernevig}\ \emph {et~al.}(2021)\citenamefont {Bernevig}, \citenamefont {Song}, \citenamefont {Regnault},\ and\ \citenamefont {Lian}}]{PhysRevB.103.205413}%
  \BibitemOpen
  \bibfield  {author} {\bibinfo {author} {\bibfnamefont {B.~A.}\ \bibnamefont {Bernevig}}, \bibinfo {author} {\bibfnamefont {Z.-D.}\ \bibnamefont {Song}}, \bibinfo {author} {\bibfnamefont {N.}~\bibnamefont {Regnault}},\ and\ \bibinfo {author} {\bibfnamefont {B.}~\bibnamefont {Lian}},\ }\bibfield  {title} {\bibinfo {title} {Twisted bilayer graphene. iii. interacting hamiltonian and exact symmetries},\ }\href {https://doi.org/10.1103/PhysRevB.103.205413} {\bibfield  {journal} {\bibinfo  {journal} {Phys. Rev. B}\ }\textbf {\bibinfo {volume} {103}},\ \bibinfo {pages} {205413} (\bibinfo {year} {2021})}\BibitemShut {NoStop}%
\bibitem [{\citenamefont {Trambly~de Laissardière}\ \emph {et~al.}(2010)\citenamefont {Trambly~de Laissardière}, \citenamefont {Mayou},\ and\ \citenamefont {Magaud}}]{Trambly2010}%
  \BibitemOpen
  \bibfield  {author} {\bibinfo {author} {\bibfnamefont {G.}~\bibnamefont {Trambly~de Laissardière}}, \bibinfo {author} {\bibfnamefont {D.}~\bibnamefont {Mayou}},\ and\ \bibinfo {author} {\bibfnamefont {L.}~\bibnamefont {Magaud}},\ }\bibfield  {title} {\bibinfo {title} {Localization of dirac electrons in rotated graphene bilayers},\ }\href {https://doi.org/10.1021/nl902948m} {\bibfield  {journal} {\bibinfo  {journal} {Nano Letters}\ }\textbf {\bibinfo {volume} {10}},\ \bibinfo {pages} {804} (\bibinfo {year} {2010})},\ \bibinfo {note} {pMID: 20121163}\BibitemShut {NoStop}%
\bibitem [{\citenamefont {Shallcross}\ \emph {et~al.}(2010)\citenamefont {Shallcross}, \citenamefont {Sharma}, \citenamefont {Kandelaki},\ and\ \citenamefont {Pankratov}}]{PhysRevB.81.165105}%
  \BibitemOpen
  \bibfield  {author} {\bibinfo {author} {\bibfnamefont {S.}~\bibnamefont {Shallcross}}, \bibinfo {author} {\bibfnamefont {S.}~\bibnamefont {Sharma}}, \bibinfo {author} {\bibfnamefont {E.}~\bibnamefont {Kandelaki}},\ and\ \bibinfo {author} {\bibfnamefont {O.~A.}\ \bibnamefont {Pankratov}},\ }\bibfield  {title} {\bibinfo {title} {Electronic structure of turbostratic graphene},\ }\href {https://doi.org/10.1103/PhysRevB.81.165105} {\bibfield  {journal} {\bibinfo  {journal} {Phys. Rev. B}\ }\textbf {\bibinfo {volume} {81}},\ \bibinfo {pages} {165105} (\bibinfo {year} {2010})}\BibitemShut {NoStop}%
\bibitem [{\citenamefont {Su\'arez~Morell}\ \emph {et~al.}(2010)\citenamefont {Su\'arez~Morell}, \citenamefont {Correa}, \citenamefont {Vargas}, \citenamefont {Pacheco},\ and\ \citenamefont {Barticevic}}]{PhysRevB.82.121407}%
  \BibitemOpen
  \bibfield  {author} {\bibinfo {author} {\bibfnamefont {E.}~\bibnamefont {Su\'arez~Morell}}, \bibinfo {author} {\bibfnamefont {J.~D.}\ \bibnamefont {Correa}}, \bibinfo {author} {\bibfnamefont {P.}~\bibnamefont {Vargas}}, \bibinfo {author} {\bibfnamefont {M.}~\bibnamefont {Pacheco}},\ and\ \bibinfo {author} {\bibfnamefont {Z.}~\bibnamefont {Barticevic}},\ }\bibfield  {title} {\bibinfo {title} {Flat bands in slightly twisted bilayer graphene: Tight-binding calculations},\ }\href {https://doi.org/10.1103/PhysRevB.82.121407} {\bibfield  {journal} {\bibinfo  {journal} {Phys. Rev. B}\ }\textbf {\bibinfo {volume} {82}},\ \bibinfo {pages} {121407} (\bibinfo {year} {2010})}\BibitemShut {NoStop}%
\bibitem [{\citenamefont {Kindermann}\ and\ \citenamefont {First}(2011)}]{PhysRevB.83.045425}%
  \BibitemOpen
  \bibfield  {author} {\bibinfo {author} {\bibfnamefont {M.}~\bibnamefont {Kindermann}}\ and\ \bibinfo {author} {\bibfnamefont {P.~N.}\ \bibnamefont {First}},\ }\bibfield  {title} {\bibinfo {title} {Local sublattice-symmetry breaking in rotationally faulted multilayer graphene},\ }\href {https://doi.org/10.1103/PhysRevB.83.045425} {\bibfield  {journal} {\bibinfo  {journal} {Phys. Rev. B}\ }\textbf {\bibinfo {volume} {83}},\ \bibinfo {pages} {045425} (\bibinfo {year} {2011})}\BibitemShut {NoStop}%
\bibitem [{\citenamefont {San-Jose}\ \emph {et~al.}(2012)\citenamefont {San-Jose}, \citenamefont {Gonz\'alez},\ and\ \citenamefont {Guinea}}]{PhysRevLett.108.216802}%
  \BibitemOpen
  \bibfield  {author} {\bibinfo {author} {\bibfnamefont {P.}~\bibnamefont {San-Jose}}, \bibinfo {author} {\bibfnamefont {J.}~\bibnamefont {Gonz\'alez}},\ and\ \bibinfo {author} {\bibfnamefont {F.}~\bibnamefont {Guinea}},\ }\bibfield  {title} {\bibinfo {title} {Non-abelian gauge potentials in graphene bilayers},\ }\href {https://doi.org/10.1103/PhysRevLett.108.216802} {\bibfield  {journal} {\bibinfo  {journal} {Phys. Rev. Lett.}\ }\textbf {\bibinfo {volume} {108}},\ \bibinfo {pages} {216802} (\bibinfo {year} {2012})}\BibitemShut {NoStop}%
\bibitem [{\citenamefont {Lopes~dos Santos}\ \emph {et~al.}(2012)\citenamefont {Lopes~dos Santos}, \citenamefont {Peres},\ and\ \citenamefont {Castro~Neto}}]{PhysRevB.86.155449}%
  \BibitemOpen
  \bibfield  {author} {\bibinfo {author} {\bibfnamefont {J.~M.~B.}\ \bibnamefont {Lopes~dos Santos}}, \bibinfo {author} {\bibfnamefont {N.~M.~R.}\ \bibnamefont {Peres}},\ and\ \bibinfo {author} {\bibfnamefont {A.~H.}\ \bibnamefont {Castro~Neto}},\ }\bibfield  {title} {\bibinfo {title} {Continuum model of the twisted graphene bilayer},\ }\href {https://doi.org/10.1103/PhysRevB.86.155449} {\bibfield  {journal} {\bibinfo  {journal} {Phys. Rev. B}\ }\textbf {\bibinfo {volume} {86}},\ \bibinfo {pages} {155449} (\bibinfo {year} {2012})}\BibitemShut {NoStop}%
\bibitem [{\citenamefont {Xie}\ \emph {et~al.}(2019)\citenamefont {Xie}, \citenamefont {Lian}, \citenamefont {J{\"a}ck}, \citenamefont {Liu}, \citenamefont {Chiu}, \citenamefont {Watanabe}, \citenamefont {Taniguchi}, \citenamefont {Bernevig},\ and\ \citenamefont {Yazdani}}]{xie2019spectroscopic}%
  \BibitemOpen
  \bibfield  {author} {\bibinfo {author} {\bibfnamefont {Y.}~\bibnamefont {Xie}}, \bibinfo {author} {\bibfnamefont {B.}~\bibnamefont {Lian}}, \bibinfo {author} {\bibfnamefont {B.}~\bibnamefont {J{\"a}ck}}, \bibinfo {author} {\bibfnamefont {X.}~\bibnamefont {Liu}}, \bibinfo {author} {\bibfnamefont {C.-L.}\ \bibnamefont {Chiu}}, \bibinfo {author} {\bibfnamefont {K.}~\bibnamefont {Watanabe}}, \bibinfo {author} {\bibfnamefont {T.}~\bibnamefont {Taniguchi}}, \bibinfo {author} {\bibfnamefont {B.~A.}\ \bibnamefont {Bernevig}},\ and\ \bibinfo {author} {\bibfnamefont {A.}~\bibnamefont {Yazdani}},\ }\bibfield  {title} {\bibinfo {title} {Spectroscopic signatures of many-body correlations in magic-angle twisted bilayer graphene},\ }\href {https://doi.org/10.1038/s41586-019-1422-x} {\bibfield  {journal} {\bibinfo  {journal} {Nature}\ }\textbf {\bibinfo {volume} {572}},\ \bibinfo {pages} {101} (\bibinfo {year} {2019})}\BibitemShut {NoStop}%
\bibitem [{\citenamefont {Kerelsky}\ \emph {et~al.}(2019)\citenamefont {Kerelsky}, \citenamefont {McGilly}, \citenamefont {Kennes}, \citenamefont {Xian}, \citenamefont {Yankowitz}, \citenamefont {Chen}, \citenamefont {Watanabe}, \citenamefont {Taniguchi}, \citenamefont {Hone}, \citenamefont {Dean} \emph {et~al.}}]{kerelsky2019maximized}%
  \BibitemOpen
  \bibfield  {author} {\bibinfo {author} {\bibfnamefont {A.}~\bibnamefont {Kerelsky}}, \bibinfo {author} {\bibfnamefont {L.~J.}\ \bibnamefont {McGilly}}, \bibinfo {author} {\bibfnamefont {D.~M.}\ \bibnamefont {Kennes}}, \bibinfo {author} {\bibfnamefont {L.}~\bibnamefont {Xian}}, \bibinfo {author} {\bibfnamefont {M.}~\bibnamefont {Yankowitz}}, \bibinfo {author} {\bibfnamefont {S.}~\bibnamefont {Chen}}, \bibinfo {author} {\bibfnamefont {K.}~\bibnamefont {Watanabe}}, \bibinfo {author} {\bibfnamefont {T.}~\bibnamefont {Taniguchi}}, \bibinfo {author} {\bibfnamefont {J.}~\bibnamefont {Hone}}, \bibinfo {author} {\bibfnamefont {C.}~\bibnamefont {Dean}}, \emph {et~al.},\ }\bibfield  {title} {\bibinfo {title} {Maximized electron interactions at the magic angle in twisted bilayer graphene},\ }\href {https://doi.org/10.1038/s41586-019-1431-9} {\bibfield  {journal} {\bibinfo  {journal} {Nature}\ }\textbf {\bibinfo {volume} {572}},\ \bibinfo {pages} {95} (\bibinfo {year} {2019})}\BibitemShut {NoStop}%
\bibitem [{\citenamefont {Tilak}\ \emph {et~al.}(2021)\citenamefont {Tilak}, \citenamefont {Lai}, \citenamefont {Wu}, \citenamefont {Zhang}, \citenamefont {Xu}, \citenamefont {Ribeiro}, \citenamefont {Canfield},\ and\ \citenamefont {Andrei}}]{tilak2021flat}%
  \BibitemOpen
  \bibfield  {author} {\bibinfo {author} {\bibfnamefont {N.}~\bibnamefont {Tilak}}, \bibinfo {author} {\bibfnamefont {X.}~\bibnamefont {Lai}}, \bibinfo {author} {\bibfnamefont {S.}~\bibnamefont {Wu}}, \bibinfo {author} {\bibfnamefont {Z.}~\bibnamefont {Zhang}}, \bibinfo {author} {\bibfnamefont {M.}~\bibnamefont {Xu}}, \bibinfo {author} {\bibfnamefont {R.~d.~A.}\ \bibnamefont {Ribeiro}}, \bibinfo {author} {\bibfnamefont {P.~C.}\ \bibnamefont {Canfield}},\ and\ \bibinfo {author} {\bibfnamefont {E.~Y.}\ \bibnamefont {Andrei}},\ }\bibfield  {title} {\bibinfo {title} {Flat band carrier confinement in magic-angle twisted bilayer graphene},\ }\href {https://doi.org/10.1038/s41467-021-24480-3} {\bibfield  {journal} {\bibinfo  {journal} {Nature communications}\ }\textbf {\bibinfo {volume} {12}},\ \bibinfo {pages} {4180} (\bibinfo {year} {2021})}\BibitemShut {NoStop}%
\bibitem [{\citenamefont {Polshyn}\ \emph {et~al.}(2022)\citenamefont {Polshyn}, \citenamefont {Zhang}, \citenamefont {Kumar}, \citenamefont {Soejima}, \citenamefont {Ledwith}, \citenamefont {Watanabe}, \citenamefont {Taniguchi}, \citenamefont {Vishwanath}, \citenamefont {Zaletel},\ and\ \citenamefont {Young}}]{polshyn2022topological}%
  \BibitemOpen
  \bibfield  {author} {\bibinfo {author} {\bibfnamefont {H.}~\bibnamefont {Polshyn}}, \bibinfo {author} {\bibfnamefont {Y.}~\bibnamefont {Zhang}}, \bibinfo {author} {\bibfnamefont {M.~A.}\ \bibnamefont {Kumar}}, \bibinfo {author} {\bibfnamefont {T.}~\bibnamefont {Soejima}}, \bibinfo {author} {\bibfnamefont {P.}~\bibnamefont {Ledwith}}, \bibinfo {author} {\bibfnamefont {K.}~\bibnamefont {Watanabe}}, \bibinfo {author} {\bibfnamefont {T.}~\bibnamefont {Taniguchi}}, \bibinfo {author} {\bibfnamefont {A.}~\bibnamefont {Vishwanath}}, \bibinfo {author} {\bibfnamefont {M.~P.}\ \bibnamefont {Zaletel}},\ and\ \bibinfo {author} {\bibfnamefont {A.~F.}\ \bibnamefont {Young}},\ }\bibfield  {title} {\bibinfo {title} {Topological charge density waves at half-integer filling of a moir{¥'e} superlattice},\ }\href {https://doi.org/10.1038/s41567-021-01418-6} {\bibfield  {journal} {\bibinfo  {journal} {Nature Physics}\ }\textbf {\bibinfo {volume} {18}},\ \bibinfo {pages} {42} (\bibinfo {year} {2022})}\BibitemShut {NoStop}%
\bibitem [{\citenamefont {Su}\ \emph {et~al.}(2025)\citenamefont {Su}, \citenamefont {Waters}, \citenamefont {Zhou}, \citenamefont {Watanabe}, \citenamefont {Taniguchi}, \citenamefont {Zhang}, \citenamefont {Yankowitz},\ and\ \citenamefont {Folk}}]{su2025moire}%
  \BibitemOpen
  \bibfield  {author} {\bibinfo {author} {\bibfnamefont {R.}~\bibnamefont {Su}}, \bibinfo {author} {\bibfnamefont {D.}~\bibnamefont {Waters}}, \bibinfo {author} {\bibfnamefont {B.}~\bibnamefont {Zhou}}, \bibinfo {author} {\bibfnamefont {K.}~\bibnamefont {Watanabe}}, \bibinfo {author} {\bibfnamefont {T.}~\bibnamefont {Taniguchi}}, \bibinfo {author} {\bibfnamefont {Y.-H.}\ \bibnamefont {Zhang}}, \bibinfo {author} {\bibfnamefont {M.}~\bibnamefont {Yankowitz}},\ and\ \bibinfo {author} {\bibfnamefont {J.}~\bibnamefont {Folk}},\ }\bibfield  {title} {\bibinfo {title} {Moir{¥'e}-driven topological electronic crystals in twisted graphene},\ }\href {https://doi.org/10.1038/s41586-024-08239-6} {\bibfield  {journal} {\bibinfo  {journal} {Nature}\ }\textbf {\bibinfo {volume} {637}},\ \bibinfo {pages} {1084} (\bibinfo {year} {2025})}\BibitemShut {NoStop}%
\bibitem [{\citenamefont {Waters}\ \emph {et~al.}(2025)\citenamefont {Waters}, \citenamefont {Okounkova}, \citenamefont {Su}, \citenamefont {Zhou}, \citenamefont {Yao}, \citenamefont {Watanabe}, \citenamefont {Taniguchi}, \citenamefont {Xu}, \citenamefont {Zhang}, \citenamefont {Folk},\ and\ \citenamefont {Yankowitz}}]{PhysRevX.15.011045}%
  \BibitemOpen
  \bibfield  {author} {\bibinfo {author} {\bibfnamefont {D.}~\bibnamefont {Waters}}, \bibinfo {author} {\bibfnamefont {A.}~\bibnamefont {Okounkova}}, \bibinfo {author} {\bibfnamefont {R.}~\bibnamefont {Su}}, \bibinfo {author} {\bibfnamefont {B.}~\bibnamefont {Zhou}}, \bibinfo {author} {\bibfnamefont {J.}~\bibnamefont {Yao}}, \bibinfo {author} {\bibfnamefont {K.}~\bibnamefont {Watanabe}}, \bibinfo {author} {\bibfnamefont {T.}~\bibnamefont {Taniguchi}}, \bibinfo {author} {\bibfnamefont {X.}~\bibnamefont {Xu}}, \bibinfo {author} {\bibfnamefont {Y.-H.}\ \bibnamefont {Zhang}}, \bibinfo {author} {\bibfnamefont {J.}~\bibnamefont {Folk}},\ and\ \bibinfo {author} {\bibfnamefont {M.}~\bibnamefont {Yankowitz}},\ }\bibfield  {title} {\bibinfo {title} {Chern insulators at integer and fractional filling in moir¥'e pentalayer graphene},\ }\href {https://doi.org/10.1103/PhysRevX.15.011045} {\bibfield  {journal} {\bibinfo  {journal} {Phys. Rev. X}\ }\textbf {\bibinfo {volume} {15}},\ \bibinfo {pages} {011045} (\bibinfo
  {year} {2025})}\BibitemShut {NoStop}%
\bibitem [{\citenamefont {Spanton}\ \emph {et~al.}(2018)\citenamefont {Spanton}, \citenamefont {Zibrov}, \citenamefont {Zhou}, \citenamefont {Taniguchi}, \citenamefont {Watanabe}, \citenamefont {Zaletel},\ and\ \citenamefont {Young}}]{spanton2018observation}%
  \BibitemOpen
  \bibfield  {author} {\bibinfo {author} {\bibfnamefont {E.~M.}\ \bibnamefont {Spanton}}, \bibinfo {author} {\bibfnamefont {A.~A.}\ \bibnamefont {Zibrov}}, \bibinfo {author} {\bibfnamefont {H.}~\bibnamefont {Zhou}}, \bibinfo {author} {\bibfnamefont {T.}~\bibnamefont {Taniguchi}}, \bibinfo {author} {\bibfnamefont {K.}~\bibnamefont {Watanabe}}, \bibinfo {author} {\bibfnamefont {M.~P.}\ \bibnamefont {Zaletel}},\ and\ \bibinfo {author} {\bibfnamefont {A.~F.}\ \bibnamefont {Young}},\ }\bibfield  {title} {\bibinfo {title} {Observation of fractional chern insulators in a van der waals heterostructure},\ }\href {https://doi.org/10.1126/science.aan8458} {\bibfield  {journal} {\bibinfo  {journal} {Science}\ }\textbf {\bibinfo {volume} {360}},\ \bibinfo {pages} {62} (\bibinfo {year} {2018})}\BibitemShut {NoStop}%
\bibitem [{\citenamefont {Cai}\ \emph {et~al.}(2023)\citenamefont {Cai}, \citenamefont {Anderson}, \citenamefont {Wang}, \citenamefont {Zhang}, \citenamefont {Liu}, \citenamefont {Holtzmann}, \citenamefont {Zhang}, \citenamefont {Fan}, \citenamefont {Taniguchi}, \citenamefont {Watanabe} \emph {et~al.}}]{cai2023signatures}%
  \BibitemOpen
  \bibfield  {author} {\bibinfo {author} {\bibfnamefont {J.}~\bibnamefont {Cai}}, \bibinfo {author} {\bibfnamefont {E.}~\bibnamefont {Anderson}}, \bibinfo {author} {\bibfnamefont {C.}~\bibnamefont {Wang}}, \bibinfo {author} {\bibfnamefont {X.}~\bibnamefont {Zhang}}, \bibinfo {author} {\bibfnamefont {X.}~\bibnamefont {Liu}}, \bibinfo {author} {\bibfnamefont {W.}~\bibnamefont {Holtzmann}}, \bibinfo {author} {\bibfnamefont {Y.}~\bibnamefont {Zhang}}, \bibinfo {author} {\bibfnamefont {F.}~\bibnamefont {Fan}}, \bibinfo {author} {\bibfnamefont {T.}~\bibnamefont {Taniguchi}}, \bibinfo {author} {\bibfnamefont {K.}~\bibnamefont {Watanabe}}, \emph {et~al.},\ }\bibfield  {title} {\bibinfo {title} {Signatures of fractional quantum anomalous hall states in twisted \ce{MoTe2}},\ }\href {https://doi.org/10.1038/s41586-023-06289-w} {\bibfield  {journal} {\bibinfo  {journal} {Nature}\ }\textbf {\bibinfo {volume} {622}},\ \bibinfo {pages} {63} (\bibinfo {year} {2023})}\BibitemShut {NoStop}%
\bibitem [{\citenamefont {Zeng}\ \emph {et~al.}(2023)\citenamefont {Zeng}, \citenamefont {Xia}, \citenamefont {Kang}, \citenamefont {Zhu}, \citenamefont {Kn{\"u}ppel}, \citenamefont {Vaswani}, \citenamefont {Watanabe}, \citenamefont {Taniguchi}, \citenamefont {Mak},\ and\ \citenamefont {Shan}}]{zeng2023thermodynamic}%
  \BibitemOpen
  \bibfield  {author} {\bibinfo {author} {\bibfnamefont {Y.}~\bibnamefont {Zeng}}, \bibinfo {author} {\bibfnamefont {Z.}~\bibnamefont {Xia}}, \bibinfo {author} {\bibfnamefont {K.}~\bibnamefont {Kang}}, \bibinfo {author} {\bibfnamefont {J.}~\bibnamefont {Zhu}}, \bibinfo {author} {\bibfnamefont {P.}~\bibnamefont {Kn{\"u}ppel}}, \bibinfo {author} {\bibfnamefont {C.}~\bibnamefont {Vaswani}}, \bibinfo {author} {\bibfnamefont {K.}~\bibnamefont {Watanabe}}, \bibinfo {author} {\bibfnamefont {T.}~\bibnamefont {Taniguchi}}, \bibinfo {author} {\bibfnamefont {K.~F.}\ \bibnamefont {Mak}},\ and\ \bibinfo {author} {\bibfnamefont {J.}~\bibnamefont {Shan}},\ }\bibfield  {title} {\bibinfo {title} {Thermodynamic evidence of fractional chern insulator in moir{\'e} \ce{MoTe2}},\ }\href {https://doi.org/10.1038/s41586-023-06452-3} {\bibfield  {journal} {\bibinfo  {journal} {Nature}\ }\textbf {\bibinfo {volume} {622}},\ \bibinfo {pages} {69} (\bibinfo {year} {2023})}\BibitemShut {NoStop}%
\bibitem [{\citenamefont {Park}\ \emph {et~al.}(2023)\citenamefont {Park}, \citenamefont {Cai}, \citenamefont {Anderson}, \citenamefont {Zhang}, \citenamefont {Zhu}, \citenamefont {Liu}, \citenamefont {Wang}, \citenamefont {Holtzmann}, \citenamefont {Hu}, \citenamefont {Liu} \emph {et~al.}}]{park2023observation}%
  \BibitemOpen
  \bibfield  {author} {\bibinfo {author} {\bibfnamefont {H.}~\bibnamefont {Park}}, \bibinfo {author} {\bibfnamefont {J.}~\bibnamefont {Cai}}, \bibinfo {author} {\bibfnamefont {E.}~\bibnamefont {Anderson}}, \bibinfo {author} {\bibfnamefont {Y.}~\bibnamefont {Zhang}}, \bibinfo {author} {\bibfnamefont {J.}~\bibnamefont {Zhu}}, \bibinfo {author} {\bibfnamefont {X.}~\bibnamefont {Liu}}, \bibinfo {author} {\bibfnamefont {C.}~\bibnamefont {Wang}}, \bibinfo {author} {\bibfnamefont {W.}~\bibnamefont {Holtzmann}}, \bibinfo {author} {\bibfnamefont {C.}~\bibnamefont {Hu}}, \bibinfo {author} {\bibfnamefont {Z.}~\bibnamefont {Liu}}, \emph {et~al.},\ }\bibfield  {title} {\bibinfo {title} {Observation of fractionally quantized anomalous hall effect},\ }\href {https://doi.org/10.1038/s41586-023-06536-0} {\bibfield  {journal} {\bibinfo  {journal} {Nature}\ }\textbf {\bibinfo {volume} {622}},\ \bibinfo {pages} {74} (\bibinfo {year} {2023})}\BibitemShut {NoStop}%
\bibitem [{\citenamefont {Xu}\ \emph {et~al.}(2023)\citenamefont {Xu}, \citenamefont {Sun}, \citenamefont {Jia}, \citenamefont {Liu}, \citenamefont {Xu}, \citenamefont {Li}, \citenamefont {Gu}, \citenamefont {Watanabe}, \citenamefont {Taniguchi}, \citenamefont {Tong} \emph {et~al.}}]{xu2023observation}%
  \BibitemOpen
  \bibfield  {author} {\bibinfo {author} {\bibfnamefont {F.}~\bibnamefont {Xu}}, \bibinfo {author} {\bibfnamefont {Z.}~\bibnamefont {Sun}}, \bibinfo {author} {\bibfnamefont {T.}~\bibnamefont {Jia}}, \bibinfo {author} {\bibfnamefont {C.}~\bibnamefont {Liu}}, \bibinfo {author} {\bibfnamefont {C.}~\bibnamefont {Xu}}, \bibinfo {author} {\bibfnamefont {C.}~\bibnamefont {Li}}, \bibinfo {author} {\bibfnamefont {Y.}~\bibnamefont {Gu}}, \bibinfo {author} {\bibfnamefont {K.}~\bibnamefont {Watanabe}}, \bibinfo {author} {\bibfnamefont {T.}~\bibnamefont {Taniguchi}}, \bibinfo {author} {\bibfnamefont {B.}~\bibnamefont {Tong}}, \emph {et~al.},\ }\bibfield  {title} {\bibinfo {title} {Observation of integer and fractional quantum anomalous hall effects in twisted bilayer \ce{MoTe2}},\ }\href {https://doi.org/10.1038/s41586-023-06536-0} {\bibfield  {journal} {\bibinfo  {journal} {Physical Review X}\ }\textbf {\bibinfo {volume} {13}},\ \bibinfo {pages} {031037} (\bibinfo {year} {2023})}\BibitemShut {NoStop}%
\bibitem [{\citenamefont {Lu}\ \emph {et~al.}(2023)\citenamefont {Lu}, \citenamefont {Han}, \citenamefont {Yao}, \citenamefont {Reddy}, \citenamefont {Yang}, \citenamefont {Seo}, \citenamefont {Watanabe}, \citenamefont {Taniguchi}, \citenamefont {Fu},\ and\ \citenamefont {Ju}}]{lu2023fractional}%
  \BibitemOpen
  \bibfield  {author} {\bibinfo {author} {\bibfnamefont {Z.}~\bibnamefont {Lu}}, \bibinfo {author} {\bibfnamefont {T.}~\bibnamefont {Han}}, \bibinfo {author} {\bibfnamefont {Y.}~\bibnamefont {Yao}}, \bibinfo {author} {\bibfnamefont {A.~P.}\ \bibnamefont {Reddy}}, \bibinfo {author} {\bibfnamefont {J.}~\bibnamefont {Yang}}, \bibinfo {author} {\bibfnamefont {J.}~\bibnamefont {Seo}}, \bibinfo {author} {\bibfnamefont {K.}~\bibnamefont {Watanabe}}, \bibinfo {author} {\bibfnamefont {T.}~\bibnamefont {Taniguchi}}, \bibinfo {author} {\bibfnamefont {L.}~\bibnamefont {Fu}},\ and\ \bibinfo {author} {\bibfnamefont {L.}~\bibnamefont {Ju}},\ }\bibfield  {title} {\bibinfo {title} {Fractional quantum anomalous hall effect in a graphene moire superlattice},\ }\bibfield  {journal} {\bibinfo  {journal} {arXiv preprint arXiv:2309.17436}\ }\href {https://doi.org/10.48550/arXiv.2309.17436} {10.48550/arXiv.2309.17436} (\bibinfo {year} {2023})\BibitemShut {NoStop}%
\bibitem [{\citenamefont {Parameswaran}\ \emph {et~al.}(2013)\citenamefont {Parameswaran}, \citenamefont {Roy},\ and\ \citenamefont {Sondhi}}]{parameswaran2013fractional}%
  \BibitemOpen
  \bibfield  {author} {\bibinfo {author} {\bibfnamefont {S.~A.}\ \bibnamefont {Parameswaran}}, \bibinfo {author} {\bibfnamefont {R.}~\bibnamefont {Roy}},\ and\ \bibinfo {author} {\bibfnamefont {S.~L.}\ \bibnamefont {Sondhi}},\ }\bibfield  {title} {\bibinfo {title} {Fractional quantum hall physics in topological flat bands},\ }\href {https://comptes-rendus.academie-sciences.fr/physique/articles/10.1016/j.crhy.2013.04.003/} {\bibfield  {journal} {\bibinfo  {journal} {C. R. Phys.}\ }\textbf {\bibinfo {volume} {14}},\ \bibinfo {pages} {816} (\bibinfo {year} {2013})}\BibitemShut {NoStop}%
\bibitem [{\citenamefont {Bergholtz}\ and\ \citenamefont {Liu}(2013)}]{bergholtzTOPOLOGICALFLATBAND2013}%
  \BibitemOpen
  \bibfield  {author} {\bibinfo {author} {\bibfnamefont {E.~J.}\ \bibnamefont {Bergholtz}}\ and\ \bibinfo {author} {\bibfnamefont {Z.}~\bibnamefont {Liu}},\ }\bibfield  {title} {\bibinfo {title} {Topological flat band models and fractional chern insulators},\ }\href {https://doi.org/10.1142/S021797921330017X} {\bibfield  {journal} {\bibinfo  {journal} {International Journal of Modern Physics B}\ }\textbf {\bibinfo {volume} {27}},\ \bibinfo {pages} {1330017} (\bibinfo {year} {2013})}\BibitemShut {NoStop}%
\bibitem [{\citenamefont {Liu}\ and\ \citenamefont {Bergholtz}(2022)}]{liuRecentDevelopmentsFractional2022}%
  \BibitemOpen
  \bibfield  {author} {\bibinfo {author} {\bibfnamefont {Z.}~\bibnamefont {Liu}}\ and\ \bibinfo {author} {\bibfnamefont {E.~J.}\ \bibnamefont {Bergholtz}},\ }\href {https://doi.org/10.48550/arXiv.2208.08449} {\bibinfo {title} {Recent {{Developments}} in {{Fractional Chern Insulators}}}} (\bibinfo {year} {2022}),\ \Eprint {https://arxiv.org/abs/2208.08449} {arxiv:2208.08449 [cond-mat, physics:math-ph, physics:quant-ph]} \BibitemShut {NoStop}%
\bibitem [{\citenamefont {Ledwith}\ \emph {et~al.}(2020{\natexlab{a}})\citenamefont {Ledwith}, \citenamefont {Tarnopolsky}, \citenamefont {Khalaf},\ and\ \citenamefont {Vishwanath}}]{ledwith2020fractional}%
  \BibitemOpen
  \bibfield  {author} {\bibinfo {author} {\bibfnamefont {P.~J.}\ \bibnamefont {Ledwith}}, \bibinfo {author} {\bibfnamefont {G.}~\bibnamefont {Tarnopolsky}}, \bibinfo {author} {\bibfnamefont {E.}~\bibnamefont {Khalaf}},\ and\ \bibinfo {author} {\bibfnamefont {A.}~\bibnamefont {Vishwanath}},\ }\bibfield  {title} {\bibinfo {title} {Fractional chern insulator states in twisted bilayer graphene: An analytical approach},\ }\href {https://doi.org/10.1103/PhysRevResearch.2.023237} {\bibfield  {journal} {\bibinfo  {journal} {Physical Review Research}\ }\textbf {\bibinfo {volume} {2}},\ \bibinfo {pages} {023237} (\bibinfo {year} {2020}{\natexlab{a}})}\BibitemShut {NoStop}%
\bibitem [{\citenamefont {Wang}\ \emph {et~al.}(2021)\citenamefont {Wang}, \citenamefont {Cano}, \citenamefont {Millis}, \citenamefont {Liu},\ and\ \citenamefont {Yang}}]{PhysRevLett.127.246403}%
  \BibitemOpen
  \bibfield  {author} {\bibinfo {author} {\bibfnamefont {J.}~\bibnamefont {Wang}}, \bibinfo {author} {\bibfnamefont {J.}~\bibnamefont {Cano}}, \bibinfo {author} {\bibfnamefont {A.~J.}\ \bibnamefont {Millis}}, \bibinfo {author} {\bibfnamefont {Z.}~\bibnamefont {Liu}},\ and\ \bibinfo {author} {\bibfnamefont {B.}~\bibnamefont {Yang}},\ }\bibfield  {title} {\bibinfo {title} {Exact landau level description of geometry and interaction in a flatband},\ }\href {https://doi.org/10.1103/PhysRevLett.127.246403} {\bibfield  {journal} {\bibinfo  {journal} {Phys. Rev. Lett.}\ }\textbf {\bibinfo {volume} {127}},\ \bibinfo {pages} {246403} (\bibinfo {year} {2021})}\BibitemShut {NoStop}%
\bibitem [{\citenamefont {Ledwith}\ \emph {et~al.}(2022)\citenamefont {Ledwith}, \citenamefont {Vishwanath},\ and\ \citenamefont {Khalaf}}]{PhysRevLett.128.176404}%
  \BibitemOpen
  \bibfield  {author} {\bibinfo {author} {\bibfnamefont {P.~J.}\ \bibnamefont {Ledwith}}, \bibinfo {author} {\bibfnamefont {A.}~\bibnamefont {Vishwanath}},\ and\ \bibinfo {author} {\bibfnamefont {E.}~\bibnamefont {Khalaf}},\ }\bibfield  {title} {\bibinfo {title} {Family of ideal chern flatbands with arbitrary chern number in chiral twisted graphene multilayers},\ }\href {https://doi.org/10.1103/PhysRevLett.128.176404} {\bibfield  {journal} {\bibinfo  {journal} {Phys. Rev. Lett.}\ }\textbf {\bibinfo {volume} {128}},\ \bibinfo {pages} {176404} (\bibinfo {year} {2022})}\BibitemShut {NoStop}%
\bibitem [{\citenamefont {Wang}\ and\ \citenamefont {Liu}(2022)}]{wang2022hierarchy}%
  \BibitemOpen
  \bibfield  {author} {\bibinfo {author} {\bibfnamefont {J.}~\bibnamefont {Wang}}\ and\ \bibinfo {author} {\bibfnamefont {Z.}~\bibnamefont {Liu}},\ }\bibfield  {title} {\bibinfo {title} {Hierarchy of ideal flatbands in chiral twisted multilayer graphene models},\ }\href {https://doi.org/10.1103/PhysRevLett.128.176403} {\bibfield  {journal} {\bibinfo  {journal} {Physical Review Letters}\ }\textbf {\bibinfo {volume} {128}},\ \bibinfo {pages} {176403} (\bibinfo {year} {2022})}\BibitemShut {NoStop}%
\bibitem [{\citenamefont {Ledwith}\ \emph {et~al.}(2023)\citenamefont {Ledwith}, \citenamefont {Vishwanath},\ and\ \citenamefont {Parker}}]{ledwith2023vortexability}%
  \BibitemOpen
  \bibfield  {author} {\bibinfo {author} {\bibfnamefont {P.~J.}\ \bibnamefont {Ledwith}}, \bibinfo {author} {\bibfnamefont {A.}~\bibnamefont {Vishwanath}},\ and\ \bibinfo {author} {\bibfnamefont {D.~E.}\ \bibnamefont {Parker}},\ }\bibfield  {title} {\bibinfo {title} {Vortexability: A unifying criterion for ideal fractional chern insulators},\ }\href {https://doi.org/10.1103/PhysRevB.108.205144} {\bibfield  {journal} {\bibinfo  {journal} {Physical Review B}\ }\textbf {\bibinfo {volume} {108}},\ \bibinfo {pages} {205144} (\bibinfo {year} {2023})}\BibitemShut {NoStop}%
\bibitem [{\citenamefont {Okuma}(2024)}]{Okuma}%
  \BibitemOpen
  \bibfield  {author} {\bibinfo {author} {\bibfnamefont {N.}~\bibnamefont {Okuma}},\ }\bibfield  {title} {\bibinfo {title} {Constructing vortex functions and basis states of chern insulators: Ideal condition, inequality from index theorem, and coherentlike states on the von neumann lattice},\ }\href {https://doi.org/10.1103/PhysRevB.110.245112} {\bibfield  {journal} {\bibinfo  {journal} {Phys. Rev. B}\ }\textbf {\bibinfo {volume} {110}},\ \bibinfo {pages} {245112} (\bibinfo {year} {2024})}\BibitemShut {NoStop}%
\bibitem [{\citenamefont {Estienne}\ \emph {et~al.}(2023)\citenamefont {Estienne}, \citenamefont {Regnault},\ and\ \citenamefont {Cr\'epel}}]{PhysRevResearch.5.L032048}%
  \BibitemOpen
  \bibfield  {author} {\bibinfo {author} {\bibfnamefont {B.}~\bibnamefont {Estienne}}, \bibinfo {author} {\bibfnamefont {N.}~\bibnamefont {Regnault}},\ and\ \bibinfo {author} {\bibfnamefont {V.}~\bibnamefont {Cr\'epel}},\ }\bibfield  {title} {\bibinfo {title} {Ideal chern bands as landau levels in curved space},\ }\href {https://doi.org/10.1103/PhysRevResearch.5.L032048} {\bibfield  {journal} {\bibinfo  {journal} {Phys. Rev. Res.}\ }\textbf {\bibinfo {volume} {5}},\ \bibinfo {pages} {L032048} (\bibinfo {year} {2023})}\BibitemShut {NoStop}%
\bibitem [{\citenamefont {Trugman}\ and\ \citenamefont {Kivelson}(1985)}]{TrugmanKivelson1985}%
  \BibitemOpen
  \bibfield  {author} {\bibinfo {author} {\bibfnamefont {S.~A.}\ \bibnamefont {Trugman}}\ and\ \bibinfo {author} {\bibfnamefont {S.}~\bibnamefont {Kivelson}},\ }\bibfield  {title} {\bibinfo {title} {Exact results for the fractional quantum hall effect with general interactions},\ }\href {https://doi.org/10.1103/PhysRevB.31.5280} {\bibfield  {journal} {\bibinfo  {journal} {Phys. Rev. B}\ }\textbf {\bibinfo {volume} {31}},\ \bibinfo {pages} {5280} (\bibinfo {year} {1985})}\BibitemShut {NoStop}%
\bibitem [{\citenamefont {Pokrovsky}\ and\ \citenamefont {Talapov}(1985)}]{VLPokrovsky_1985}%
  \BibitemOpen
  \bibfield  {author} {\bibinfo {author} {\bibfnamefont {V.~L.}\ \bibnamefont {Pokrovsky}}\ and\ \bibinfo {author} {\bibfnamefont {A.~L.}\ \bibnamefont {Talapov}},\ }\bibfield  {title} {\bibinfo {title} {A simple model for fractional hall effect},\ }\href {https://doi.org/10.1088/0022-3719/18/23/002} {\bibfield  {journal} {\bibinfo  {journal} {Journal of Physics C: Solid State Physics}\ }\textbf {\bibinfo {volume} {18}},\ \bibinfo {pages} {L691} (\bibinfo {year} {1985})}\BibitemShut {NoStop}%
\bibitem [{\citenamefont {Barkeshli}\ and\ \citenamefont {Qi}(2012)}]{barkeshliTopologicalNematicStates2012}%
  \BibitemOpen
  \bibfield  {author} {\bibinfo {author} {\bibfnamefont {M.}~\bibnamefont {Barkeshli}}\ and\ \bibinfo {author} {\bibfnamefont {X.-L.}\ \bibnamefont {Qi}},\ }\bibfield  {title} {\bibinfo {title} {Topological nematic states and non-abelian lattice dislocations},\ }\href {https://doi.org/10.1103/PhysRevX.2.031013} {\bibfield  {journal} {\bibinfo  {journal} {Physical Review X}\ }\textbf {\bibinfo {volume} {2}},\ \bibinfo {pages} {031013} (\bibinfo {year} {2012})}\BibitemShut {NoStop}%
\bibitem [{\citenamefont {Wu}\ \emph {et~al.}(2013)\citenamefont {Wu}, \citenamefont {Regnault},\ and\ \citenamefont {Bernevig}}]{wuBlochModelWave2013}%
  \BibitemOpen
  \bibfield  {author} {\bibinfo {author} {\bibfnamefont {Y.-L.}\ \bibnamefont {Wu}}, \bibinfo {author} {\bibfnamefont {N.}~\bibnamefont {Regnault}},\ and\ \bibinfo {author} {\bibfnamefont {B.~A.}\ \bibnamefont {Bernevig}},\ }\bibfield  {title} {\bibinfo {title} {Bloch model wave functions and pseudopotentials for all fractional chern insulators},\ }\href {https://doi.org/10.1103/PhysRevLett.110.106802} {\bibfield  {journal} {\bibinfo  {journal} {Physical Review Letters}\ }\textbf {\bibinfo {volume} {110}},\ \bibinfo {pages} {106802} (\bibinfo {year} {2013})}\BibitemShut {NoStop}%
\bibitem [{\citenamefont {Kumar}\ \emph {et~al.}(2014)\citenamefont {Kumar}, \citenamefont {Roy},\ and\ \citenamefont {Sondhi}}]{Kumar2014Generalizing}%
  \BibitemOpen
  \bibfield  {author} {\bibinfo {author} {\bibfnamefont {A.}~\bibnamefont {Kumar}}, \bibinfo {author} {\bibfnamefont {R.}~\bibnamefont {Roy}},\ and\ \bibinfo {author} {\bibfnamefont {S.~L.}\ \bibnamefont {Sondhi}},\ }\bibfield  {title} {\bibinfo {title} {Generalizing quantum hall ferromagnetism to fractional chern bands},\ }\href {https://doi.org/10.1103/PhysRevB.90.245106} {\bibfield  {journal} {\bibinfo  {journal} {Phys. Rev. B}\ }\textbf {\bibinfo {volume} {90}},\ \bibinfo {pages} {245106} (\bibinfo {year} {2014})}\BibitemShut {NoStop}%
\bibitem [{\citenamefont {Dong}\ \emph {et~al.}(2023)\citenamefont {Dong}, \citenamefont {Ledwith}, \citenamefont {Khalaf}, \citenamefont {Lee},\ and\ \citenamefont {Vishwanath}}]{PhysRevResearch.5.023166}%
  \BibitemOpen
  \bibfield  {author} {\bibinfo {author} {\bibfnamefont {J.}~\bibnamefont {Dong}}, \bibinfo {author} {\bibfnamefont {P.~J.}\ \bibnamefont {Ledwith}}, \bibinfo {author} {\bibfnamefont {E.}~\bibnamefont {Khalaf}}, \bibinfo {author} {\bibfnamefont {J.~Y.}\ \bibnamefont {Lee}},\ and\ \bibinfo {author} {\bibfnamefont {A.}~\bibnamefont {Vishwanath}},\ }\bibfield  {title} {\bibinfo {title} {Many-body ground states from decomposition of ideal higher chern bands: Applications to chirally twisted graphene multilayers},\ }\href {https://doi.org/10.1103/PhysRevResearch.5.023166} {\bibfield  {journal} {\bibinfo  {journal} {Phys. Rev. Res.}\ }\textbf {\bibinfo {volume} {5}},\ \bibinfo {pages} {023166} (\bibinfo {year} {2023})}\BibitemShut {NoStop}%
\bibitem [{\citenamefont {Wang}\ \emph {et~al.}(2023{\natexlab{b}})\citenamefont {Wang}, \citenamefont {Klevtsov},\ and\ \citenamefont {Liu}}]{PhysRevResearch.5.023167}%
  \BibitemOpen
  \bibfield  {author} {\bibinfo {author} {\bibfnamefont {J.}~\bibnamefont {Wang}}, \bibinfo {author} {\bibfnamefont {S.}~\bibnamefont {Klevtsov}},\ and\ \bibinfo {author} {\bibfnamefont {Z.}~\bibnamefont {Liu}},\ }\bibfield  {title} {\bibinfo {title} {Origin of model fractional chern insulators in all topological ideal flatbands: Explicit color-entangled wave function and exact density algebra},\ }\href {https://doi.org/10.1103/PhysRevResearch.5.023167} {\bibfield  {journal} {\bibinfo  {journal} {Phys. Rev. Res.}\ }\textbf {\bibinfo {volume} {5}},\ \bibinfo {pages} {023167} (\bibinfo {year} {2023}{\natexlab{b}})}\BibitemShut {NoStop}%
\bibitem [{\citenamefont {Liu}\ \emph {et~al.}(2021{\natexlab{b}})\citenamefont {Liu}, \citenamefont {Abouelkomsan},\ and\ \citenamefont {Bergholtz}}]{PhysRevLett.126.026801}%
  \BibitemOpen
  \bibfield  {author} {\bibinfo {author} {\bibfnamefont {Z.}~\bibnamefont {Liu}}, \bibinfo {author} {\bibfnamefont {A.}~\bibnamefont {Abouelkomsan}},\ and\ \bibinfo {author} {\bibfnamefont {E.~J.}\ \bibnamefont {Bergholtz}},\ }\bibfield  {title} {\bibinfo {title} {Gate-tunable fractional chern insulators in twisted double bilayer graphene},\ }\href {https://doi.org/10.1103/PhysRevLett.126.026801} {\bibfield  {journal} {\bibinfo  {journal} {Phys. Rev. Lett.}\ }\textbf {\bibinfo {volume} {126}},\ \bibinfo {pages} {026801} (\bibinfo {year} {2021}{\natexlab{b}})}\BibitemShut {NoStop}%
\bibitem [{\citenamefont {Eisenstein}\ \emph {et~al.}(1990)\citenamefont {Eisenstein}, \citenamefont {Stormer}, \citenamefont {Pfeiffer},\ and\ \citenamefont {West}}]{PhysRevB.41.7910}%
  \BibitemOpen
  \bibfield  {author} {\bibinfo {author} {\bibfnamefont {J.~P.}\ \bibnamefont {Eisenstein}}, \bibinfo {author} {\bibfnamefont {H.~L.}\ \bibnamefont {Stormer}}, \bibinfo {author} {\bibfnamefont {L.~N.}\ \bibnamefont {Pfeiffer}},\ and\ \bibinfo {author} {\bibfnamefont {K.~W.}\ \bibnamefont {West}},\ }\bibfield  {title} {\bibinfo {title} {Evidence for a spin transition in the ¥ensuremath{¥nu}=2/3 fractional quantum hall effect},\ }\href {https://doi.org/10.1103/PhysRevB.41.7910} {\bibfield  {journal} {\bibinfo  {journal} {Phys. Rev. B}\ }\textbf {\bibinfo {volume} {41}},\ \bibinfo {pages} {7910} (\bibinfo {year} {1990})}\BibitemShut {NoStop}%
\bibitem [{\citenamefont {Engel}\ \emph {et~al.}(1992)\citenamefont {Engel}, \citenamefont {Hwang}, \citenamefont {Sajoto}, \citenamefont {Tsui},\ and\ \citenamefont {Shayegan}}]{PhysRevB.45.3418}%
  \BibitemOpen
  \bibfield  {author} {\bibinfo {author} {\bibfnamefont {L.~W.}\ \bibnamefont {Engel}}, \bibinfo {author} {\bibfnamefont {S.~W.}\ \bibnamefont {Hwang}}, \bibinfo {author} {\bibfnamefont {T.}~\bibnamefont {Sajoto}}, \bibinfo {author} {\bibfnamefont {D.~C.}\ \bibnamefont {Tsui}},\ and\ \bibinfo {author} {\bibfnamefont {M.}~\bibnamefont {Shayegan}},\ }\bibfield  {title} {\bibinfo {title} {Fractional quantum hall effect at ¥ensuremath{¥nu}=2/3 and 3/5 in tilted magnetic fields},\ }\href {https://doi.org/10.1103/PhysRevB.45.3418} {\bibfield  {journal} {\bibinfo  {journal} {Phys. Rev. B}\ }\textbf {\bibinfo {volume} {45}},\ \bibinfo {pages} {3418} (\bibinfo {year} {1992})}\BibitemShut {NoStop}%
\bibitem [{\citenamefont {Wu}\ and\ \citenamefont {Jain}(1994)}]{PhysRevB.49.7515}%
  \BibitemOpen
  \bibfield  {author} {\bibinfo {author} {\bibfnamefont {X.~G.}\ \bibnamefont {Wu}}\ and\ \bibinfo {author} {\bibfnamefont {J.~K.}\ \bibnamefont {Jain}},\ }\bibfield  {title} {\bibinfo {title} {Fractional quantum hall states in the low-zeeman-energy limit},\ }\href {https://doi.org/10.1103/PhysRevB.49.7515} {\bibfield  {journal} {\bibinfo  {journal} {Phys. Rev. B}\ }\textbf {\bibinfo {volume} {49}},\ \bibinfo {pages} {7515} (\bibinfo {year} {1994})}\BibitemShut {NoStop}%
\bibitem [{\citenamefont {Verdene}\ \emph {et~al.}(2007)\citenamefont {Verdene}, \citenamefont {Martin}, \citenamefont {Gamez}, \citenamefont {Smet}, \citenamefont {Von~Klitzing}, \citenamefont {Mahalu}, \citenamefont {Schuh}, \citenamefont {Abstreiter},\ and\ \citenamefont {Yacoby}}]{verdene2007microscopic}%
  \BibitemOpen
  \bibfield  {author} {\bibinfo {author} {\bibfnamefont {B.}~\bibnamefont {Verdene}}, \bibinfo {author} {\bibfnamefont {J.}~\bibnamefont {Martin}}, \bibinfo {author} {\bibfnamefont {G.}~\bibnamefont {Gamez}}, \bibinfo {author} {\bibfnamefont {J.}~\bibnamefont {Smet}}, \bibinfo {author} {\bibfnamefont {K.}~\bibnamefont {Von~Klitzing}}, \bibinfo {author} {\bibfnamefont {D.}~\bibnamefont {Mahalu}}, \bibinfo {author} {\bibfnamefont {D.}~\bibnamefont {Schuh}}, \bibinfo {author} {\bibfnamefont {G.}~\bibnamefont {Abstreiter}},\ and\ \bibinfo {author} {\bibfnamefont {A.}~\bibnamefont {Yacoby}},\ }\bibfield  {title} {\bibinfo {title} {Microscopic manifestation of the spin phase transition at filling factor 2/3},\ }\href {https://doi.org/10.1038/nphys588} {\bibfield  {journal} {\bibinfo  {journal} {Nature Physics}\ }\textbf {\bibinfo {volume} {3}},\ \bibinfo {pages} {392} (\bibinfo {year} {2007})}\BibitemShut {NoStop}%
\bibitem [{\citenamefont {Wu}\ \emph {et~al.}(1993)\citenamefont {Wu}, \citenamefont {Dev},\ and\ \citenamefont {Jain}}]{PhysRevLett.71.153}%
  \BibitemOpen
  \bibfield  {author} {\bibinfo {author} {\bibfnamefont {X.~G.}\ \bibnamefont {Wu}}, \bibinfo {author} {\bibfnamefont {G.}~\bibnamefont {Dev}},\ and\ \bibinfo {author} {\bibfnamefont {J.~K.}\ \bibnamefont {Jain}},\ }\bibfield  {title} {\bibinfo {title} {Mixed-spin incompressible states in the fractional quantum hall effect},\ }\href {https://doi.org/10.1103/PhysRevLett.71.153} {\bibfield  {journal} {\bibinfo  {journal} {Phys. Rev. Lett.}\ }\textbf {\bibinfo {volume} {71}},\ \bibinfo {pages} {153} (\bibinfo {year} {1993})}\BibitemShut {NoStop}%
\bibitem [{\citenamefont {Kukushkin}\ \emph {et~al.}(1999)\citenamefont {Kukushkin}, \citenamefont {v.~Klitzing},\ and\ \citenamefont {Eberl}}]{PhysRevLett.82.3665}%
  \BibitemOpen
  \bibfield  {author} {\bibinfo {author} {\bibfnamefont {I.~V.}\ \bibnamefont {Kukushkin}}, \bibinfo {author} {\bibfnamefont {K.}~\bibnamefont {v.~Klitzing}},\ and\ \bibinfo {author} {\bibfnamefont {K.}~\bibnamefont {Eberl}},\ }\bibfield  {title} {\bibinfo {title} {Spin polarization of composite fermions: Measurements of the fermi energy},\ }\href {https://doi.org/10.1103/PhysRevLett.82.3665} {\bibfield  {journal} {\bibinfo  {journal} {Phys. Rev. Lett.}\ }\textbf {\bibinfo {volume} {82}},\ \bibinfo {pages} {3665} (\bibinfo {year} {1999})}\BibitemShut {NoStop}%
\bibitem [{\citenamefont {Tarnopolsky}\ \emph {et~al.}(2019{\natexlab{a}})\citenamefont {Tarnopolsky}, \citenamefont {Kruchkov},\ and\ \citenamefont {Vishwanath}}]{PhysRevLett.122.106405}%
  \BibitemOpen
  \bibfield  {author} {\bibinfo {author} {\bibfnamefont {G.}~\bibnamefont {Tarnopolsky}}, \bibinfo {author} {\bibfnamefont {A.~J.}\ \bibnamefont {Kruchkov}},\ and\ \bibinfo {author} {\bibfnamefont {A.}~\bibnamefont {Vishwanath}},\ }\bibfield  {title} {\bibinfo {title} {Origin of magic angles in twisted bilayer graphene},\ }\href {https://doi.org/10.1103/PhysRevLett.122.106405} {\bibfield  {journal} {\bibinfo  {journal} {Phys. Rev. Lett.}\ }\textbf {\bibinfo {volume} {122}},\ \bibinfo {pages} {106405} (\bibinfo {year} {2019}{\natexlab{a}})}\BibitemShut {NoStop}%
\bibitem [{\citenamefont {Ledwith}\ \emph {et~al.}(2020{\natexlab{b}})\citenamefont {Ledwith}, \citenamefont {Tarnopolsky}, \citenamefont {Khalaf},\ and\ \citenamefont {Vishwanath}}]{PhysRevResearch.2.023237}%
  \BibitemOpen
  \bibfield  {author} {\bibinfo {author} {\bibfnamefont {P.~J.}\ \bibnamefont {Ledwith}}, \bibinfo {author} {\bibfnamefont {G.}~\bibnamefont {Tarnopolsky}}, \bibinfo {author} {\bibfnamefont {E.}~\bibnamefont {Khalaf}},\ and\ \bibinfo {author} {\bibfnamefont {A.}~\bibnamefont {Vishwanath}},\ }\bibfield  {title} {\bibinfo {title} {Fractional chern insulator states in twisted bilayer graphene: An analytical approach},\ }\href {https://doi.org/10.1103/PhysRevResearch.2.023237} {\bibfield  {journal} {\bibinfo  {journal} {Phys. Rev. Res.}\ }\textbf {\bibinfo {volume} {2}},\ \bibinfo {pages} {023237} (\bibinfo {year} {2020}{\natexlab{b}})}\BibitemShut {NoStop}%
\bibitem [{\citenamefont {Varjas}\ \emph {et~al.}(2022)\citenamefont {Varjas}, \citenamefont {Abouelkomsan}, \citenamefont {Yang},\ and\ \citenamefont {Bergholtz}}]{varjas2022topological}%
  \BibitemOpen
  \bibfield  {author} {\bibinfo {author} {\bibfnamefont {D.}~\bibnamefont {Varjas}}, \bibinfo {author} {\bibfnamefont {A.}~\bibnamefont {Abouelkomsan}}, \bibinfo {author} {\bibfnamefont {K.}~\bibnamefont {Yang}},\ and\ \bibinfo {author} {\bibfnamefont {E.}~\bibnamefont {Bergholtz}},\ }\bibfield  {title} {\bibinfo {title} {Topological lattice models with constant berry curvature},\ }\href {https://scipost.org/10.21468/SciPostPhys.12.4.118} {\bibfield  {journal} {\bibinfo  {journal} {SciPost Physics}\ }\textbf {\bibinfo {volume} {12}},\ \bibinfo {pages} {118} (\bibinfo {year} {2022})}\BibitemShut {NoStop}%
\bibitem [{\citenamefont {Mera}\ and\ \citenamefont {Ozawa}(2021)}]{Mera2021b}%
  \BibitemOpen
  \bibfield  {author} {\bibinfo {author} {\bibfnamefont {B.}~\bibnamefont {Mera}}\ and\ \bibinfo {author} {\bibfnamefont {T.}~\bibnamefont {Ozawa}},\ }\bibfield  {title} {\bibinfo {title} {K\"ahler geometry and chern insulators: Relations between topology and the quantum metric},\ }\href {https://doi.org/10.1103/PhysRevB.104.045104} {\bibfield  {journal} {\bibinfo  {journal} {Phys. Rev. B}\ }\textbf {\bibinfo {volume} {104}},\ \bibinfo {pages} {045104} (\bibinfo {year} {2021})}\BibitemShut {NoStop}%
\bibitem [{\citenamefont {Fujimoto}\ \emph {et~al.}(2025)\citenamefont {Fujimoto}, \citenamefont {Parker}, \citenamefont {Dong}, \citenamefont {Khalaf}, \citenamefont {Vishwanath},\ and\ \citenamefont {Ledwith}}]{PhysRevLett.134.106502}%
  \BibitemOpen
  \bibfield  {author} {\bibinfo {author} {\bibfnamefont {M.}~\bibnamefont {Fujimoto}}, \bibinfo {author} {\bibfnamefont {D.~E.}\ \bibnamefont {Parker}}, \bibinfo {author} {\bibfnamefont {J.}~\bibnamefont {Dong}}, \bibinfo {author} {\bibfnamefont {E.}~\bibnamefont {Khalaf}}, \bibinfo {author} {\bibfnamefont {A.}~\bibnamefont {Vishwanath}},\ and\ \bibinfo {author} {\bibfnamefont {P.}~\bibnamefont {Ledwith}},\ }\bibfield  {title} {\bibinfo {title} {Higher vortexability: Zero-field realization of higher landau levels},\ }\href {https://doi.org/10.1103/PhysRevLett.134.106502} {\bibfield  {journal} {\bibinfo  {journal} {Phys. Rev. Lett.}\ }\textbf {\bibinfo {volume} {134}},\ \bibinfo {pages} {106502} (\bibinfo {year} {2025})}\BibitemShut {NoStop}%
\bibitem [{\citenamefont {Liu}\ \emph {et~al.}(2025)\citenamefont {Liu}, \citenamefont {Mera}, \citenamefont {Fujimoto}, \citenamefont {Ozawa},\ and\ \citenamefont {Wang}}]{1zg9-qbd6}%
  \BibitemOpen
  \bibfield  {author} {\bibinfo {author} {\bibfnamefont {Z.}~\bibnamefont {Liu}}, \bibinfo {author} {\bibfnamefont {B.}~\bibnamefont {Mera}}, \bibinfo {author} {\bibfnamefont {M.}~\bibnamefont {Fujimoto}}, \bibinfo {author} {\bibfnamefont {T.}~\bibnamefont {Ozawa}},\ and\ \bibinfo {author} {\bibfnamefont {J.}~\bibnamefont {Wang}},\ }\bibfield  {title} {\bibinfo {title} {Theory of generalized landau levels and its implications for non-abelian states},\ }\href {https://doi.org/10.1103/1zg9-qbd6} {\bibfield  {journal} {\bibinfo  {journal} {Phys. Rev. X}\ }\textbf {\bibinfo {volume} {15}},\ \bibinfo {pages} {031019} (\bibinfo {year} {2025})}\BibitemShut {NoStop}%
\bibitem [{\citenamefont {Repellin}\ \emph {et~al.}(2020)\citenamefont {Repellin}, \citenamefont {Dong}, \citenamefont {Zhang},\ and\ \citenamefont {Senthil}}]{Repellin19}%
  \BibitemOpen
  \bibfield  {author} {\bibinfo {author} {\bibfnamefont {C.}~\bibnamefont {Repellin}}, \bibinfo {author} {\bibfnamefont {Z.}~\bibnamefont {Dong}}, \bibinfo {author} {\bibfnamefont {Y.-H.}\ \bibnamefont {Zhang}},\ and\ \bibinfo {author} {\bibfnamefont {T.}~\bibnamefont {Senthil}},\ }\bibfield  {title} {\bibinfo {title} {Ferromagnetism in narrow bands of moir\'e superlattices},\ }\href {https://doi.org/10.1103/PhysRevLett.124.187601} {\bibfield  {journal} {\bibinfo  {journal} {Phys. Rev. Lett.}\ }\textbf {\bibinfo {volume} {124}},\ \bibinfo {pages} {187601} (\bibinfo {year} {2020})}\BibitemShut {NoStop}%
\bibitem [{\citenamefont {Lian}\ \emph {et~al.}(2021)\citenamefont {Lian}, \citenamefont {Song}, \citenamefont {Regnault}, \citenamefont {Efetov}, \citenamefont {Yazdani},\ and\ \citenamefont {Bernevig}}]{TBGIVGroundState}%
  \BibitemOpen
  \bibfield  {author} {\bibinfo {author} {\bibfnamefont {B.}~\bibnamefont {Lian}}, \bibinfo {author} {\bibfnamefont {Z.-D.}\ \bibnamefont {Song}}, \bibinfo {author} {\bibfnamefont {N.}~\bibnamefont {Regnault}}, \bibinfo {author} {\bibfnamefont {D.~K.}\ \bibnamefont {Efetov}}, \bibinfo {author} {\bibfnamefont {A.}~\bibnamefont {Yazdani}},\ and\ \bibinfo {author} {\bibfnamefont {B.~A.}\ \bibnamefont {Bernevig}},\ }\bibfield  {title} {\bibinfo {title} {Twisted bilayer graphene. iv. exact insulator ground states and phase diagram},\ }\href {https://doi.org/10.1103/PhysRevB.103.205414} {\bibfield  {journal} {\bibinfo  {journal} {Phys. Rev. B}\ }\textbf {\bibinfo {volume} {103}},\ \bibinfo {pages} {205414} (\bibinfo {year} {2021})}\BibitemShut {NoStop}%
\bibitem [{\citenamefont {Ledwith}\ \emph {et~al.}(2021{\natexlab{b}})\citenamefont {Ledwith}, \citenamefont {Khalaf},\ and\ \citenamefont {Vishwanath}}]{ledwith2021168646}%
  \BibitemOpen
  \bibfield  {author} {\bibinfo {author} {\bibfnamefont {P.~J.}\ \bibnamefont {Ledwith}}, \bibinfo {author} {\bibfnamefont {E.}~\bibnamefont {Khalaf}},\ and\ \bibinfo {author} {\bibfnamefont {A.}~\bibnamefont {Vishwanath}},\ }\bibfield  {title} {\bibinfo {title} {Strong coupling theory of magic-angle graphene: A pedagogical introduction},\ }\href {https://doi.org/https://doi.org/10.1016/j.aop.2021.168646} {\bibfield  {journal} {\bibinfo  {journal} {Annals of Physics}\ }\textbf {\bibinfo {volume} {435}},\ \bibinfo {pages} {168646} (\bibinfo {year} {2021}{\natexlab{b}})},\ \bibinfo {note} {special issue on Philip W. Anderson}\BibitemShut {NoStop}%
\bibitem [{\citenamefont {Lin}\ \emph {et~al.}(2025)\citenamefont {Lin}, \citenamefont {Yang}, \citenamefont {Lu}, \citenamefont {Zhai},\ and\ \citenamefont {Yao}}]{lin_fractional_2025}%
  \BibitemOpen
  \bibfield  {author} {\bibinfo {author} {\bibfnamefont {Z.}~\bibnamefont {Lin}}, \bibinfo {author} {\bibfnamefont {W.}~\bibnamefont {Yang}}, \bibinfo {author} {\bibfnamefont {H.}~\bibnamefont {Lu}}, \bibinfo {author} {\bibfnamefont {D.}~\bibnamefont {Zhai}},\ and\ \bibinfo {author} {\bibfnamefont {W.}~\bibnamefont {Yao}},\ }\bibfield  {title} {\bibinfo {title} {Fractional {Chern} insulator states in an isolated flat band of zero {Chern} number},\ }\href {http://arxiv.org/abs/2505.09009} {\bibfield  {journal} {\bibinfo  {journal} {arXiv:2505.09009}\ } (\bibinfo {year} {2025})}\BibitemShut {NoStop}%
\bibitem [{\citenamefont {Kwan}\ \emph {et~al.}(2024{\natexlab{c}})\citenamefont {Kwan}, \citenamefont {Wagner}, \citenamefont {Bultinck}, \citenamefont {Simon}, \citenamefont {Berg},\ and\ \citenamefont {Parameswaran}}]{kwanElectronphononCouplingCompeting2024}%
  \BibitemOpen
  \bibfield  {author} {\bibinfo {author} {\bibfnamefont {Y.~H.}\ \bibnamefont {Kwan}}, \bibinfo {author} {\bibfnamefont {G.}~\bibnamefont {Wagner}}, \bibinfo {author} {\bibfnamefont {N.}~\bibnamefont {Bultinck}}, \bibinfo {author} {\bibfnamefont {S.~H.}\ \bibnamefont {Simon}}, \bibinfo {author} {\bibfnamefont {E.}~\bibnamefont {Berg}},\ and\ \bibinfo {author} {\bibfnamefont {S.~A.}\ \bibnamefont {Parameswaran}},\ }\bibfield  {title} {\bibinfo {title} {Electron-phonon coupling and competing {{Kekul}}{\textbackslash}'e orders in twisted bilayer graphene},\ }\href {https://doi.org/10.1103/PhysRevB.110.085160} {\bibfield  {journal} {\bibinfo  {journal} {Physical Review B}\ }\textbf {\bibinfo {volume} {110}},\ \bibinfo {pages} {085160} (\bibinfo {year} {2024}{\natexlab{c}})}\BibitemShut {NoStop}%
\bibitem [{\citenamefont {Khalaf}\ \emph {et~al.}(2021)\citenamefont {Khalaf}, \citenamefont {Chatterjee}, \citenamefont {Bultinck}, \citenamefont {Zaletel},\ and\ \citenamefont {Vishwanath}}]{Khalaf_sciadv2021}%
  \BibitemOpen
  \bibfield  {author} {\bibinfo {author} {\bibfnamefont {E.}~\bibnamefont {Khalaf}}, \bibinfo {author} {\bibfnamefont {S.}~\bibnamefont {Chatterjee}}, \bibinfo {author} {\bibfnamefont {N.}~\bibnamefont {Bultinck}}, \bibinfo {author} {\bibfnamefont {M.~P.}\ \bibnamefont {Zaletel}},\ and\ \bibinfo {author} {\bibfnamefont {A.}~\bibnamefont {Vishwanath}},\ }\bibfield  {title} {\bibinfo {title} {Charged skyrmions and topological origin of superconductivity in magic-angle graphene},\ }\href {https://doi.org/10.1126/sciadv.abf5299} {\bibfield  {journal} {\bibinfo  {journal} {Science Advances}\ }\textbf {\bibinfo {volume} {7}},\ \bibinfo {pages} {eabf5299} (\bibinfo {year} {2021})}\BibitemShut {NoStop}%
\bibitem [{\citenamefont {Chatterjee}\ \emph {et~al.}(2020)\citenamefont {Chatterjee}, \citenamefont {Bultinck},\ and\ \citenamefont {Zaletel}}]{PhysRevB.101.165141}%
  \BibitemOpen
  \bibfield  {author} {\bibinfo {author} {\bibfnamefont {S.}~\bibnamefont {Chatterjee}}, \bibinfo {author} {\bibfnamefont {N.}~\bibnamefont {Bultinck}},\ and\ \bibinfo {author} {\bibfnamefont {M.~P.}\ \bibnamefont {Zaletel}},\ }\bibfield  {title} {\bibinfo {title} {Symmetry breaking and skyrmionic transport in twisted bilayer graphene},\ }\href {https://doi.org/10.1103/PhysRevB.101.165141} {\bibfield  {journal} {\bibinfo  {journal} {Phys. Rev. B}\ }\textbf {\bibinfo {volume} {101}},\ \bibinfo {pages} {165141} (\bibinfo {year} {2020})}\BibitemShut {NoStop}%
\bibitem [{\citenamefont {Chatterjee}\ \emph {et~al.}(2022)\citenamefont {Chatterjee}, \citenamefont {Ippoliti},\ and\ \citenamefont {Zaletel}}]{PhysRevB.106.035421}%
  \BibitemOpen
  \bibfield  {author} {\bibinfo {author} {\bibfnamefont {S.}~\bibnamefont {Chatterjee}}, \bibinfo {author} {\bibfnamefont {M.}~\bibnamefont {Ippoliti}},\ and\ \bibinfo {author} {\bibfnamefont {M.~P.}\ \bibnamefont {Zaletel}},\ }\bibfield  {title} {\bibinfo {title} {Skyrmion superconductivity: Dmrg evidence for a topological route to superconductivity},\ }\href {https://doi.org/10.1103/PhysRevB.106.035421} {\bibfield  {journal} {\bibinfo  {journal} {Phys. Rev. B}\ }\textbf {\bibinfo {volume} {106}},\ \bibinfo {pages} {035421} (\bibinfo {year} {2022})}\BibitemShut {NoStop}%
\bibitem [{\citenamefont {Christos}\ \emph {et~al.}(2020)\citenamefont {Christos}, \citenamefont {Sachdev},\ and\ \citenamefont {Scheurer}}]{Christos_pnas2020}%
  \BibitemOpen
  \bibfield  {author} {\bibinfo {author} {\bibfnamefont {M.}~\bibnamefont {Christos}}, \bibinfo {author} {\bibfnamefont {S.}~\bibnamefont {Sachdev}},\ and\ \bibinfo {author} {\bibfnamefont {M.~S.}\ \bibnamefont {Scheurer}},\ }\bibfield  {title} {\bibinfo {title} {Superconductivity, correlated insulators, and wess–zumino–witten terms in twisted bilayer graphene},\ }\href {https://doi.org/10.1073/pnas.2014691117} {\bibfield  {journal} {\bibinfo  {journal} {Proceedings of the National Academy of Sciences}\ }\textbf {\bibinfo {volume} {117}},\ \bibinfo {pages} {29543} (\bibinfo {year} {2020})}\BibitemShut {NoStop}%
\bibitem [{\citenamefont {Kwan}\ \emph {et~al.}(2022)\citenamefont {Kwan}, \citenamefont {Wagner}, \citenamefont {Bultinck}, \citenamefont {Simon},\ and\ \citenamefont {Parameswaran}}]{PhysRevX.12.031020}%
  \BibitemOpen
  \bibfield  {author} {\bibinfo {author} {\bibfnamefont {Y.~H.}\ \bibnamefont {Kwan}}, \bibinfo {author} {\bibfnamefont {G.}~\bibnamefont {Wagner}}, \bibinfo {author} {\bibfnamefont {N.}~\bibnamefont {Bultinck}}, \bibinfo {author} {\bibfnamefont {S.~H.}\ \bibnamefont {Simon}},\ and\ \bibinfo {author} {\bibfnamefont {S.~A.}\ \bibnamefont {Parameswaran}},\ }\bibfield  {title} {\bibinfo {title} {Skyrmions in twisted bilayer graphene: Stability, pairing, and crystallization},\ }\href {https://doi.org/10.1103/PhysRevX.12.031020} {\bibfield  {journal} {\bibinfo  {journal} {Phys. Rev. X}\ }\textbf {\bibinfo {volume} {12}},\ \bibinfo {pages} {031020} (\bibinfo {year} {2022})}\BibitemShut {NoStop}%
\bibitem [{\citenamefont {Khalaf}\ \emph {et~al.}(2022)\citenamefont {Khalaf}, \citenamefont {Ledwith},\ and\ \citenamefont {Vishwanath}}]{PhysRevB.105.224508}%
  \BibitemOpen
  \bibfield  {author} {\bibinfo {author} {\bibfnamefont {E.}~\bibnamefont {Khalaf}}, \bibinfo {author} {\bibfnamefont {P.}~\bibnamefont {Ledwith}},\ and\ \bibinfo {author} {\bibfnamefont {A.}~\bibnamefont {Vishwanath}},\ }\bibfield  {title} {\bibinfo {title} {Symmetry constraints on superconductivity in twisted bilayer graphene: Fractional vortices, $4e$ condensates, or nonunitary pairing},\ }\href {https://doi.org/10.1103/PhysRevB.105.224508} {\bibfield  {journal} {\bibinfo  {journal} {Phys. Rev. B}\ }\textbf {\bibinfo {volume} {105}},\ \bibinfo {pages} {224508} (\bibinfo {year} {2022})}\BibitemShut {NoStop}%
\bibitem [{\citenamefont {Sondhi}\ \emph {et~al.}(1993)\citenamefont {Sondhi}, \citenamefont {Karlhede}, \citenamefont {Kivelson},\ and\ \citenamefont {Rezayi}}]{sondhiSkyrmionsCrossoverInteger1993}%
  \BibitemOpen
  \bibfield  {author} {\bibinfo {author} {\bibfnamefont {S.~L.}\ \bibnamefont {Sondhi}}, \bibinfo {author} {\bibfnamefont {A.}~\bibnamefont {Karlhede}}, \bibinfo {author} {\bibfnamefont {S.~A.}\ \bibnamefont {Kivelson}},\ and\ \bibinfo {author} {\bibfnamefont {E.~H.}\ \bibnamefont {Rezayi}},\ }\bibfield  {title} {\bibinfo {title} {Skyrmions and the crossover from the integer to fractional quantum {{Hall}} effect at small {{Zeeman}} energies},\ }\href {https://doi.org/10.1103/PhysRevB.47.16419} {\bibfield  {journal} {\bibinfo  {journal} {Physical Review B}\ }\textbf {\bibinfo {volume} {47}},\ \bibinfo {pages} {16419} (\bibinfo {year} {1993})}\BibitemShut {NoStop}%
\bibitem [{\citenamefont {Moon}\ \emph {et~al.}(1995)\citenamefont {Moon}, \citenamefont {Mori}, \citenamefont {Yang}, \citenamefont {Girvin}, \citenamefont {MacDonald}, \citenamefont {Zheng}, \citenamefont {Yoshioka},\ and\ \citenamefont {Zhang}}]{moonSpontaneousInterlayerCoherence1995a}%
  \BibitemOpen
  \bibfield  {author} {\bibinfo {author} {\bibfnamefont {K.}~\bibnamefont {Moon}}, \bibinfo {author} {\bibfnamefont {H.}~\bibnamefont {Mori}}, \bibinfo {author} {\bibfnamefont {K.}~\bibnamefont {Yang}}, \bibinfo {author} {\bibfnamefont {S.~M.}\ \bibnamefont {Girvin}}, \bibinfo {author} {\bibfnamefont {A.~H.}\ \bibnamefont {MacDonald}}, \bibinfo {author} {\bibfnamefont {L.}~\bibnamefont {Zheng}}, \bibinfo {author} {\bibfnamefont {D.}~\bibnamefont {Yoshioka}},\ and\ \bibinfo {author} {\bibfnamefont {S.-C.}\ \bibnamefont {Zhang}},\ }\bibfield  {title} {\bibinfo {title} {Spontaneous interlayer coherence in double-layer quantum {{Hall}} systems: {{Charged}} vortices and {{Kosterlitz-Thouless}} phase transitions},\ }\href {https://doi.org/10.1103/PhysRevB.51.5138} {\bibfield  {journal} {\bibinfo  {journal} {Physical Review B}\ }\textbf {\bibinfo {volume} {51}},\ \bibinfo {pages} {5138} (\bibinfo {year} {1995})}\BibitemShut {NoStop}%
\bibitem [{\citenamefont {Sahay}\ \emph {et~al.}(2024)\citenamefont {Sahay}, \citenamefont {Divic}, \citenamefont {Parker}, \citenamefont {Soejima}, \citenamefont {Anand}, \citenamefont {Hauschild}, \citenamefont {Aidelsburger}, \citenamefont {Vishwanath}, \citenamefont {Chatterjee}, \citenamefont {Yao},\ and\ \citenamefont {Zaletel}}]{sahaySuperconductivityTopologicalLattice2024}%
  \BibitemOpen
  \bibfield  {author} {\bibinfo {author} {\bibfnamefont {R.}~\bibnamefont {Sahay}}, \bibinfo {author} {\bibfnamefont {S.}~\bibnamefont {Divic}}, \bibinfo {author} {\bibfnamefont {D.~E.}\ \bibnamefont {Parker}}, \bibinfo {author} {\bibfnamefont {T.}~\bibnamefont {Soejima}}, \bibinfo {author} {\bibfnamefont {S.}~\bibnamefont {Anand}}, \bibinfo {author} {\bibfnamefont {J.}~\bibnamefont {Hauschild}}, \bibinfo {author} {\bibfnamefont {M.}~\bibnamefont {Aidelsburger}}, \bibinfo {author} {\bibfnamefont {A.}~\bibnamefont {Vishwanath}}, \bibinfo {author} {\bibfnamefont {S.}~\bibnamefont {Chatterjee}}, \bibinfo {author} {\bibfnamefont {N.~Y.}\ \bibnamefont {Yao}},\ and\ \bibinfo {author} {\bibfnamefont {M.~P.}\ \bibnamefont {Zaletel}},\ }\bibfield  {title} {\bibinfo {title} {Superconductivity in a topological lattice model with strong repulsion},\ }\href {https://doi.org/10.1103/PhysRevB.110.195126} {\bibfield  {journal} {\bibinfo  {journal} {Physical Review B}\ }\textbf {\bibinfo {volume} {110}},\ \bibinfo {pages}
  {195126} (\bibinfo {year} {2024})}\BibitemShut {NoStop}%
\bibitem [{\citenamefont {Suzuura}\ and\ \citenamefont {Ando}(2002)}]{PhysRevB.65.235412}%
  \BibitemOpen
  \bibfield  {author} {\bibinfo {author} {\bibfnamefont {H.}~\bibnamefont {Suzuura}}\ and\ \bibinfo {author} {\bibfnamefont {T.}~\bibnamefont {Ando}},\ }\bibfield  {title} {\bibinfo {title} {Phonons and electron-phonon scattering in carbon nanotubes},\ }\href {https://doi.org/10.1103/PhysRevB.65.235412} {\bibfield  {journal} {\bibinfo  {journal} {Phys. Rev. B}\ }\textbf {\bibinfo {volume} {65}},\ \bibinfo {pages} {235412} (\bibinfo {year} {2002})}\BibitemShut {NoStop}%
\bibitem [{\citenamefont {San-Jose}\ \emph {et~al.}(2014)\citenamefont {San-Jose}, \citenamefont {Guti\'errez-Rubio}, \citenamefont {Sturla},\ and\ \citenamefont {Guinea}}]{PhysRevB.90.115152}%
  \BibitemOpen
  \bibfield  {author} {\bibinfo {author} {\bibfnamefont {P.}~\bibnamefont {San-Jose}}, \bibinfo {author} {\bibfnamefont {A.}~\bibnamefont {Guti\'errez-Rubio}}, \bibinfo {author} {\bibfnamefont {M.}~\bibnamefont {Sturla}},\ and\ \bibinfo {author} {\bibfnamefont {F.}~\bibnamefont {Guinea}},\ }\bibfield  {title} {\bibinfo {title} {Electronic structure of spontaneously strained graphene on hexagonal boron nitride},\ }\href {https://doi.org/10.1103/PhysRevB.90.115152} {\bibfield  {journal} {\bibinfo  {journal} {Phys. Rev. B}\ }\textbf {\bibinfo {volume} {90}},\ \bibinfo {pages} {115152} (\bibinfo {year} {2014})}\BibitemShut {NoStop}%
\bibitem [{\citenamefont {Tarnopolsky}\ \emph {et~al.}(2019{\natexlab{b}})\citenamefont {Tarnopolsky}, \citenamefont {Kruchkov},\ and\ \citenamefont {Vishwanath}}]{tarnopolskyorigin2019}%
  \BibitemOpen
  \bibfield  {author} {\bibinfo {author} {\bibfnamefont {G.}~\bibnamefont {Tarnopolsky}}, \bibinfo {author} {\bibfnamefont {A.~J.}\ \bibnamefont {Kruchkov}},\ and\ \bibinfo {author} {\bibfnamefont {A.}~\bibnamefont {Vishwanath}},\ }\bibfield  {title} {\bibinfo {title} {Origin of magic angles in twisted bilayer graphene},\ }\href {https://doi.org/10.1103/PhysRevLett.122.106405} {\bibfield  {journal} {\bibinfo  {journal} {Phys. Rev. Lett.}\ }\textbf {\bibinfo {volume} {122}},\ \bibinfo {pages} {106405} (\bibinfo {year} {2019}{\natexlab{b}})}\BibitemShut {NoStop}%
\bibitem [{\citenamefont {Becker}\ \emph {et~al.}(2021)\citenamefont {Becker}, \citenamefont {Embree}, \citenamefont {Wittsten},\ and\ \citenamefont {Zworski}}]{PhysRevB.103.165113}%
  \BibitemOpen
  \bibfield  {author} {\bibinfo {author} {\bibfnamefont {S.}~\bibnamefont {Becker}}, \bibinfo {author} {\bibfnamefont {M.}~\bibnamefont {Embree}}, \bibinfo {author} {\bibfnamefont {J.}~\bibnamefont {Wittsten}},\ and\ \bibinfo {author} {\bibfnamefont {M.}~\bibnamefont {Zworski}},\ }\bibfield  {title} {\bibinfo {title} {Spectral characterization of magic angles in twisted bilayer graphene},\ }\href {https://doi.org/10.1103/PhysRevB.103.165113} {\bibfield  {journal} {\bibinfo  {journal} {Phys. Rev. B}\ }\textbf {\bibinfo {volume} {103}},\ \bibinfo {pages} {165113} (\bibinfo {year} {2021})}\BibitemShut {NoStop}%
\bibitem [{\citenamefont {Sarkar}\ \emph {et~al.}(2024)\citenamefont {Sarkar}, \citenamefont {Wan}, \citenamefont {Zhang},\ and\ \citenamefont {Sun}}]{sarkar2024ideal}%
  \BibitemOpen
  \bibfield  {author} {\bibinfo {author} {\bibfnamefont {S.}~\bibnamefont {Sarkar}}, \bibinfo {author} {\bibfnamefont {X.}~\bibnamefont {Wan}}, \bibinfo {author} {\bibfnamefont {Y.}~\bibnamefont {Zhang}},\ and\ \bibinfo {author} {\bibfnamefont {K.}~\bibnamefont {Sun}},\ }\bibfield  {title} {\bibinfo {title} {Ideal topological flat bands in chiral symmetric moir$\backslash$'e systems from non-holomorphic functions},\ }\href {https://doi.org/10.48550/arXiv.2408.12555} {\bibfield  {journal} {\bibinfo  {journal} {arXiv preprint arXiv:2408.12555}\ } (\bibinfo {year} {2024})}\BibitemShut {NoStop}%
\end{thebibliography}%


\end{document}